\documentclass[a4paper]{article}
\pdfoutput=1
\usepackage[T1]{fontenc}
\usepackage{graphicx}
\usepackage{amsmath}
\usepackage{xspace}
\usepackage[numbers,sort&compress]{natbib}
\usepackage{subfig}
\usepackage[shortlabels]{enumitem}

\usepackage{pstricks}
\usepackage{graphicx}
\usepackage{xspace}

\usepackage{jheppub}
\hypersetup{
  pdfauthor={S. Kallweit, J. M. Lindert, P. Maierhoefer, S. Pozzorini, M. Schoenherr},
  pdftitle={NLO QCD+EW predictions for V+jets including off-shell vector-boson decays and multijet merging}
}

\newcommand{\Mt}{m_{\Pt}}

\newcommand{\GF}{{G_\mu}}

\newcommand{\mix}{\text{mix}\xspace}
\newcommand{\LO}{\text{LO}\xspace}
\newcommand{\NLO}{\text{NLO}\xspace}
\newcommand{\QED}{\text{QED}\xspace}
\newcommand{\QCD}{\text{QCD}\xspace}
\newcommand{\QCDpQED}{\text{\QCD{}+\QED}\xspace}
\newcommand{\EW}{\text{EW}\xspace}
\newcommand{\EWv}{\ensuremath{\EW_\text{virt}}\xspace}
\newcommand{\EWvirt}{\ensuremath{\EW_\text{virt}}\xspace}
\newcommand{\QCDmEW}{\text{\QCD--\EW}\xspace}
\newcommand{\QCDpEW}{\text{\QCD{}+\EW}\xspace}
\newcommand{\QCDpEWpLOmix}{\text{\QCD{}+\EW{}+\LO \mix}\xspace}
\newcommand{\QCDpEWvirt}{\text{\QCD{}+\EW{}$_\mathrm{virt}$}\xspace}

\newcommand{\QCDtEW}{\text{\QCD{}\ensuremath{\times}\EW}\xspace}

\newcommand{\Comix}{{\rmfamily\scshape Comix}\xspace}
\newcommand{\Amegic}{{\rmfamily\scshape Amegic}\xspace}
\newcommand{\Collier}{{\rmfamily\scshape Collier}\xspace}
\newcommand{\Munich}{{\rmfamily \scshape Munich}\xspace}
\newcommand{\Sherpa}{{\rmfamily\scshape Sherpa}\xspace}
\newcommand{\SherpaOpenLoops}{{\rmfamily\scshape Sherpa+OpenLoops}\xspace}
\newcommand{\OpenLoops}{{\rmfamily\scshape OpenLoops}\xspace}
\newcommand{\MunichOpenLoops}{{\rmfamily \scshape Munich+OpenLoops}\xspace}
\newcommand{\Rivet}{{\rmfamily\scshape Rivet}\xspace}
\newcommand{\LHAPDF}{{\rmfamily\scshape Lhapdf}\xspace}

\newcommand{\Pt}{\ensuremath{\mathrm{t}}\xspace}

\newcommand{\PV}{\ensuremath{V}\xspace}
\newcommand{\PW}{\ensuremath{\mathrm{W}}\xspace}
\newcommand{\PWp}{\ensuremath{\mathrm{W}^+}\xspace}
\newcommand{\PWm}{\ensuremath{\mathrm{W}^-}\xspace}
\newcommand{\PWpm}{\ensuremath{\mathrm{W}^\pm}\xspace}
\newcommand{\PZ}{\ensuremath{\mathrm{Z}}\xspace}
\newcommand{\PH}{\ensuremath{\mathrm{H}}\xspace}

\newcommand{\lmn}{\ensuremath{\ell^-\bar{\nu}_{\ell}}\xspace}
\newcommand{\lpn}{\ensuremath{\ell^+\nu_{\ell}}\xspace}
\newcommand{\lplm}{\ensuremath{\ell^+\ell^-}\xspace}
\newcommand{\nn}{\ensuremath{\nu_\ell\bar\nu_\ell}\xspace}

\newcommand{\jet}{\ensuremath{j}\xspace}

\newcommand{\mr}[1]{\mathrm{#1}}
\newcommand{\mc}[1]{\mathcal{#1}}
\newcommand{\ri}{\mathrm{i}}
\newcommand{\rF}{\mathrm{F}}
\newcommand{\rR}{\mathrm{R}}

\newcommand{\rT}{\mathrm{T}}

\newcommand{\rS}{\mathrm{S}}

\newcommand{\MW}{M_\mathrm{W}}
\newcommand{\MZ}{M_\mathrm{Z}}
\newcommand{\MH}{M_\mathrm{H}}

\newcommand{\muq}{\mu_{Q}}

\newcommand{\qcut}{Q_{\mathrm{cut}}}
\newcommand{\rratio}{r_{2/1}}
\newcommand{\rcut}{\rratio^{\mathrm{cut}}}

\newcommand{\nmax}{n_{\mathrm{max}}}
\newcommand{\core}{\text{core}}
\newcommand{\CKKW}{\text{CKKW}}
\newcommand{\MeV}{\text{MeV}\xspace}
\newcommand{\GeV}{\text{GeV}\xspace}
\newcommand{\TeV}{\text{TeV}\xspace}

\newcommand{\alphaS}{\alpha_{\rS}}
\newcommand{\ord}{\mathcal{O}}

\newcommand{\HTprimehat}{\hat{H}_{\mathrm{T}}'}

\newcommand{\HTtot}{H_{\mathrm{T}}^{\mathrm{tot}}}

\newcommand{\HTvis}{H_{\mathrm{T}}^{\mathrm{vis}}}

\newcommand{\kT}{k_{\mathrm{T}}}
\newcommand{\pT}{\ensuremath{p_{\mathrm{T}}}\xspace}

\newcommand{\pTji}{p_{\mathrm{T},\jet_i}}
\newcommand{\pTjone}{p_{\mathrm{T},\jet_1}}
\newcommand{\pTjtwo}{p_{\mathrm{T},\jet_2}}

\newcommand{\pTw}{p_{\mathrm{T,W}}}
\newcommand{\pTV}{p_{\mathrm{T,V}}}

\newcommand{\deltajj}{\Delta\phi_{\jet_1\jet_2}}
\newcommand{\mTW}{m^{\mathrm{W}}_{\mathrm{T}}}
\newcommand{\mll}{m_{\mathrm{\ell\ell}}}
\newcommand{\mjj}{m_{\mathrm{\jet_1\jet_2}}}
\newcommand{\missingET}{\ensuremath{\displaystyle{\not}E_{\rT}}\xspace}
\newcommand{\done}{\mr{d}}

\newcommand{\beqar}{\begin{eqnarray}}
\newcommand{\eeqar}{\end{eqnarray}}
\newcommand{\beq}{\begin{equation}}
\newcommand{\eeq}{\end{equation}}
\newcommand{\bit}{\begin{itemize}}
\newcommand{\eit}{\end{itemize}}

\def\refeq#1{\mbox{(\ref{#1})}}
\def\refeqs#1#2{\mbox{(\ref{#1})--(\ref{#2})}}
\def\reffi#1{\mbox{Fig.~\ref{#1}}}
\def\reffis#1#2{\mbox{Figures~\ref{#1}--\ref{#2}}}
\def\refta#1{\mbox{Table~\ref{#1}}}

\def\refse#1{\mbox{Section~\ref{#1}}}
\def\refses#1#2{\mbox{Sections~\ref{#1}--\ref{#2}}}
\def\refapp#1{\mbox{Appendix~\ref{#1}}}

\def\ie{i.e.\ }
\def\eg{e.g.\ }

\newcommand{\MCatNLO}{M\protect\scalebox{0.8}{C}@N\protect\scalebox{0.8}{LO}\xspace}
\newcommand{\SMCatNLO}{S--M\protect\scalebox{0.8}{C}@N\protect\scalebox{0.8}{LO}\xspace}

\newcommand{\MEPS}{M\scalebox{0.8}{E}P\scalebox{0.8}{S}\xspace}

\newcommand{\MEPSatLO}{M\protect\scalebox{0.8}{E}P\protect\scalebox{0.8}{S}@L\protect\scalebox{0.8}{O}\xspace}
\newcommand{\MEPSatNLO}{M\protect\scalebox{0.8}{E}P\protect\scalebox{0.8}{S}@N\protect\scalebox{0.8}{LO}\xspace}

\def\zgammathr{z_{\mathrm{thr}}}

\def\sw{s_{\mathrm{w}}}
\def\cw{c_{\mathrm{w}}}

\def\relplotwidth{0.47}
\def\relplotwidthsmall{0.47}

\preprint{
\begin{flushright}
DCPT/15/140 \\ FR-PHENO-2015-014 \\ IPPP/15/70  \\ MCNET-15-23 \\ MITP/15-108 \\  ZU-TH 41/15   
\end{flushright}
}

\author[a]{S.~Kallweit,}
\author[b]{J.~M.~Lindert,}
\author[c,d]{P.~Maierh\"ofer,}
\author[b]{S.~Pozzorini,}
\author[b]{and M.~Sch{\"o}nherr}

\affiliation[a]{Institut f\"ur Physik \& PRISMA Cluster of Excellence, 
Johannes Gutenberg Universit\"at, 55099 Mainz, Germany}
\affiliation[b]{Physik-Institut, Universit\"at Z\"urich,
Winterthurerstrasse 190, 
	CH-8057 Z\"urich,
	Switzerland }
\affiliation[c]{Institute for Particle Physics Phenomenology, 
           Durham University, 
           Durham DH1 3LE, 
           UK}
\affiliation[d]{
Physikalisches Institut, Albert-Ludwigs-Universit\"at Freiburg,
79104 Freiburg, Germany}

\emailAdd{kallweit@uni-mainz.de}
\emailAdd{lindert@physik.uzh.ch}
\emailAdd{philipp.maierhoefer@physik.uni-freiburg.de}
\emailAdd{pozzorin@physik.uzh.ch}
\emailAdd{marek.schoenherr@physik.uzh.ch}

\title{NLO \protect\QCDpEW predictions for 
$\boldsymbol{\PV+}\,$jets including
off-shell vector-boson decays and multijet merging}

\abstract{
We present next-to-leading order~(\NLO)  predictions including \QCD and electroweak (\EW) corrections
for the production and decay of off-shell electroweak vector bosons
in association with up to two jets at the 13\,TeV LHC.
All possible dilepton final states with zero, one or two charged leptons 
that can arise from off-shell \PW and \PZ bosons or photons
are considered.
All predictions are obtained using the automated implementation
of NLO \QCDpEW corrections in the 
\OpenLoops matrix-element generator combined with the \Munich and \Sherpa 
Monte Carlo frameworks. 
Electroweak corrections
play an especially important role in the context of BSM searches,
due to the presence of large EW Sudakov logarithms at the TeV scale.
In this kinematic regime,
important observables such as the jet transverse momentum or the total
transverse energy are strongly sensitive to multijet emissions. As a result,
fixed-order NLO \QCDpEW predictions
are plagued by huge \QCD corrections and poor theoretical precision.
To remedy this problem we present an approximate method that allows for a
simple and reliable implementation of NLO \EW corrections in the \MEPSatNLO
multijet merging framework.  Using this general approach we present an
inclusive simulation of vector-boson production in association with jets that
guarantees NLO \QCDpEW accuracy in all phase-space regions involving up to two
resolved jets.


}

\begin{document}
\maketitle
\flushbottom

\section{Introduction}

The production of electroweak (\EW) vector bosons in association with jets plays a key role 
in the physics programme of the  Large Hadron Collider (LHC).
Inclusive and differential measurements of vector-boson
plus multijet cross sections \cite{Aad:2010ab,Aad:2012en,Aad:2014qxa,
  Chatrchyan:2011ne,Chatrchyan:2012jra,Khachatryan:2014uva}  
can be performed for a wide range of jet multiplicities exploiting 
various clean final states that arise from the
leptonic decays of \PW and \PZ bosons or off-shell photons. This offers
unique opportunities to test the Standard Model at high precision and to
validate fundamental aspects of theoretical simulations at hadron colliders. Associated $\PV+\,$multijet production 
($\PV=\PW,\PZ$) 
represents also an important background to a large variety of 
analyses based on signatures with leptons, missing energy and jets.  In
particular, it is a prominent background in searches for physics beyond the
Standard Model (BSM) at the TeV scale.
 In this context, the availability of precise theoretical predictions 
for $\PV+\,$multijet production can play a critical role
for the sensitivity to new phenomena and for the interpretation of possible
discoveries.

Predictions  for 
$\PV+$\,multijet production at next-to-leading order (NLO) in \QCD~\cite{Arnold:1988dp,Arnold:1989ub,Campbell:2002tg,FebresCordero:2006sj,
Campbell:2008hh,Ellis:2011cr,Ita:2011hi,Berger:2009zg,Ellis:2009zw,
KeithEllis:2009bu,Berger:2009ep,Berger:2010zx,Bern:2013gka} are widely available,
and the precision of higher-order \QCD calculations has already
reached the next-to-next-to-leading order (NNLO) for $pp\to \PV+1$\,jet~\cite{Boughezal:2015dva,Ridder:2015dxa}. 
Also \EW corrections can play an important role. Their inclusion 
is mandatory for
any precision measurement. Moreover, \EW corrections are especially relevant 
at the TeV scale,
where large logarithms of Sudakov
type~\cite{Fadin:1999bq,Kuhn:1999nn,Denner:2000jv,Denner:2001gw,
Ciafaloni:2000df,Baur:2006sn,Mishra:2013una}
can lead to NLO \EW effects of tens of percent.
While NLO predictions for electroweak-boson production in association with a
single jet~\cite{
Kuhn:2004em,
Kuhn:2005az,
Kuhn:2005gv,
Kuhn:2007qc,
Kuhn:2007cv,
Denner:2009gj,Denner:2011vu,Denner:2012ts,Kallweit:2015fta}
have been available for a while,
thanks to the recent progress in NLO automation also
$\PV+\,$multijet calculations at NLO \EW became feasible.
In particular, various algorithms for the 
automated generation of one-loop scattering amplitudes
have proven to possess the degree of flexibility that is required 
in order to address NLO \EW calculations~\cite{Actis:2012qn,Actis:2013dfa,
Cascioli:2011va,Kallweit:2014xda,
Frixione:2014qaa,
Chiesa:2015mya}.
Predictions for vector-boson plus multijet production at NLO \EW are
motivated by the large impact of \EW Sudakov effects on BSM signatures with
multiple jets~\cite{Chiesa:2013yma} and, more generally, by the abundance of multijet emissions 
in $pp\to\PV+$\,jets at high energy.
First NLO \EW predictions for vector-boson production in association with 
more than one jet have been presented for 
$pp\to \lplm\jet\jet$~\cite{Denner:2014ina}
and for on-shell \PWp-boson production with 
up to three associated jets at NLO \QCDpEW~\cite{Kallweit:2014xda}.
Independent NLO \EW results for $pp\to\PW+2$\,jets have been reported in~\cite{Chiesa:2015mya}.

In this paper we present new NLO \QCDpEW results for $pp\to\PV+$\,jets that
involve up to two jets and cover all possible signatures resulting from
off-shell vector-boson decays into charged leptons or neutrinos, \ie we perform full
$2\to 3$ and $2\to 4$ calculations for  $pp\to\lpn+1,2$\,jets,
$pp\to\lmn+1,2$\,jets, $pp\to\lplm+1,2$\,jets and $pp\to\nn+1,2$\,jets.  
For convenience, the above mentioned processes will often be denoted as $\PV+$\,jet(s) production, 
while all results in this paper correspond to off-shell 
$\ell\ell/\ell\nu/\nu\nu+$jet(s) production.

Our predictions are obtained within the fully automated NLO \QCDpEW
framework~\cite{Kallweit:2014xda} provided by the
\OpenLoops~\cite{Cascioli:2011va,hepforge} generator in combination with the
\Munich~\cite{munich} and
\Sherpa~\cite{sherpaqedbrems,Gleisberg:2007md,Gleisberg:2008ta} Monte Carlo
programs. Off-shell effects in vector-boson decays are fully taken into account
thanks to a general implementation of the complex-mass
scheme~\cite{Denner:2005fg} at NLO \QCDpEW in \OpenLoops. This is
applicable to any process that involves the production and decay of intermediate
electroweak vector bosons, top quarks and Higgs bosons.

Higher-order calculations for $pp\to\PV+n$\,jets are obviously relevant for
physical observables that involve at least $n$ hard jets, but they can play
a very important role also for more inclusive observables where less
than $n$ hard jets are explicitly required.  Prominent examples are provided
by the inclusive distributions in the transverse momentum~(\pT) of the leading jet
and in the total transverse energy.  As is well known, the tails of such
distributions receive huge contributions from multijet emissions that
tend to saturate the recoil induced by the leading jet, while the 
vector boson remains relatively soft.  As a result, NLO \QCD predictions
for $pp\to\PV+1$\,jet at high jet \pT are plagued by giant
$K$-factors~\cite{Rubin:2010xp}, and their accuracy is effectively reduced
to leading order due to the dominance of $n$-jet final states with $n\ge 2$. 
In this situation it is clear that also NLO \EW parton-level results for
$pp\to\PV+1$\,jet are not applicable as they entirely miss the dominant
source of \EW higher-order effects, namely Sudakov-type \EW corrections to
$\PV+\,$multijet production.  
At fixed order in perturbation theory, the natural remedy 
would be given by $pp\to\PV+1$\,jet calculations with NNLO \QCD and mixed NNLO
\QCDmEW corrections. Very recently, mixed QCD--EW corrections of 
$\ord(\alphaS\alpha)$ to Drell--Yan processes in the resonance region have been presented in
\cite{Dittmaier:2014qza,Dittmaier:2015rxo}. 
However, a corresponding calculation for $pp\to\PV+1$\,jet is clearly out of reach. 
Thus, as a viable alternative, in this paper we will adopt the 
multijet merging approach at 
NLO~\cite{Hoeche:2012yf,Gehrmann:2012yg,Lonnblad:2012ix,Frederix:2012ps}, 
which allows one to  
combine NLO simulations of $pp\to\PV+0,1,\dots,n$\,jets matched to parton
showers in a way that guarantees parton-shower resummation and NLO accuracy
in all phase-space regions with up to $n$ resolved jets.  
While multijet merging
methods at NLO \QCD---and applications thereof to $\PV+\,$multijet
production~\cite{Hoeche:2012yf,Frederix:2015eii}---are already well
established, in this paper we address the
inclusion of NLO \EW corrections for the first time. 
To this end we exploit an approximate treatment of NLO \EW corrections, based on
exact virtual \EW contributions in combination with an appropriate cancellation
of infrared singularities. This allows us to implement NLO \EW effects in the
\MEPSatNLO multijet merging framework~\cite{Hoeche:2012yf} in a relatively
straightforward way. The proposed approach is completely general, and we
implemented it in \SherpaOpenLoops in a fully automated way. It is ideally
suited for processes and observables that receive large \EW Sudakov corrections
and involve sizable contributions from multijet emissions.

The paper is organised as follows. In~\refse{se:setup} we provide technical
aspects related to the employed tools and the setup of the calculation. Giant
$K$-factors for $\PV+1\,$jet production and related issues are recapitulated
in~\refse{se:vj}. In \refse{se:vjj} we present fixed-order NLO \QCDpEW
predictions for $pp\to\PV+2\,$jets including all channels with off-shell \PW or
$\PZ/\gamma^*$ decays to leptons and neutrinos. The merging of NLO \QCDpEW
predictions for processes with variable jet multiplicity is addressed
in~\refse{se:meps}, which starts with an illustration of NLO merging features
based on the exclusive-sums approach at parton level. In the following we
introduce and validate an approximation of NLO \EW corrections which is then
used in order to inject NLO \EW precision into the \MEPSatNLO framework. First
\MEPSatNLO predictions with NLO \QCDpEW accuracy are presented for
$pp\to\lmn+$\,jets including NLO matrix elements with up to two final-state
jets. Our conclusions can be found in \refse{se:conclusions}.

\section{Technical ingredients and setup of the simulations}
\label{se:setup}

This section deals with technical aspects of the simulations.
The reader might decide to skip it and to proceed directly to the
presentation of physics results in \refses{se:vj}{se:conclusions}.

\subsection{Considered processes and perturbative contributions}

\begin{figure*}[t]
\centering
\subfloat[][$\ord(\alphaS\alpha^2)$ LO]{
  \includegraphics[width=0.27\textwidth]{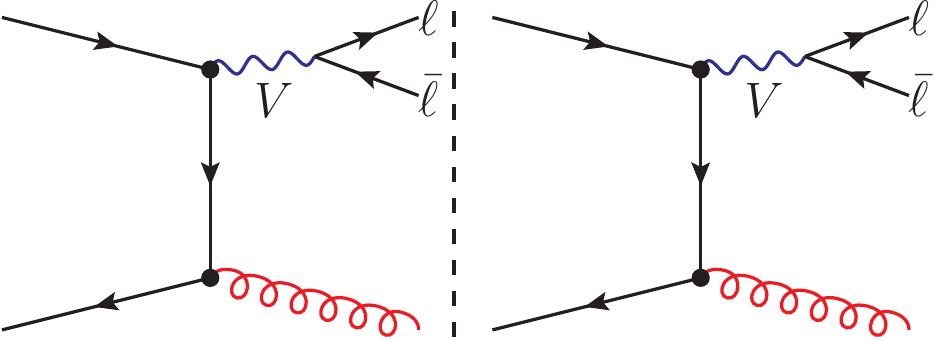}
  \label{fig:vjtreeQCD}
}\qquad
\caption{Representative LO contribution to $\PV+1$\,jet production.}
\label{fig:diag_vj_lo}
\end{figure*}

\begin{figure*}[t]
\centering
\subfloat[][Virtual $\ord(\alphaS\alpha^3)$ correction]{
  \includegraphics[width=0.27\textwidth]{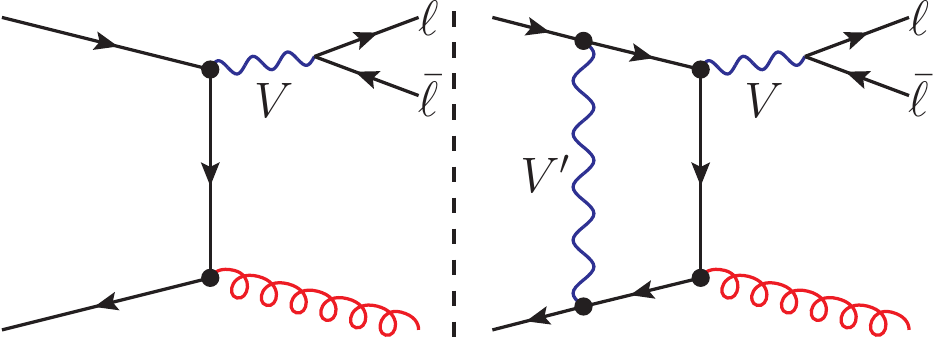}
  \label{fig:vj2qvirtEW}
}\qquad
\subfloat[][Real $\ord(\alphaS\alpha^3)$ correction]{
  \includegraphics[width=0.27\textwidth]{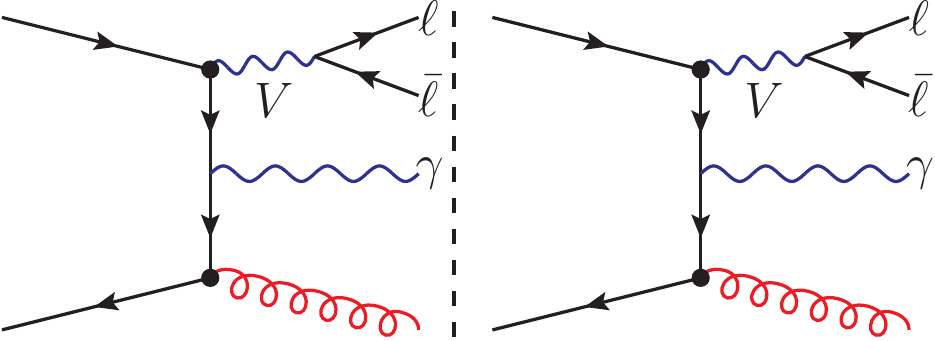}
  \label{fig:vj2qrealEW}
}\qquad
\subfloat[][Real  $\ord(\alphaS\alpha^3)$ correction]{
  \includegraphics[width=0.27\textwidth]{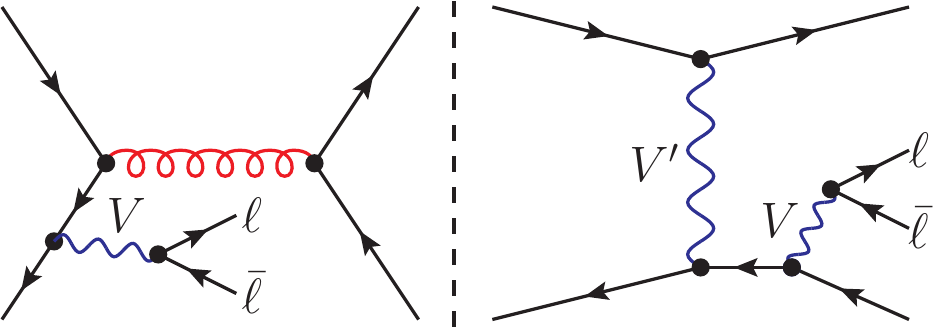}
  \label{fig:vj4qrealEW}
}
\caption{Representative virtual and real NLO \EW contributions to $\PV+1$\,jet production.}
\label{fig:diag_vj_nlo}
\end{figure*}

In this paper we study the production and 
decay of electroweak bosons ($\PV=\PWpm,\PZ/\gamma^*$)
in association with one and two jets at NLO \QCDpEW, including off-shell effects and taking 
into account all decay channels with leptons and neutrinos, \ie we address 
off-shell $2\to 3$ and $2\to 4$ processes with 
$\PWp\to\lpn$, $\PWm\to\lmn$, 
$\PZ/\gamma^*\to\lplm$ 
and $\PZ\to\nn$ final states in combination with jets.
In the case of charged leptons, only one generation is included, whereas for
invisible $\PZ$-boson decays all neutrino species ($\nu_e,\nu_\mu,\nu_\tau$)
are taken into account trivially. 

In general, NLO \QCD and \EW corrections have to be understood within a mixed
coupling expansion in $\alpha$ and $\alphaS$, where Born and one-loop scattering
amplitudes for a given process consist of towers of $\ord(\alphaS^N\alpha^{M})$
contributions with a fixed overall order $N+M$ that is distributed among \QCD
and \EW couplings in different possible $(N,M)$ combinations.

The production and off-shell decay of $\PV+1$\,jet involves a unique 
LO contribution of $\ord(\alphaS\alpha^2)$ 
and receives NLO \QCD corrections of
$\ord(\alphaS^2\alpha^2)$ and NLO \EW corrections of
$\ord(\alphaS\alpha^3)$.  Representative Feynman diagrams are illustrated in
Figs.~\ref{fig:diag_vj_lo} and \ref{fig:diag_vj_nlo}.
Here it is important to keep in mind a somewhat counter-intuitive feature of NLO
\EW corrections, namely that real emission at $\ord(\alphaS\alpha^3)$ does not
only involve photon bremsstrahlung (\reffi{fig:vj2qrealEW}) but also
$\PV+2$\,jet final states resulting from the emission of quarks through mixed
\QCDmEW interference terms (\reffi{fig:vj4qrealEW}).

\begin{figure*}[t]
\centering
\subfloat[][$\ord(\alphaS^2\alpha^2)$ LO]{
  \includegraphics[width=0.27\textwidth]{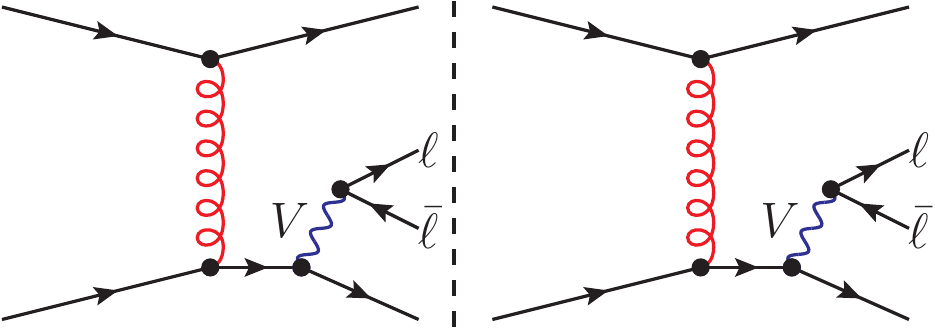}
  \label{fig:4qQCD_lo}
}\qquad
\subfloat[][$\ord(\alphaS\alpha^3)$ LO mix]{
  \includegraphics[width=0.27\textwidth]{diagrams/vjEWnlorealIII.pdf}
  \label{fig:4mix_lo}
}\qquad
\subfloat[][$\ord(\alpha^4)$ LO \EW]{
  \includegraphics[width=0.27\textwidth]{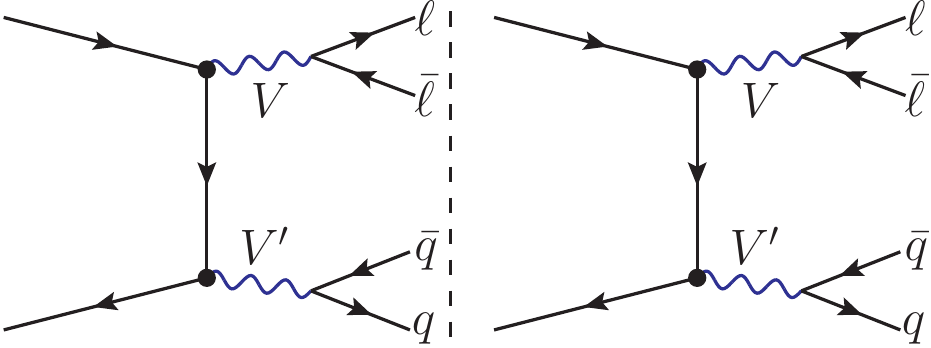}
  \label{fig:4qEW_lo}
}
\caption{Representative LO, LO mix and LO \EW contributions
to $\PV+2$\,jet production. 
}
\label{fig:diag_vjj_lo}
\end{figure*}

\begin{figure*}[t]
\centering
\subfloat[][Virtual $\ord(\alphaS^2\alpha^3)$ correction]{
  \includegraphics[width=0.27\textwidth]{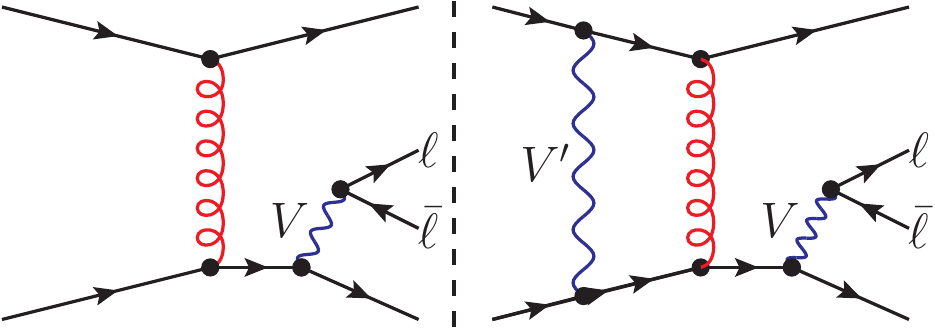}
  \label{fig:4qtreeQCD}
}\qquad
\subfloat[][Virtual $\ord(\alphaS^2\alpha^3)$ correction]{
  \includegraphics[width=0.27\textwidth]{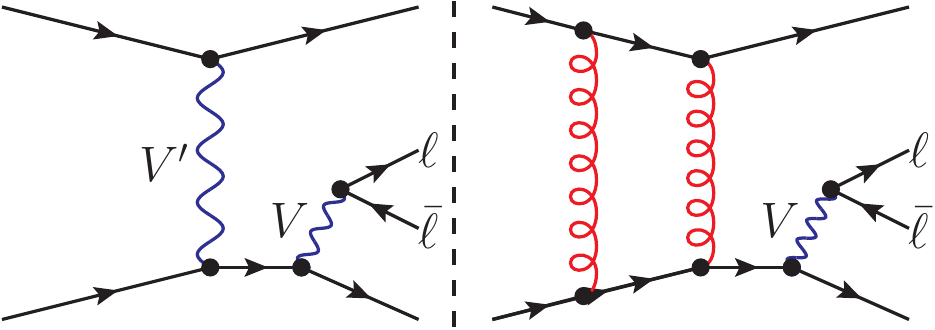}
  \label{fig:4qrealEW}
}\qquad
\subfloat[][Real  $\ord(\alphaS^2\alpha^3)$ correction]{
  \includegraphics[width=0.27\textwidth]{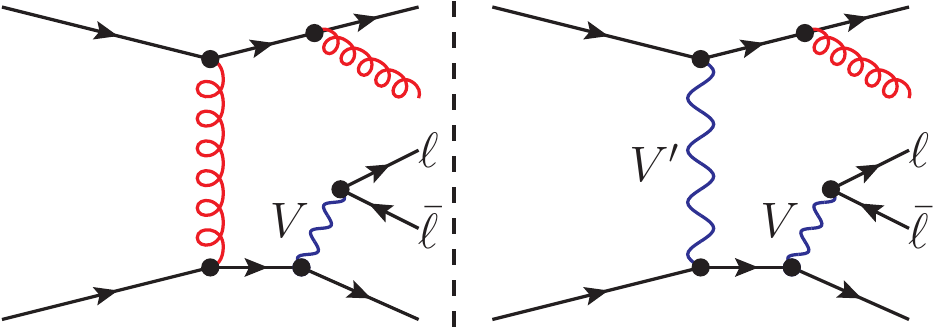}
  \label{fig:4qvirtEW}
}
\caption{Representative virtual and real NLO \EW contributions to $\PV+2$\,jet production.}
\label{fig:diag_vjj_nlo}
\end{figure*}

The LO production and off-shell decay of $\PV+2$\,jets receives
contributions from a tower of $\ord(\alphaS^k\alpha^{4-k})$ terms with powers 
$k=2,1,0$ in the strong coupling.
The contributions of 
$\ord(\alphaS^2\alpha^2)$, $\ord(\alphaS\alpha^3)$ and $\ord(\alpha^4)$ 
will be denoted as LO, LO mix and LO \EW, respectively. The 
two subleading orders contribute only via partonic channels with four 
external (anti)quark legs, and the LO \EW contribution includes, inter alia, 
the production of dibosons with semi-leptonic decays. Representative 
Feynman diagrams for $\PV+2$\,jet production are shown in Figs.~\ref{fig:diag_vjj_lo}
and \ref{fig:diag_vjj_nlo}. 
The NLO contributions of $\ord(\alphaS^3\alpha^2)$ and 
$\ord(\alphaS^2\alpha^3)$ are denoted as NLO \QCD and NLO \EW, respectively. 
They are the main subject of this paper, while subleading NLO
contributions of $\ord(\alphaS\alpha^4)$ or $\ord(\alpha^5)$ are not 
considered. Apart from the terminology, let us remind the reader that $\ord(\alphaS^2\alpha^3)$ NLO \EW 
contributions represent at the same time $\ord(\alpha)$ 
corrections with respect to LO and $\ord(\alphaS)$ corrections to 
LO mix contributions. 
Therefore, in order to cancel the $\ord(\alphaS^2\alpha^3)$
leading logarithmic  dependence on the renormalisation and factorization scales,
NLO \EW corrections should be combined with LO and LO mix terms.\footnote{
LO mix and NLO EW contributions are shown separately in the 
fixed-order analysis of \refse{se:vjj}, while 
in the merging framework of \refse{se:meps} they are systematically combined.}

For what concerns the combination of NLO \QCD and NLO \EW corrections,
\beqar
\sigma^{\NLO}_{\QCD} &=& 
\sigma^{\LO}+\delta\sigma^\NLO_\QCD,
\qquad\sigma^{\NLO}_{\EW} =
\sigma^{\LO}+\delta\sigma^\NLO_\EW,
\eeqar
as a default we adopt an additive prescription, 
\beqar
\sigma^{\NLO}_{\QCDpEW} &=& 
\sigma^{\LO}+\delta\sigma^\NLO_\QCD + \delta\sigma^\NLO_{\EW}.
\label{eq:qcdplusew}
\eeqar
Here, for the case of $\PV+n\,$jet production,
$\sigma^{\LO}$ is the $\ord(\alphaS^{n}\alpha^2)$ LO cross section, while
$\delta\sigma^\NLO_\QCD$ and
$\delta\sigma^\NLO_\EW$ correspond to the $\ord(\alphaS^{n+1}\alpha^2)$ and
$\ord(\alphaS^n\alpha^3)$ corrections, respectively. 
Alternatively, in order to identify potentially large effects due to the interplay of 
\EW and \QCD corrections beyond NLO, we present results considering the following factorised 
combination of \EW and \QCD corrections,
\beqar
\sigma^\NLO_\QCDtEW &=& 
\sigma^\NLO_\QCD\Biggl(1 + \frac{\delta\sigma^\NLO_{\EW}}{\sigma^\LO}\Biggr)
=
\sigma^\NLO_\EW\Biggl(1 + \frac{\delta\sigma^\NLO_{\QCD}}{\sigma^\LO}\Biggr)\,.
\label{eq:qcdtimesew}
\eeqar
In situations where the factorised approach can be justified by a 
clear separation of scales---such as where \QCD corrections are dominated by soft 
interactions well below the \EW scale---the factorised formula \refeq{eq:qcdtimesew} can 
be regarded as an improved prediction. However, in general, the difference between~\refeq{eq:qcdplusew} 
and~\refeq{eq:qcdtimesew} should be considered as an estimate of unknown 
higher-order corrections of \QCDmEW mixed type.

Subleading Born and photon-induced contributions of
$\ord(\alphaS^{n-1}\alpha^3)$ and $\ord(\alphaS^{n-2}\alpha^4)$ will also be
investigated and partly included in our predictions.

\subsection{Methods and tools}

Predictions presented in this paper have been obtained with the Monte Carlo
frameworks \MunichOpenLoops and \SherpaOpenLoops, which support in a fully
automated way NLO \QCDpEW simulations \cite{Kallweit:2014xda} at parton level
and particle level, respectively.
Virtual \QCD and \EW amplitudes are provided by the publicly\footnote{The
publicly available \OpenLoops amplitude library includes all relevant matrix
elements to compute NLO \QCD corrections, including colour- and
helicity-correlations and real radiation as well as loop-squared amplitudes,
for more than a hundred LHC processes.  Libraries containing NLO \EW
amplitudes will be provided soon.} available \OpenLoops
program~\cite{hepforge}, which is based on a fast numerical recursion for
the generation of one-loop scattering amplitudes~\cite{Cascioli:2011va}.
Combined with the \Collier tensor reduction library
\cite{Denner:2014gla}, which implements the Denner--Dittmaier reduction
techniques~\cite{Denner:2002ii,Denner:2005nn} and the scalar integrals
of~\cite{Denner:2010tr}, the employed recursion permits to achieve very high
CPU performance and a high degree of numerical stability.  A sophisticated
stability system is in place to rescue potential unstable phase-space points
via a re-evaluation at quadrupole precision using~{\sc
CutTools}~\cite{Ossola:2007ax}, which implements the OPP
method~\cite{Ossola:2006us}, together with the {\sc OneLOop}
library~\cite{vanHameren:2010cp}.  
As anticipated in the introduction, in order to address the production and decay of
unstable particles, the original automation 
of one-loop \EW corrections in \OpenLoops~\cite{Kallweit:2014xda} was supplemented 
by a fully general implementation of the complex-mass scheme~\cite{Denner:2005fg}.

All remaining tasks, \ie the bookkeeping of
partonic subprocesses, phase-space integration and the subtraction of \QCD
and \QED bremsstrahlung are supported by the two independent and fully
automated Monte Carlo generators \Munich~\cite{munich} and
\Sherpa~\cite{sherpaqedbrems,Gleisberg:2007md,Gleisberg:2008ta}.  The first
one, \Munich, is a fully generic and very fast parton-level Monte Carlo
integrator, which has been used, mainly in combination with \OpenLoops,
for various pioneering NLO multi-leg
\cite{Denner:2010jp, Denner:2012yc, Denner:2012dz, Cascioli:2013wga} and NNLO
applications~\cite{Grazzini:2013bna,Cascioli:2014yka,Gehrmann:2014fva,
Grazzini:2015nwa,Grazzini:2015wpa,Grazzini:2015hta}.  
\Sherpa is a particle-level Monte Carlo generator
providing all stages of hadron collider simulations,
including parton showering, hadronisation and underlying event simulations.  It was used in
the pioneering NLO \QCD calculations of vector-boson plus multijet
production~\cite{Berger:2009zg,Ellis:2009zw,KeithEllis:2009bu,
Berger:2009ep,Berger:2010zx,Bern:2013gka}, as well as for their matching to
the parton shower~\cite{Hoeche:2012ft} and the merging of multijet final
states at NLO~\cite{Hoeche:2012yf}.  
For tree amplitudes,
with all relevant colour and helicity correlations,
\Munich relies on \OpenLoops, while \Sherpa  generates them internally with
\Amegic \cite{Krauss:2001iv} and \Comix \cite{Gleisberg:2008fv}.
For the cancellation of infrared singularities both Monte Carlo tools,
\Munich and \Sherpa, employ the dipole subtraction
scheme~\cite{Catani:1996vz,Catani:2002hc}.  Both codes were extensively
checked against each other, and sub-permille level agreement was found.

\subsection{Physics objects and selection cuts}
For the definition of jets we employ the anti-$\kT$ algorithm
\cite{Cacciari:2008gp} with $R=0.4$.  More precisely, in order to guarantee
infrared safeness in presence of NLO \QCD and \EW corrections, we adopt a
democratic clustering
approach~\cite{Glover:1993xc,GehrmannDeRidder:1997wx,GehrmannDeRidder:1998ba},
where \QCD partons and photons are recombined. 
In order to ensure the cancellation of collinear singularities that arise from
collinear photon emissions off charged leptons and quarks, collinear pairs of
photons and charged fermions with $\Delta R_{\gamma f}<0.1$ are recombined via
four-momentum addition, and all observables are defined in terms of such dressed
fermions. Fermion dressing is applied prior to the jet algorithm, and photons
that have been recombined with leptons, as well as (dressed) leptons, are not
subject to jet clustering.

After jet clustering \QCD jets are separated from photons by imposing an upper
bound $\zgammathr=0.5$ to the photon energy fraction inside jets. In this case,
the cut $\zgammathr<0.5$ is applied only to photons that are inside the jet, but
outside the technical recombination cone with $\Delta R_{\gamma q}<0.1$. The
recombination of (anti)quark--photon pairs with $\Delta R_{\gamma q}<0.1$
represents a technical regularisation prescription to ensure the cancellation of
collinear photon--quark singularities. As demonstrated
in~\cite{Kallweit:2014xda}, this provides an excellent approximation to a more
rigorous approach for the cancellation of collinear singularities based on
fragmentation functions.

For the selection of signatures of type $\ell\ell/\ell\nu/\nu\nu\,+1,2$\,jets,
which result from the various vector-boson decays, we apply the
leptonic cuts listed in \refta{Tab:cuts}. 
They correspond to the ATLAS analysis of~\cite{Aad:2014rta}.

\begin{table}
\begin{center}
\begin{tabular}{|l@{\hspace{-2mm}}ll||c|c|c|}
\hline
                     && &  $\PWpm\to\lpn,\lmn$ & $\PZ\to\lplm$  & $\PZ\to\nn$ 
\\\hline
$\ell$        &&$\in$ &  $e,\mu$              & $e,\mu$        &  $e,\mu,\tau$ 
\\
$p_{\rT,\ell^\pm}$&[GeV]    &>&   25                  & 25         &                  
\\
$\missingET$&[GeV]       &>&   25                  &            & 25     
\\
$\mTW$&[GeV]            &>&   40                  &            &
\\
$m_{\lplm}$&[GeV]      &$\in$&                      & [66, 116]       &
\\
$|\eta_{\ell^\pm}|$&     &<&   2.5                 & 2.5        &
\\
$\Delta R_{\ell^\pm j}$&    &>&   0.5             & 0.5        &
\\
$\Delta R_{\ell^+\ell^-}$&    &>&                   & 0.2        &
\\[2mm]\hline
\end{tabular}
\caption{Selection cuts for the various $\PV+$\,jets production and decay processes.
The missing transverse energy $\missingET$ is calculated from the vector sum of 
neutrino momenta, and the $\PW$-boson transverse mass is defined as $\mTW=\sqrt{2p_{\rT,\ell} p_{\rT,\nu}
(1-\cos\Delta\phi_{\ell\nu}) }$. 
The lepton--jet separation cut, $\Delta R_{\ell^\pm j}>0.5$ is applied to all 
jets in the region \refeq{eq:JETcuts}.
}
\label{Tab:cuts}
\end{center}
\end{table}
Events will be categorised according to the number of anti-$\kT$ jets with $R=0.4$ in the
transverse-momentum and pseudo-rapidity region
\beqar\label{eq:JETcuts}
p_{\rT,\jet}&>&30\,\GeV, \qquad {|\eta_{\jet}|  < 4.5}.
\eeqar
Additionally, for certain observables we present results vetoing a second jet with details explained in the text.

\subsection{Input parameters, scale choices and variations}

As input parameters to simulate $pp\to\ell\ell/\ell\nu/\nu\nu\,+$\,jets
at NLO \QCDpEW we use the 
gauge-boson masses and widths~\cite{Agashe:2014kda}
\beqar\label{eq:massesew}
\MZ=91.1876~\GeV,\quad
\MW=80.385~\GeV,\quad
\Gamma_\PZ=2.4955~\GeV,\quad
\Gamma_\PW=2.0897~\GeV.
\eeqar
The latter are obtained from state-of-the art theoretical calculations.
For the top quark we use the mass reported in~\cite{Agashe:2014kda}, 
and we compute the width at NLO \QCD,
\beqar\label{eq:massestop}
\Mt=173.2~\GeV,\quad
\Gamma_\Pt=1.339~\GeV.
\eeqar
For the Higgs-boson mass and width~\cite{Heinemeyer:2013tqa} we use
\beqar\label{eq:masseshiggs}
\MH=125~\GeV,\quad
\Gamma_\PH=4.07~\MeV.
\eeqar
Electroweak contributions to $pp\to\PV+2$\,jets involve topologies with
$s$-channel top-quark and Higgs propagators that require a finite top and
Higgs width.  However, at the perturbative order considered in this paper,
such topologies arise only in interference terms that do not give rise to
Breit--Wigner resonances.  The dependence of our results on $\Gamma_t$ and $\Gamma_H$
is thus completely negligible.

All unstable particles are treated in the complex-mass scheme~\cite{Denner:2005fg},
where width effects are absorbed into the complex-valued renormalised masses
\beqar\label{eq:complexmasses}
\mu^2_i=M_i^2-\ri\Gamma_iM_i \qquad\mbox{for}\;i=\PW,\PZ,\Pt,\PH.
\eeqar 
The electroweak couplings are derived from the gauge-boson masses and the Fermi constant,
$\GF=1.16637\times10^{-5}~\GeV^{-2}$, using 
\beq\label{eq:defalpha}
\alpha=\left|\frac{\sqrt{2}\sw^2\mu^2_\PW G_\mu}{\pi}\right|,
\eeq
where the $\PW$-boson mass and the squared sine of the mixing angle, 
\beq\label{eq:defsintheta}
\sw^2=1-\cw^2=1-\frac{\mu_\PW^2}{\mu_\PZ^2},
\eeq
are complex-valued.
The $G_\mu$-scheme guarantees an  optimal description of pure SU(2) interactions
at the electroweak scale. It is the scheme of choice for
$\PW+$\,jets production, and it provides a very decent description of
$\PZ\,+$\,jets production as well. 

The CKM matrix is assumed to be diagonal, while colour effects and related
interferences are included throughout, without applying any large-$N_c$
expansion.

For the calculation of hadron-level cross sections we employ the NNPDF2.3
\QED parton distributions~\cite{Ball:2013hta} which include NLO \QCD and LO
\QED effects, and we use the PDF set corresponding to
$\alphaS(\MZ)=0.118$.\footnote{To be precise we use the
\texttt{NNPDF23\_nlo\_as\_0118\_qed} set interfaced through the \LHAPDF library 5.9.1 
(\Munich) and 6.1.5 (\Sherpa) \cite{Buckley:2014ana}.}
Matrix elements are
evaluated using the running strong coupling supported by the PDFs, and,
consistently with the variable flavour-number scheme implemented in the
NNPDFs, at the top threshold we switch from five to six active quark
flavours in the renormalisation of $\alphaS$.  All light quarks, including
bottom quarks, are treated as massless particles, and top-quark loops are included
throughout in the calculation.  The NLO PDF set is used
for LO as well as for NLO \QCD and NLO \EW predictions.

In all fixed-order results the renormalisation scale $\mu_R$ and factorisation scale $\mu_F$ are set to
\beq\label{eq:RFscales} 
\mu_{\rR,\rF}=\xi_{\rR,\rF}\mu_0,
\quad\mbox{with}\quad 
\mu_0= \HTprimehat/2
\quad\mbox{and}\quad 
\frac{1}{2}\le \xi_{\rR},\xi_{\rF}\le 2,
\eeq 
where $\HTprimehat$ is the scalar sum of the transverse energy of all parton-level
final-state objects,
\beq\label{eq:BHscale} 
\HTprimehat = \sum_{i\in \{\mathrm{quarks,gluons}\}} p_{\rT,i} + p_{\rT,\gamma} +
E_{\rT,\PV}\,. 
\eeq
Also \QCD partons and photons that are radiated at NLO are included in $\HTprimehat$,
and the vector-boson transverse energy, $E_{\rT,\PV}$,  is computed using the
total (off-shell) four-momentum of the corresponding decay products, \ie
\beq\label{ETboson}
E^2_{\rT,\PZ}=p^2_{\rT,\ell\ell}+m_{\ell\ell}^2,\qquad
E^2_{\rT,\PW}=p^2_{\rT,\ell\nu}+m_{\ell\nu}^2\,.
\eeq
In order to guarantee infrared safeness at NLO \EW, the scale
\refeq{eq:BHscale} must be insensitive to collinear photon emissions off
quarks and leptons.  To this end, all terms in 
\refeqs{eq:BHscale}{ETboson} are computed in terms of dressed leptons and quarks, while the
$p_{\rT,\gamma}$ term in \refeq{eq:BHscale} involves only photons that have not been recombined
with charged fermions.

Our default scale choice corresponds to $\xi_{\rR}=\xi_{\rF}=1$,
and theoretical fixed-order uncertainties are assessed by
applying the scale variations $(\xi_\rR,\xi_\rF)=(2,2)$,
$(2,1)$, $(1,2)$, $(1,1)$, $(1,0.5)$, $(0.5,1)$, $(0.5,0.5)$, while 
theoretical uncertainties of our \MEPS predictions are assessed by 
applying the scale variations $(\xi_\rR,\xi_\rF)=(2,2)$, $(1,1)$, $(0.5,0.5)$.
As shown in
\cite{Berger:2009zg,Ellis:2009zw,KeithEllis:2009bu,Berger:2009ep,
Berger:2010zx, Bern:2013gka} the scale choice~\refeq{eq:RFscales} guarantees
a good perturbative convergence for $\PV+$\,multijet production over a wide
range of observables and energy scales.

\section{Giant \texorpdfstring{$\boldsymbol{K}$}{K}-factors and electroweak corrections for \texorpdfstring{$\boldsymbol{\PV+1}$}{V+1} jet production}
\label{se:vj}

In this section we start our discussion of $\PV+\,$jets production at NLO
\QCDpEW by recalling some pathological features of fixed-order 
calculations for $pp\to\PV+1$\,jet.  Such observations will provide the main
motivation for the multiparticle calculations and the multijet merging
approach presented in Sections~\ref{se:vjj} and \ref{se:meps}.

It is well known that NLO \QCD predictions for $\PV+1$\,jet
production~\cite{Arnold:1988dp,Arnold:1989ub,Campbell:2002tg,Denner:2009gj}
suffer from a very poor convergence of the perturbative expansion, which
manifests itself in the form of giant $K$-factors~\cite{Rubin:2010xp} at large
jet transverse momenta. In this kinematic regime the NLO QCD cross section is
dominated by dijet configurations where the hardest jet recoils against a
similarly hard second jet, while the vector boson remains relatively soft. Such
bremsstrahlung configurations are effectively described at LO, with
correspondingly large scale uncertainties. Moreover, in this situation NLO \EW
calculations for $pp\to\PV+1$\,jet are meaningless, as they completely miss \EW
correction effects for the dominating dijet configurations.

\begin{figure*}[t]
\centering
   \includegraphics[width=\relplotwidth\textwidth]{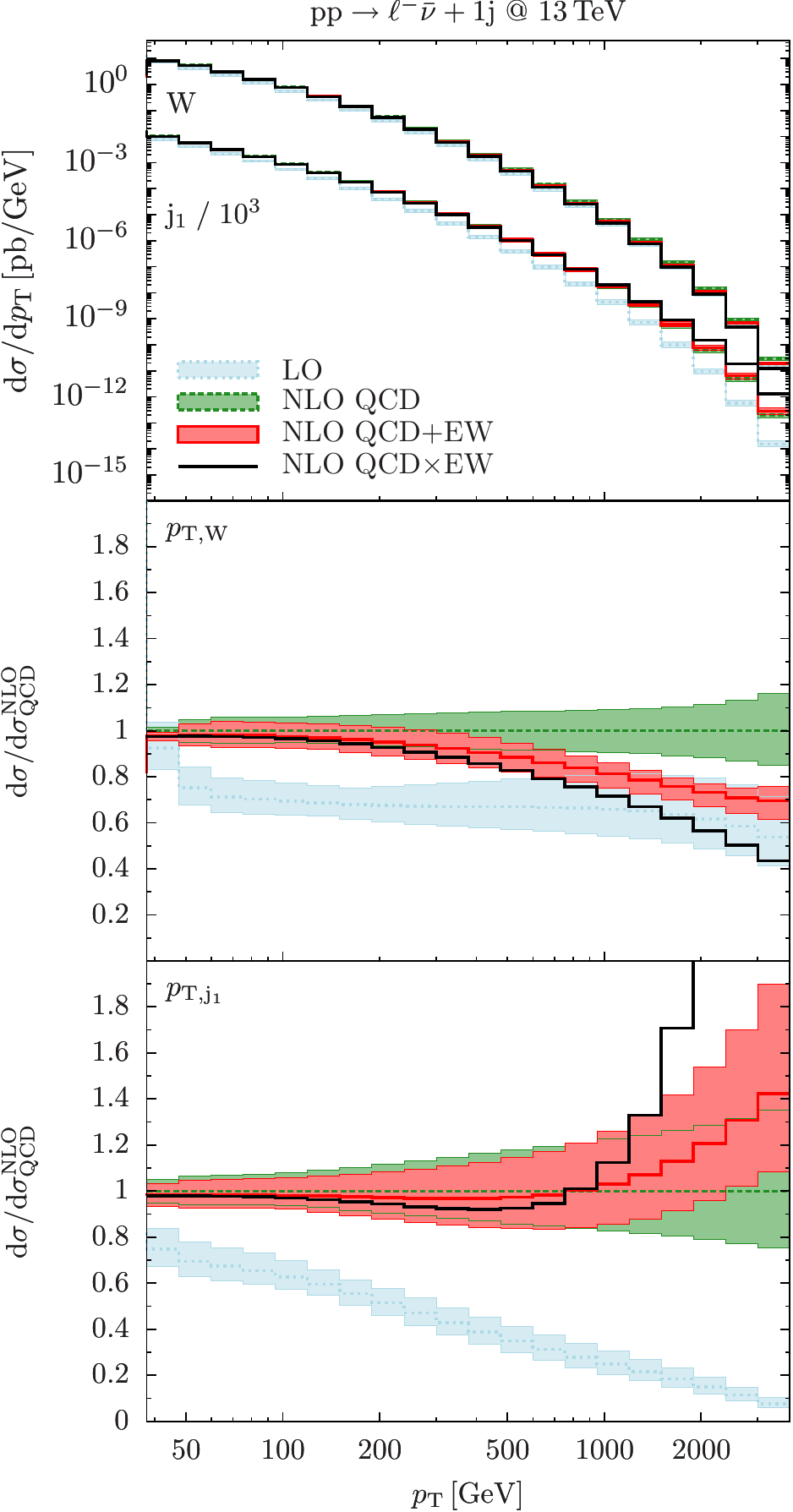}
         \qquad 
   \includegraphics[width=\relplotwidth\textwidth]{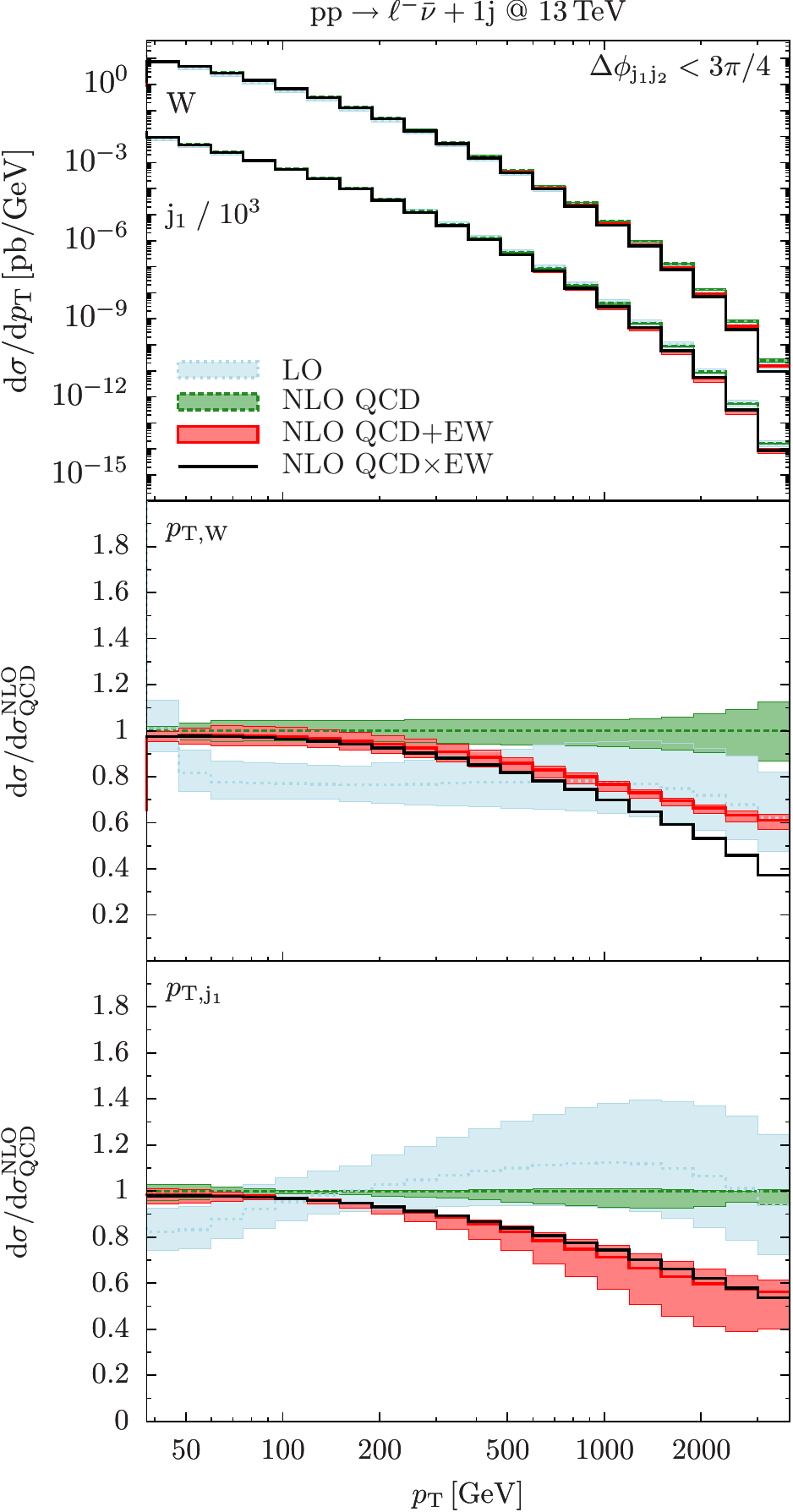}
\caption{
Distributions in the transverse momenta of the reconstructed vector boson, $\pTV$, 
and of the hardest jet, $p_{\rT,j_1}$, for $pp\to\lmn+1$\,jet at 13\,TeV 
with standard cuts
(left) and with an additional cut $\deltajj<3\pi/4$ (right).
Absolute LO (light blue), NLO \QCD (green), NLO \QCDpEW (red) and NLO
\QCDtEW (black) predictions (upper panel) and relative corrections
with respect to NLO \QCD (lower panels). The bands correspond to scale variations, 
and in the case of ratios only the numerator is varied. 
The absolute predictions in $p_{\rT,j_1}$ are rescaled by a factor $10^{-3}$.
}
\label{fig:Vj_pTV_pTj1}
\end{figure*}

The above mentioned anomalies are clearly manifest in \reffi{fig:Vj_pTV_pTj1},
where NLO \QCD and \EW effects in  $pp\to \lmn+1$\,jet\footnote{
A similar behaviour is encountered also 
in the various other channels with $\ell\ell/\ell\nu/\nu\nu+$\,jet final states.
A more detailed discussion of the interplay between \QCD and \EW corrections in the presence of
giant $K$-factors, for the case of $\PW+\,$jets production, can be found in Section 6.1
of \cite{Kallweit:2014xda}.} are plotted versus the transverse momenta of
the reconstructed vector boson, defined in terms of their
decay products, \ie $\pTV=|\vec p_{\rT,\ell_1}+\vec p_{\rT,\ell_2}|$ for $V\to \ell_1\ell_2$, and of the leading jet.
While overall QCD corrections to the \PW-boson \pT distribution are moderate (at the level of 40-50\%) they strongly increase in the tail of the distribution reaching 100\% at 3 TeV. In the case of the jet-\pT the QCD corrections show a clear pathological behaviour growing larger than several 100\% in the multi-TeV region.
In the $\pTw$ distribution, NLO \EW 
corrections present a consistent Sudakov shape,
with corrections growing negative like $\ln^2(\hat s/\MW^2)$ 
and reaching a few tens of percent in the tail.
However, as reflected in the
sizeable disparity between additive \QCDpEW
and multiplicative \QCDtEW combinations,
the large size of 
NLO \QCD and NLO \EW effects suggests the presence of important
uncontrolled mixed NNLO \QCDmEW corrections.
In the case of the jet-\pT distribution these problems become dramatic. 
Besides the explosion of NLO \QCD corrections, in the multi-TeV range we
observe a pathological NLO \EW behaviour, with large positive corrections
instead of negative Sudakov effects.  On one side, similarly as for the
giant \QCD $K$-factor, this feature can be attributed to hard dijet
configurations that enter the NLO \EW bremsstrahlung through mixed \QCDmEW
terms of $\ord(\alphaS\alpha^3)$ (see \reffi{fig:vj4qrealEW}).  On the other
side, \EW Sudakov effects are completely suppressed due to the absence of
one-loop corrections for $\PV+2$\,jet configurations.

In principle, the pathological behaviour of NLO predictions can be avoided by
imposing a cut that renders the $\PV+1$\,jet cross section sufficiently
exclusive with respect to extra jet radiation.  For instance, as shown in the
right plot of \reffi{fig:Vj_pTV_pTj1}, suppressing bremsstrahlung effects
with a veto against dijet configurations with angular separation
$\deltajj>3\pi/4$  leads to well-behaved \QCD predictions and a standard NLO
\EW Sudakov behaviour, with up to $-$40\% corrections at $\pT=2~\TeV$.

Thus, giant $K$-factors and related issues can be 
circumvented through a jet veto. However, in order to obtain a precise theoretical
description of inclusive $\PV+$\,jets production at high \pT, it is clear that
fixed-order NLO \QCDpEW calculations for one-jet final states have to be
supplemented by corresponding predictions for multijet final states.  This task,
as well as the consistent merging of NLO \QCDpEW cross sections with different
jet multiplicity, will be the subject of the rest of this paper.

\section{Fixed-order predictions for \texorpdfstring{$\boldsymbol{\PV+2}\,$}{V+2}jet production}
\label{se:vjj}

In this section we present numerical results for 
$\ell\ell/\ell\nu/\nu\nu+2$\,jet production,
including  NLO \QCD and \EW corrections, as well as
subleading Born and photon-induced contributions.

\subsection[NLO \texorpdfstring{\QCDpEW}{QCD+EW} predictions]{NLO \texorpdfstring{\QCDpEW}{QCD+EW} predictions}
\label{se:vjj-results}


\begin{figure*}[t]
\centering
   \includegraphics[width=\relplotwidth\textwidth]{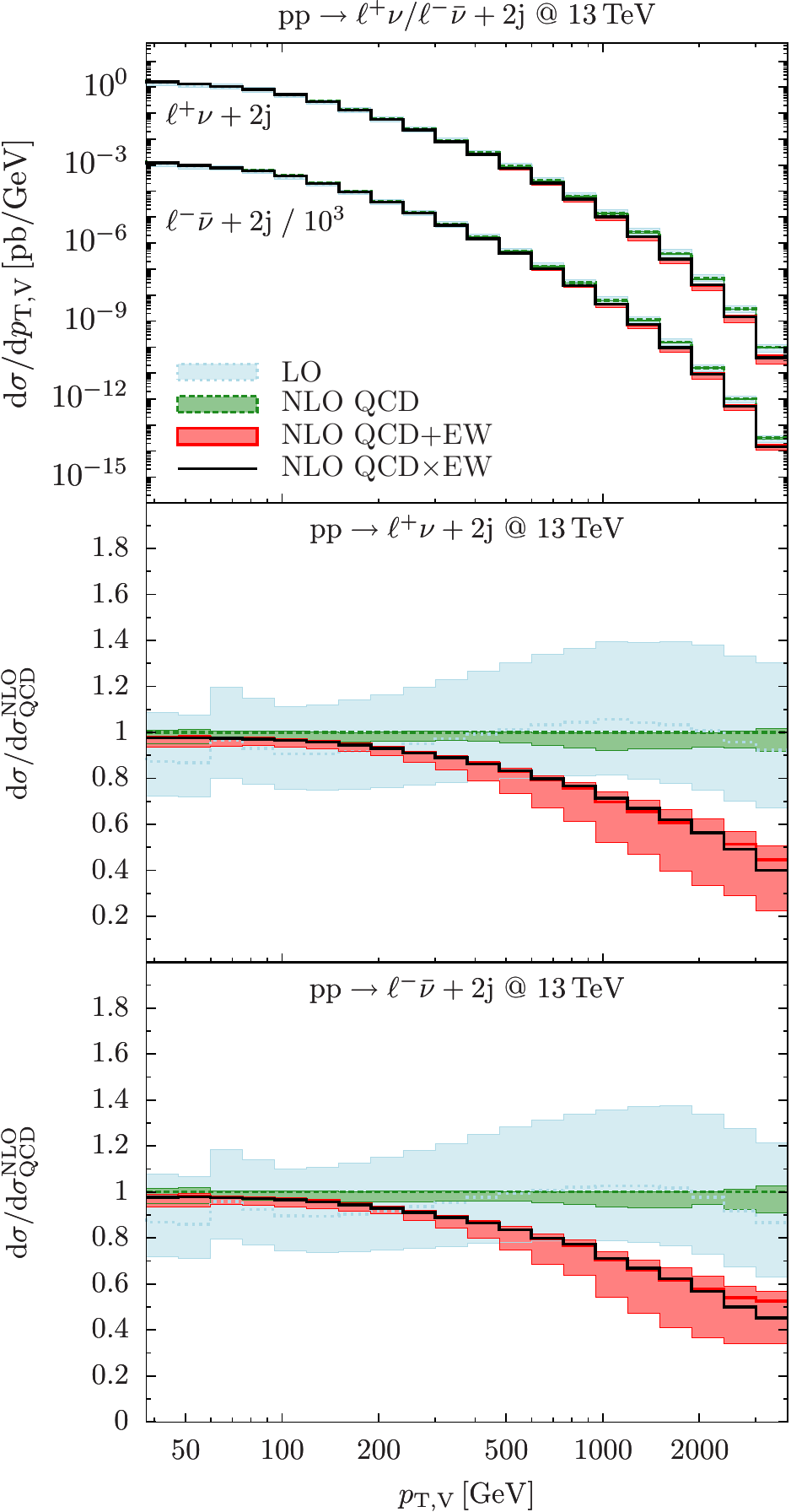}
         \qquad 
   \includegraphics[width=\relplotwidth\textwidth]{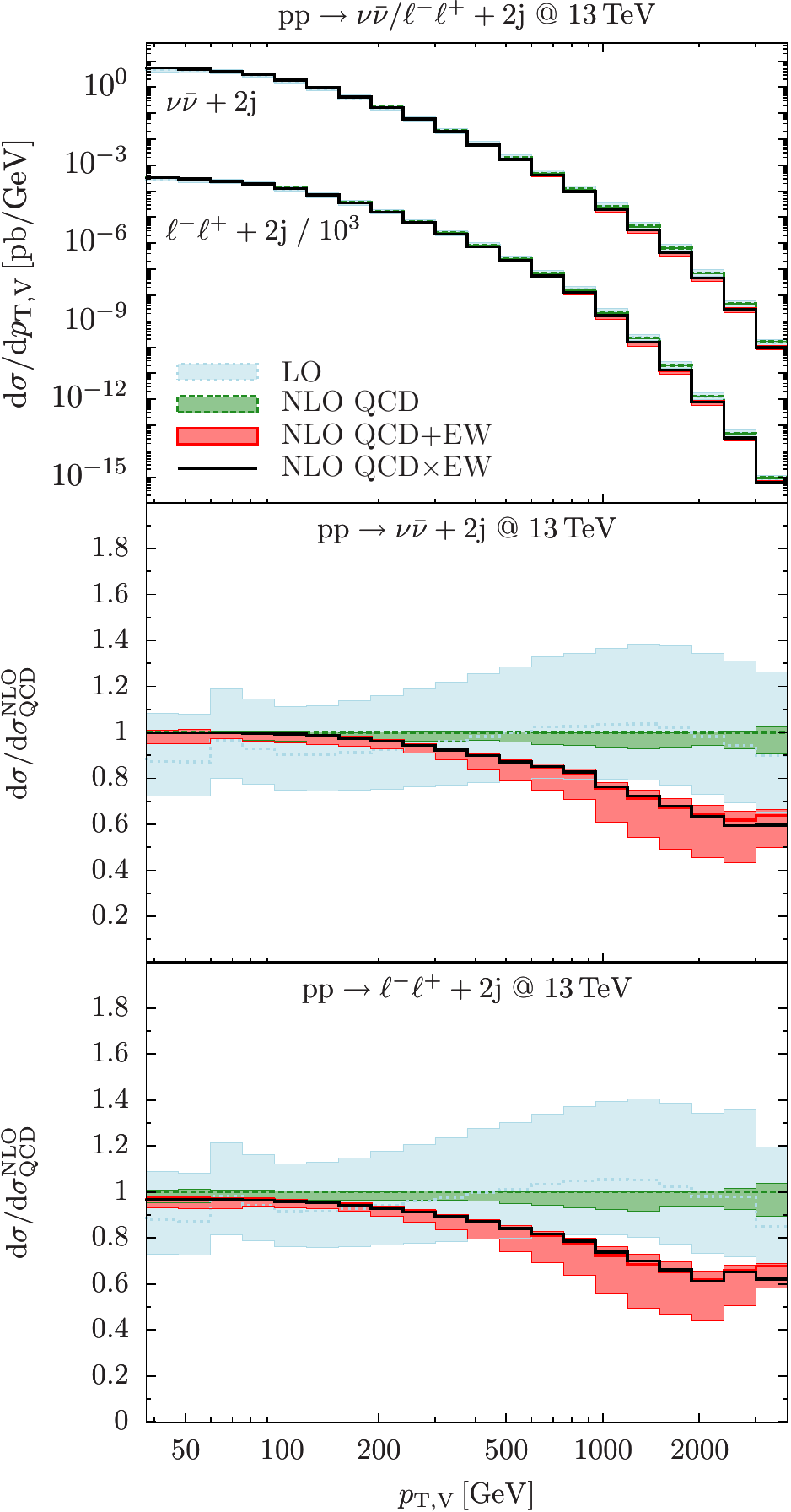}
\caption{
Distributions in the reconstructed transverse momentum of the off-shell vector boson, $\pTV$,
for $pp\to \ell\nu+2$\,jets (left) and $pp\to\ell\ell/\nu\nu+2$\,jets (right) at 13\,TeV.
Curves and bands as in \reffi{fig:Vj_pTV_pTj1}.
}
\label{fig:Vjj_pTV}
\end{figure*}

In the following, we discuss a series of fixed-order NLO \QCDpEW results for
$pp\to \PV+2$\,jets including leptonic decays, \ie we investigate the processes
$pp\to\lpn+2$\,jets, $pp\to\lmn+2$\,jets, $pp\to\lplm+2$\,jets
and $pp\to\nn+2$\,jets at 13\;TeV.
We will focus on the effect of \EW corrections on the \pT spectra of 
reconstructed vector bosons, charged leptons and jets.
Such observables are of direct relevance
as a background for many searches for new physics including dark matter at the LHC.
Instead of presenting the four processes and their higher-order corrections
independently, we will mostly show them together for the different observables in
order to highlight important similarities and investigate possible differences. 
Additionally, for $pp\to \ell\nu+2$\,jets we show distributions in the transverse
mass and missing energy, while for $pp\to \ell^+\ell^-+2$\,jets we show the
distribution in the invariant mass of the leptonic decay products. Predictions
for further kinematic observables are presented in Appendix \ref{app:vjj}.\footnote{Our NLO \EW predictions 
for $pp\to\lplm+2$\,jets have been compared in detail against the results of \cite{Denner:2014ina}. Good agreement 
was found within the 
small uncertainties due to the different treatment of photons and b-quark induced processes.}

\begin{figure*}[t]
\centering
   \includegraphics[width=\relplotwidth\textwidth]{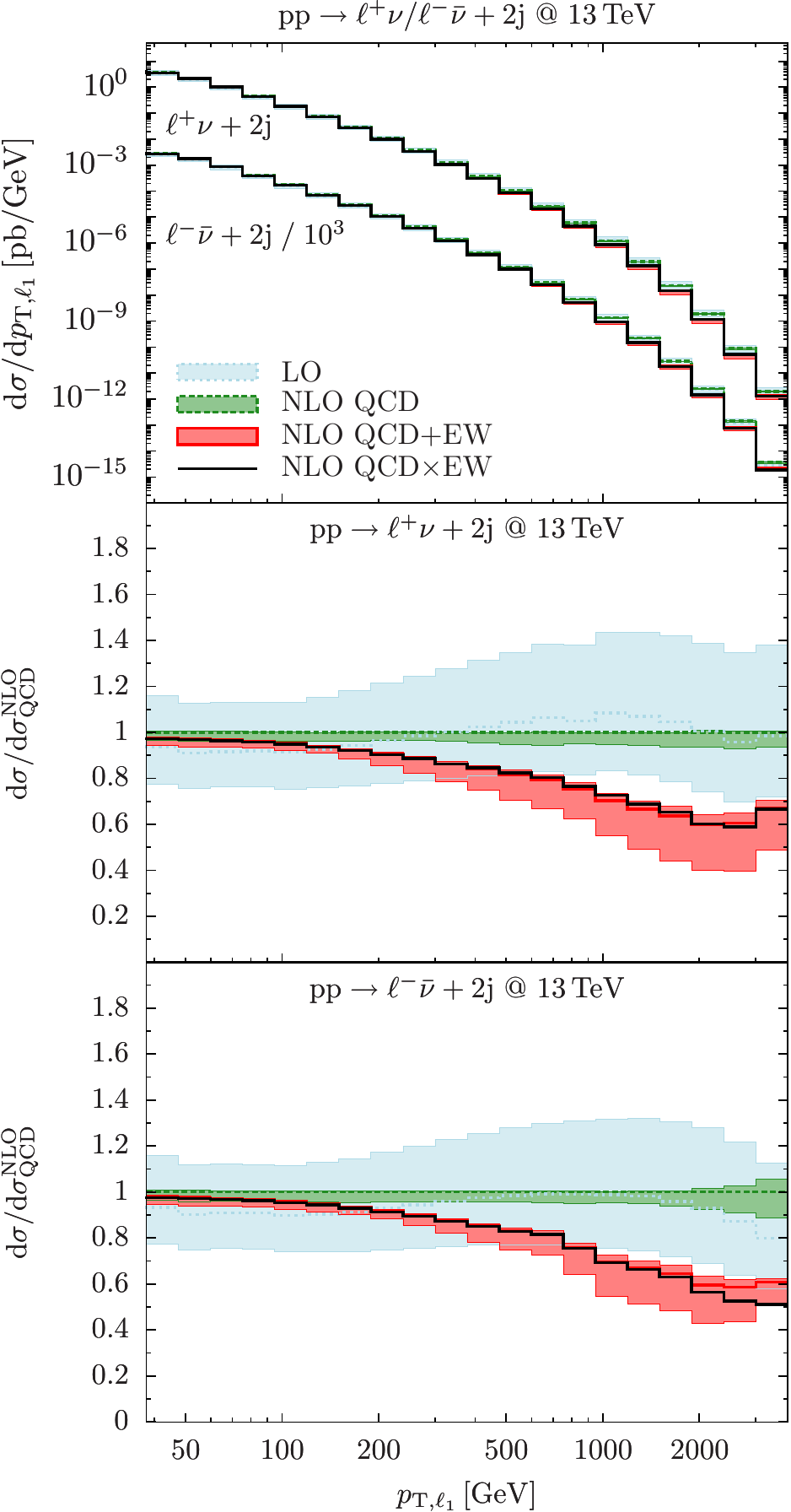}
         \qquad 
   \includegraphics[width=\relplotwidth\textwidth]{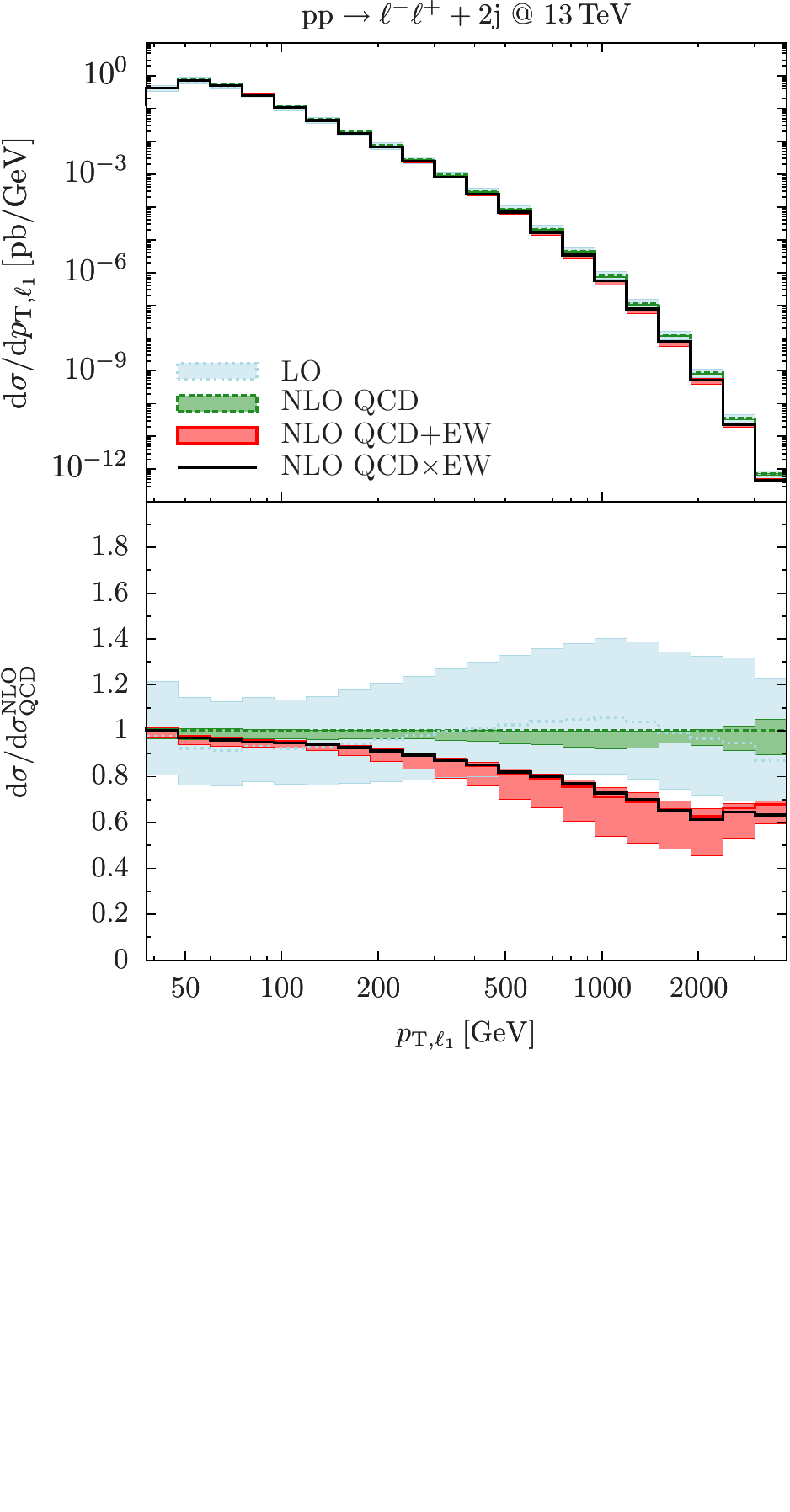}
\caption{
Distributions in the transverse momentum of the hardest charged lepton, $p_{\rT,\ell_1}$, 
for $pp\to \ell\nu+2$\,jets (left) and $pp\to\ell\ell+2$\,jets (right) at 13\,TeV.
Curves and bands as in \reffi{fig:Vj_pTV_pTj1}.
}
\label{fig:Vjj_pTl1}
\end{figure*}

Figure~\ref{fig:Vjj_pTV} displays results for the transverse-momentum 
spectra of the reconstructed (off-shell) vector bosons.
For all processes NLO \QCD corrections are remarkably small,
and even in the tails scale uncertainties hardly exceed 10\%. 
In contrast, NLO \EW corrections feature a standard
Sudakov behaviour and become very large at high \pT.
They  exceed \QCD scale
uncertainties already at a few hundred GeV
and reach about $-40\%$ at 2~TeV.
Due to the small size of \QCD corrections, 
for all processes we observe a good consistency between
NLO \QCDpEW and NLO \QCDtEW results.
As expected, \QCD and \EW corrections for $\ell^+\nu_\ell+2$\,jets turn out to be very similar to
the ones observed in the corresponding calculation of \cite{Kallweit:2014xda} 
where the $\PW$ boson was kept on-shell.

In \reffi{fig:Vjj_pTl1} we plot, where applicable, the \pT spectra of the
hardest lepton. The behaviour of the \QCD and \EW corrections is very
similar to the one observed for the \pT of the reconstructed vector bosons. Clearly, the observed
large Sudakov corrections are a result of the TeV scale dynamics that 
enter the production of a high-\pT vector boson, while they
are hardly affected by vector-boson decay processes, which occur at much smaller energy
scales.

\begin{figure*}[t]
\vspace*{23ex}
\centering
   \includegraphics[width=\relplotwidth\textwidth]{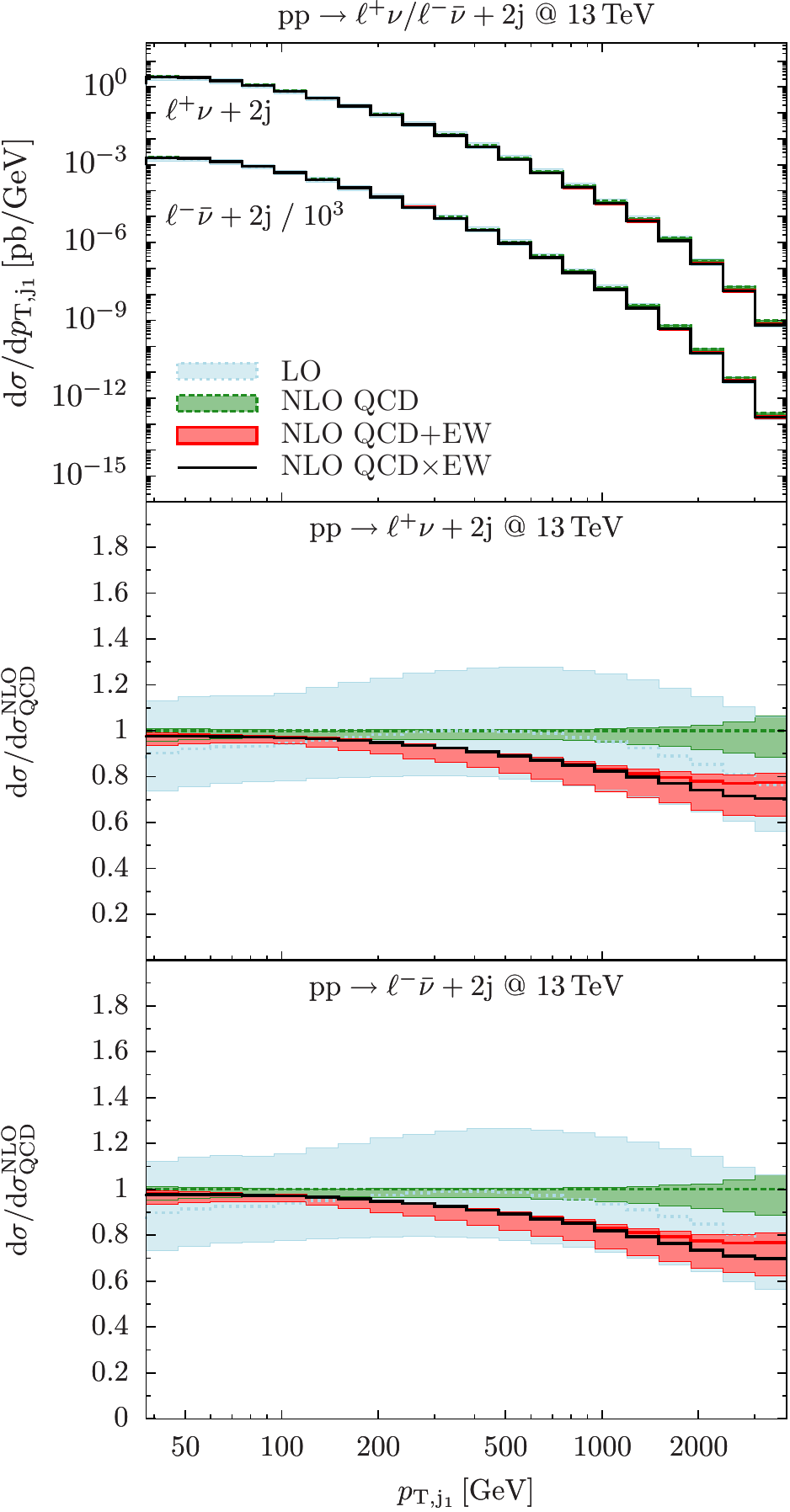}
         \qquad 
   \includegraphics[width=\relplotwidth\textwidth]{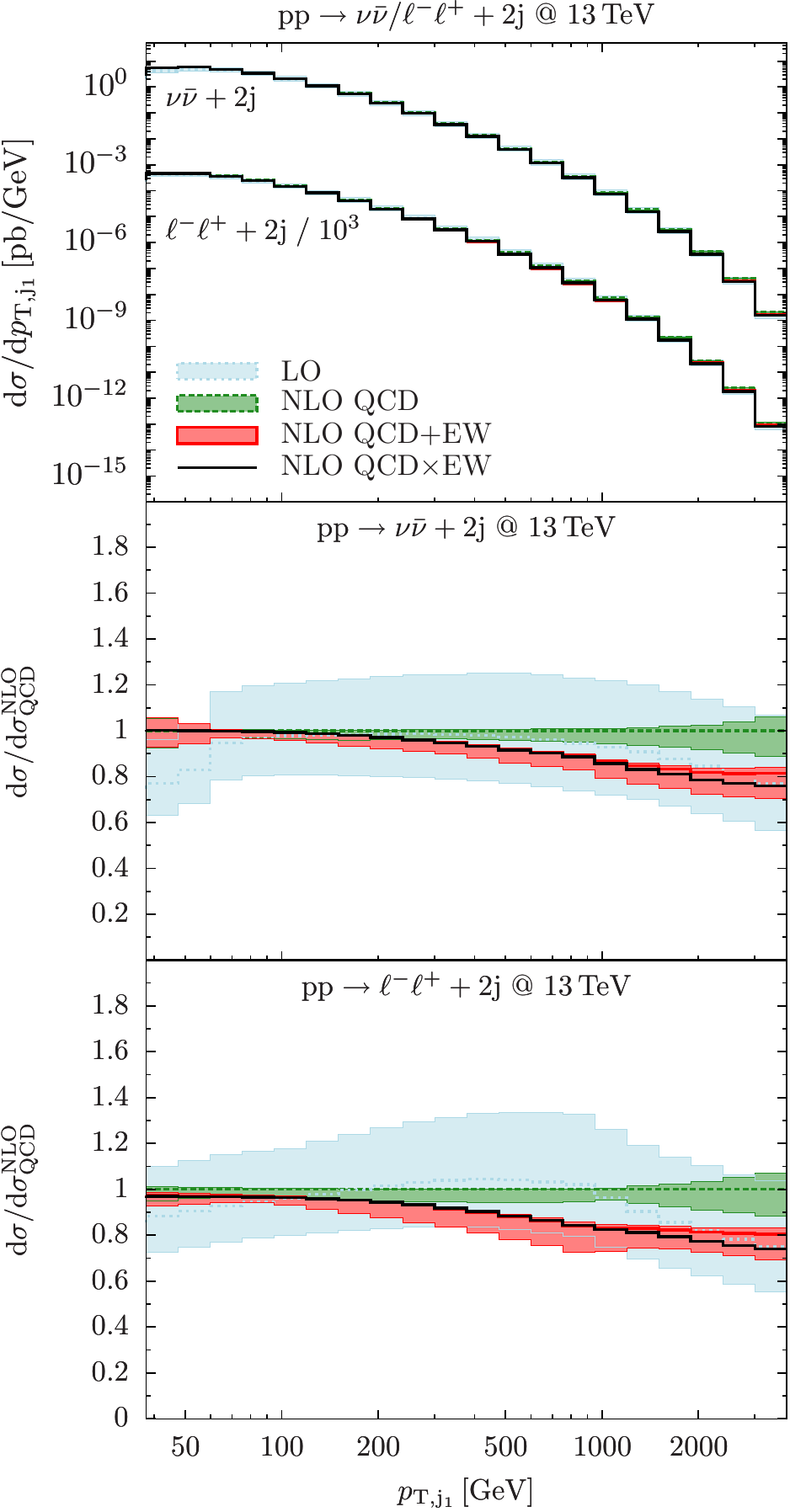}
\caption{
Distributions in the transverse momentum of the hardest
jet, $p_{\rT,j_1}$, 
for $pp\to \ell\nu+2$\,jets (left) and $pp\to\ell\ell/\nu\nu+2$\,jets (right) at 13\,TeV.
Curves and bands as in \reffi{fig:Vj_pTV_pTj1}.
}
\vspace*{22ex}
\label{fig:Vjj_pTj1}
\end{figure*}

\begin{figure*}[t]
\vspace*{23ex}
\centering
   \includegraphics[width=\relplotwidth\textwidth]{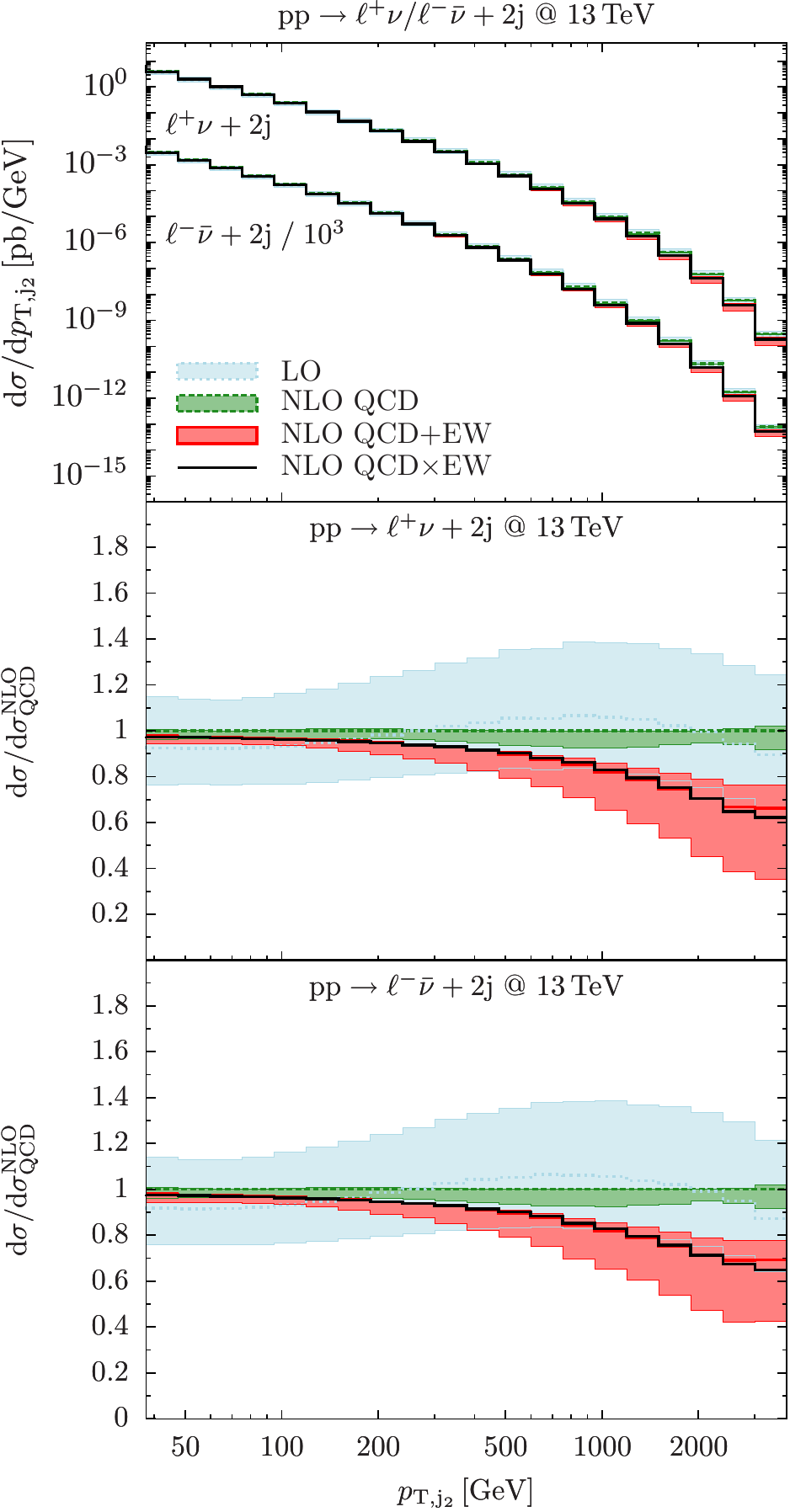}
         \qquad 
   \includegraphics[width=\relplotwidth\textwidth]{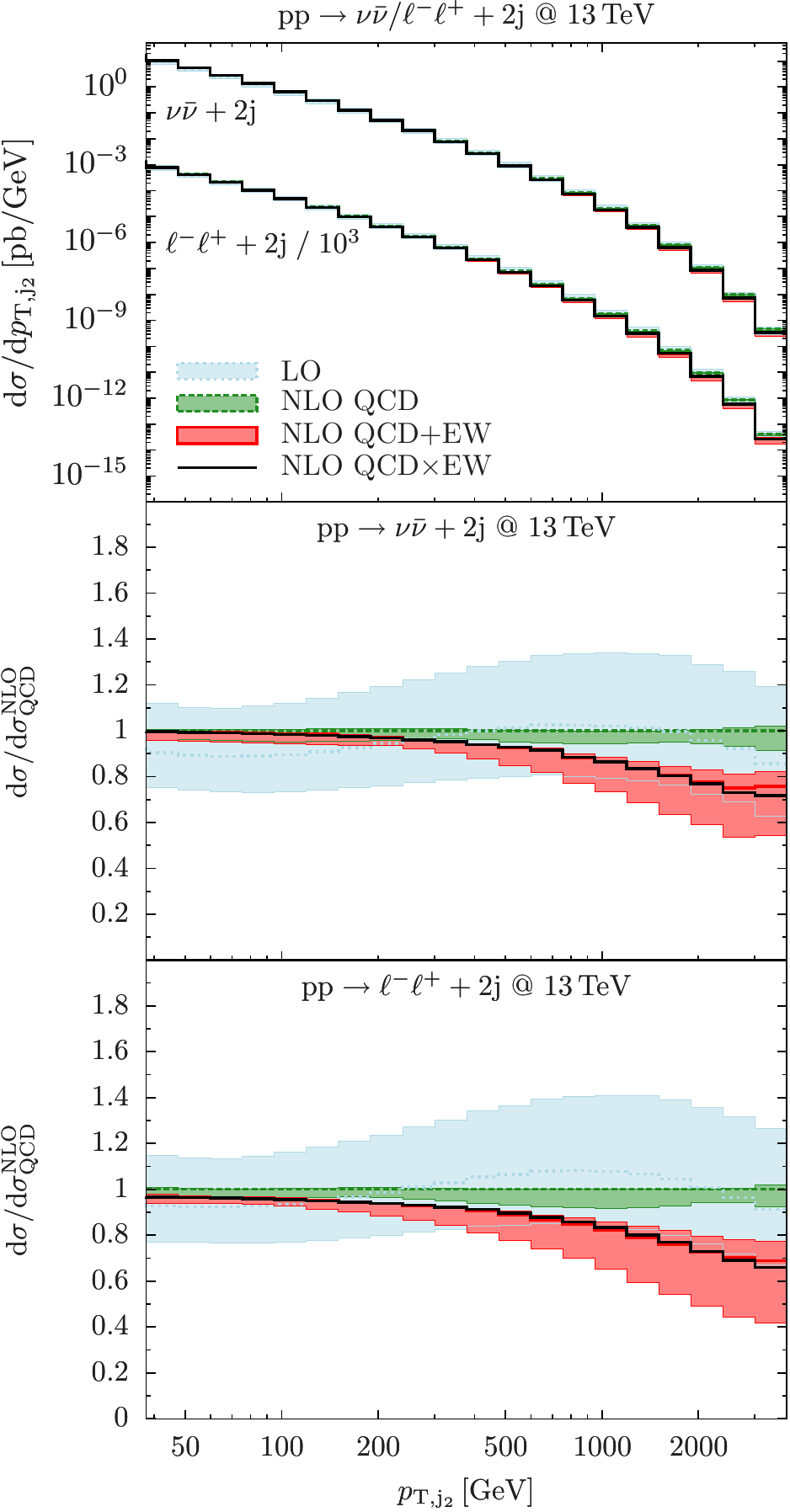}
\caption{
Distributions in the transverse momentum of the second-hardest
jet, $p_{\rT,j_2}$, 
for $pp\to \ell\nu+2$\,jets (left) and $pp\to\ell\ell/\nu\nu+2$\,jets (right) at 13\,TeV.
Curves and bands as in \reffi{fig:Vj_pTV_pTj1}.
}
\vspace*{22ex}
\label{fig:Vjj_pTj2}
\end{figure*}

Figures~\ref{fig:Vjj_pTj1} and \ref{fig:Vjj_pTj2} present distributions in
the transverse momenta of the hardest and second-hardest jet, respectively. 
Again, the perturbative \QCD expansion turns out to be very stable, with
scale uncertainties that hardly exceed 10\%.  In these jet-\pT distributions we
observe smaller NLO~\EW corrections as compared to the case of the vector-boson 
\pT spectrum. This is due to the fact that $W$ and $Z$ bosons carry
larger SU(2) charges as compared to gluons and quarks inside jets.  Thus,
the largest \EW Sudakov corrections arise when the vector-boson \pT is
highest, while very hard jets in combination with less hard vector bosons
yield less pronounced \EW Sudakov logarithms. We also find that, at a given \pT,
the second jet always receives larger \EW corrections than the first jet.
Quantitatively, the \EW corrections to the different $\PW+2$\,jet and $Z+2$\,jet
processes are rather similar. Thus, corresponding ratios are expected to be only
mildly sensitive to \EW (or \QCD) corrections.

\begin{figure*}[t]
\centering
   \includegraphics[width=\relplotwidth\textwidth]{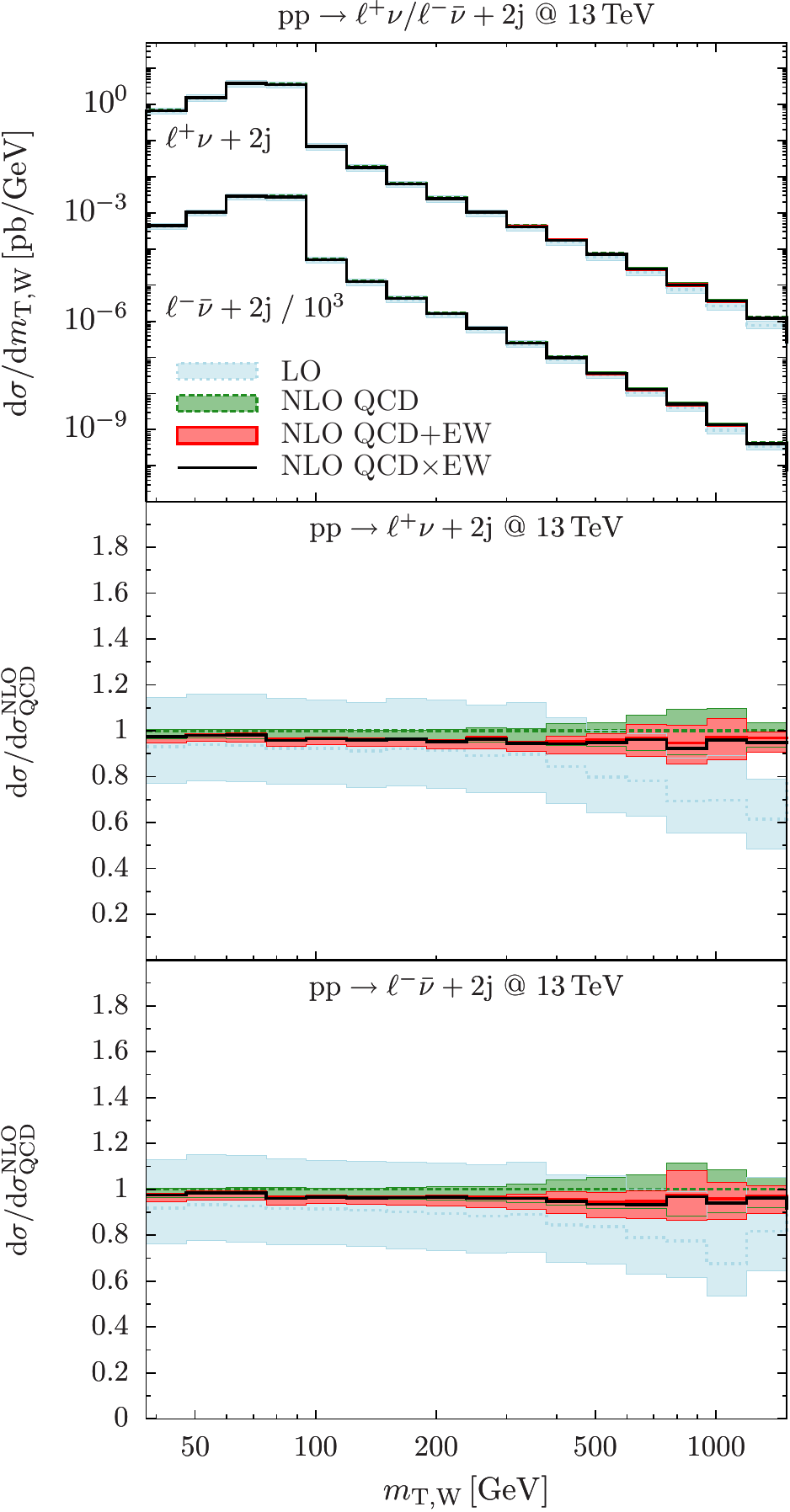}
         \qquad 
   \includegraphics[width=\relplotwidth\textwidth]{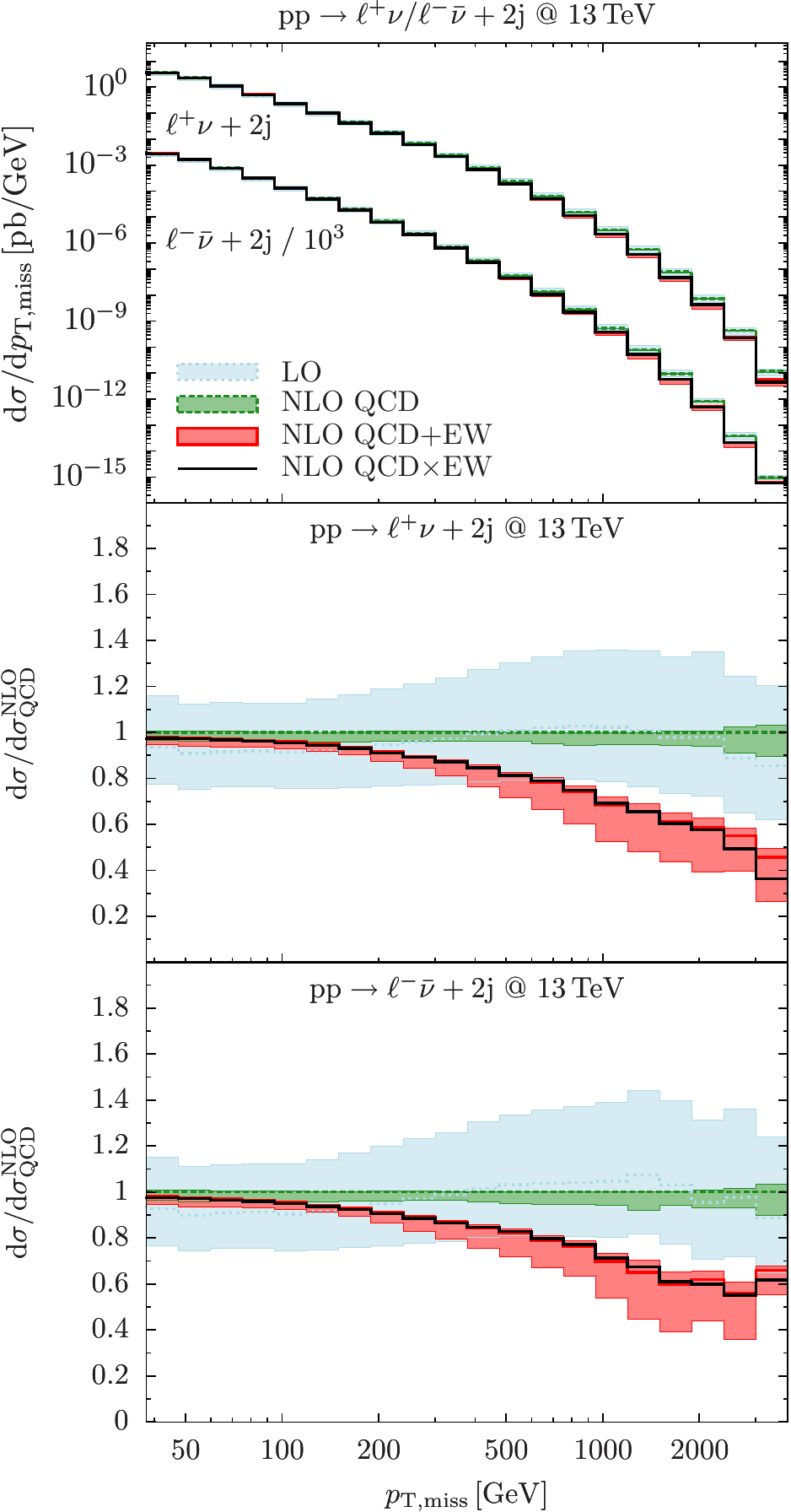}
\caption{
Distributions in the transverse mass, $\mTW=\sqrt{2p_{\rT,\ell} p_{\rT,\nu}
(1-\cos\Delta\phi_{\ell\nu}) }$, (left)
and the missing transverse energy, $p_{\rT,\text{miss}}$, (right)
for $pp\to\ell\nu+2$\,jets at 13\,TeV.
Curves and bands as in \reffi{fig:Vj_pTV_pTj1}.
}
\label{fig:Wjj_mT_pTmiss}
\end{figure*}

In Figure~\ref{fig:Wjj_mT_pTmiss} we show  distributions in the transverse
mass, $\mTW$, and in the missing transverse energy (\ie the \pT spectrum of
the neutrino) for $\PW+2$\,jet production.  Both observables are of
paramount importance in many BSM searches, especially in the high-energy regime.
Again, \QCD effects and uncertainties turn out to be rather mild.
As far as \EW corrections are concerned, at
large transverse masses we observe only a minor impact,
which does not exceed $-10\%$ and
remains at the level of \QCD scale uncertainties. 
In contrast, and as expected,
the missing-energy distributions follow the behaviour of the lepton-$\pT$
distribution shown in \reffi{fig:Vjj_pTl1},
and NLO \EW corrections reach about $-40\%$ at 2~TeV.

\begin{figure*}[t]
\centering
   \includegraphics[width=\relplotwidth\textwidth]{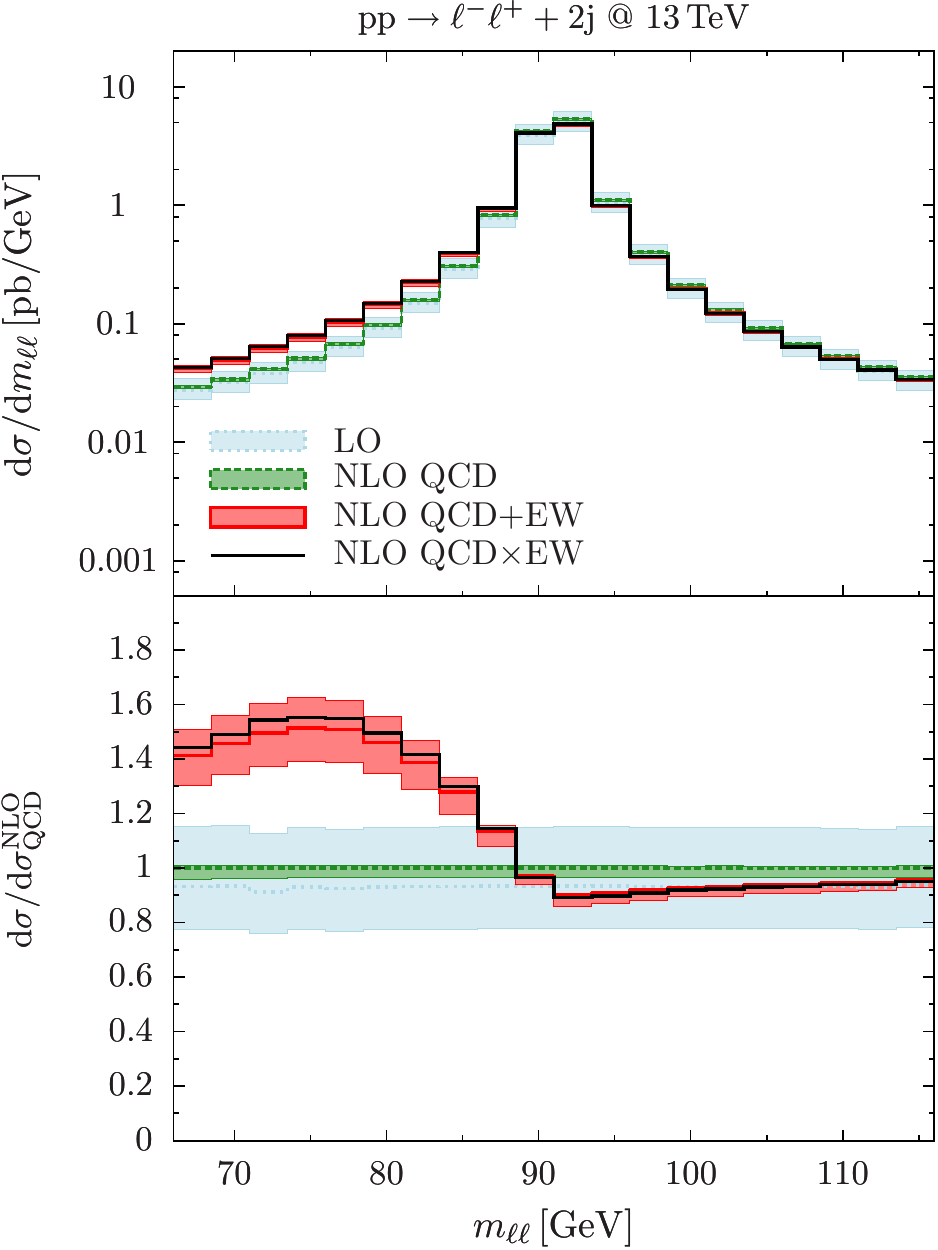}
\caption{
Distribution in the invariant mass $m_{\ell\ell}$ of the lepton pair in $pp\to \lplm+2$\,jet production
at 13\,TeV.
Curves and bands as in \reffi{fig:Vj_pTV_pTj1}.
}
\label{fig:Zjj_mll}
\end{figure*}

Finally, in \reffi{fig:Zjj_mll} we turn to the differential distribution in the
invariant mass, $\mll$, of the lepton pair produced in $pp\to\lplm+2$\,jets. The
plotted range corresponds to the event selection specified in \refta{Tab:cuts}
and does not extend up to the high-energy region, where \EW Sudakov effects
would show up. However, the NLO \EW corrections are sensitive to \QED radiation
off the charged leptons and shift parts of the cross section from above the
Breit--Wigner peak to below the peak. The observed shape of the \EW corrections
is qualitatively very similar to the well-known NLO \EW corrections to
neutral-current Drell--Yan production
\cite{CarloniCalame:2007cd,Dittmaier:2009cr}. In this kinematic regime, \QCD
corrections are very small and always below 10\%, while scale uncertainties are
as small as a few percent.

In summary, NLO \QCDpEW effects for $pp\to\PV+2$\,jets turn out to be completely
free from the perturbative instabilities that plague NLO predictions for
$\PV+1$\,jet production: the perturbative \QCD expansion is very well behaved,
and NLO \EW corrections feature, as expected, Sudakov effects that become very
large at the TeV scale, especially for $V+2$\,jet configurations where the
highest transverse momentum is carried by the electroweak vector boson.

\subsection{Subleading Born and photon-induced contributions}
\label{se:subLO}


\begin{figure*}[t]
\centering
   \includegraphics[width=\relplotwidth\textwidth]{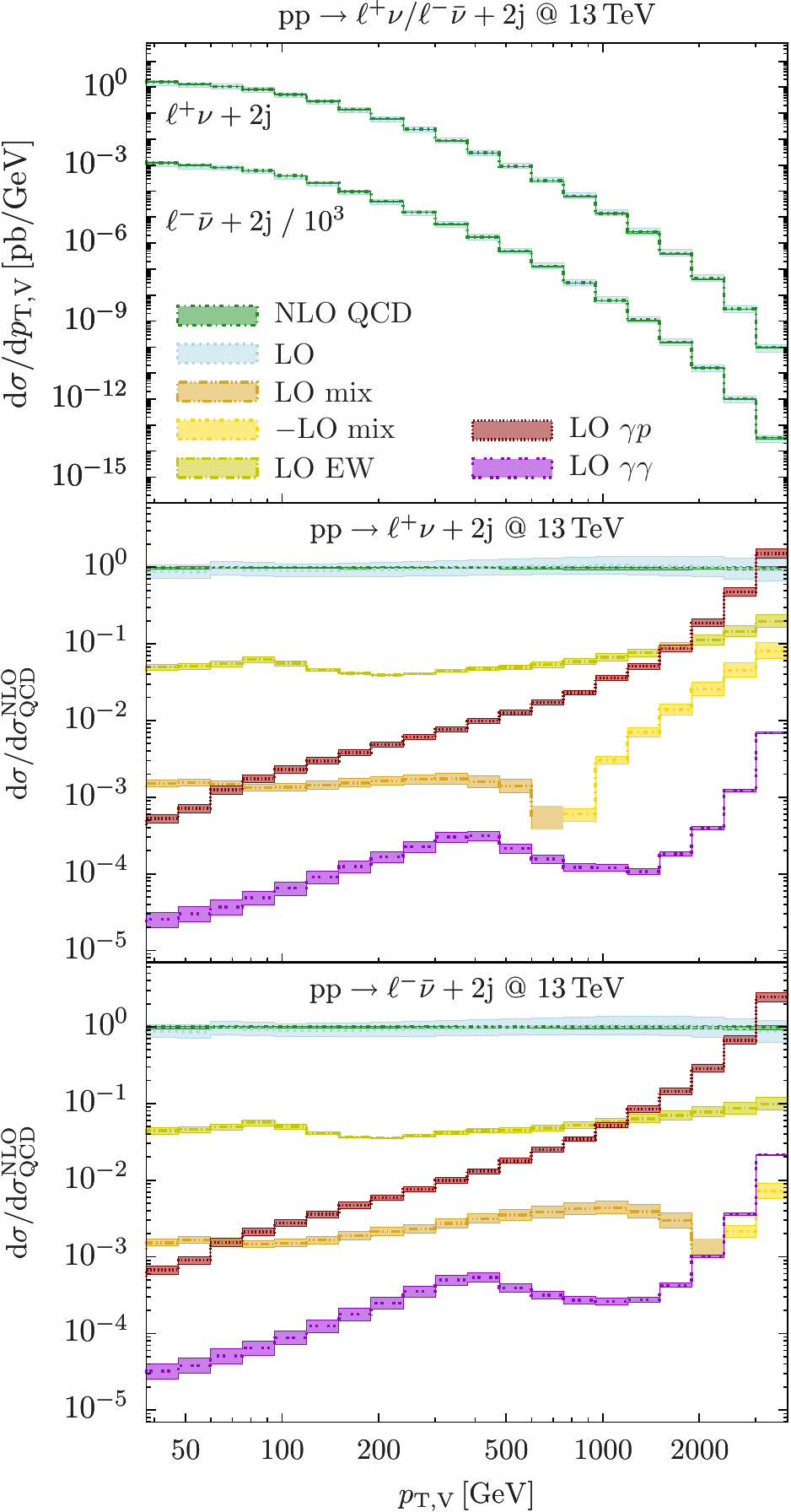}
         \qquad 
   \includegraphics[width=\relplotwidth\textwidth]{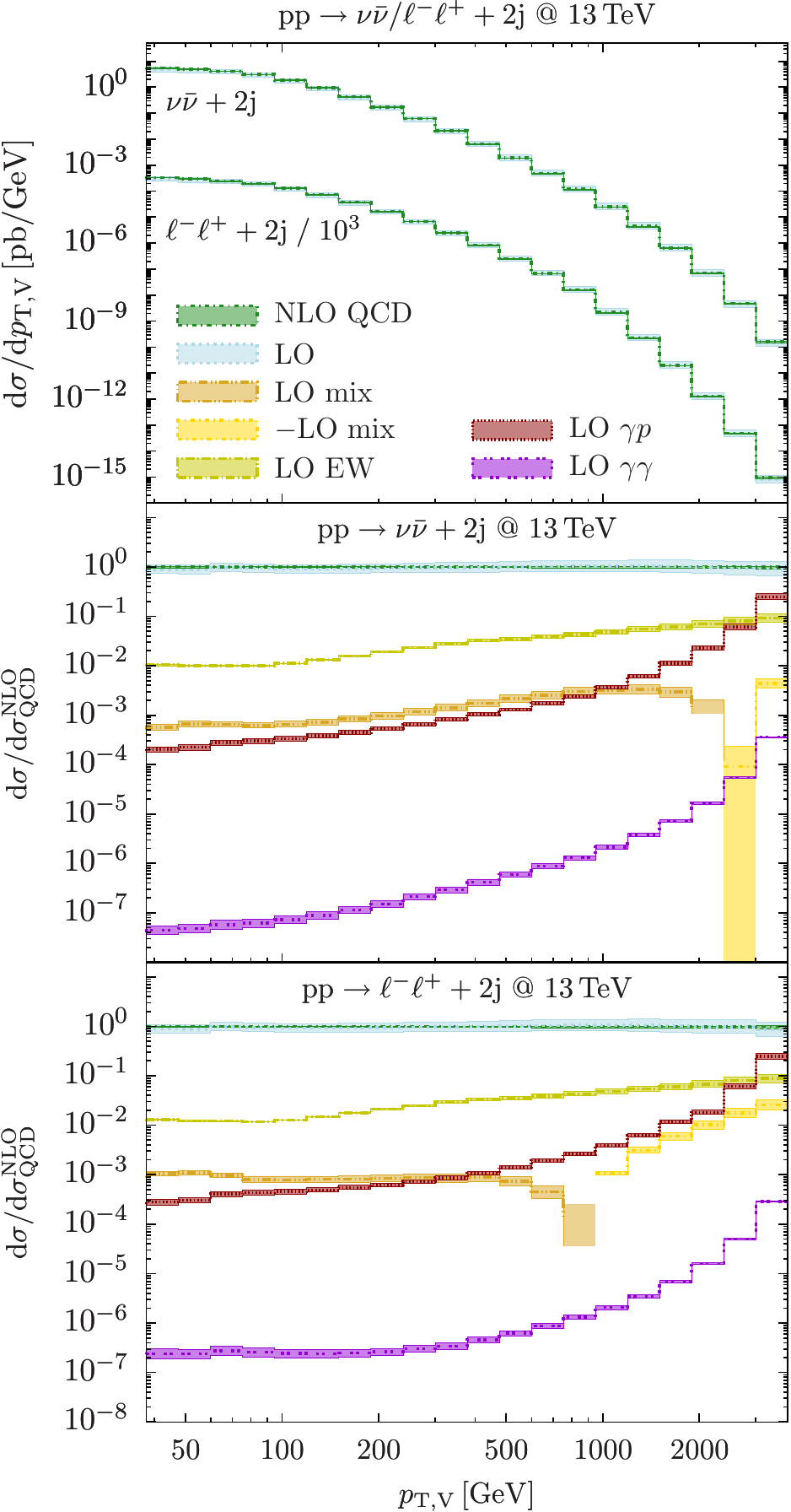}
\caption{
Distributions in the reconstructed transverse momenta of the vector boson 
for off-shell $\PWpm+2\,$jet (left) and $\PZ+2\,$jet (right) production and decay.
Absolute LO (light blue) and NLO \QCD (green) predictions (upper panel) and relative corrections
with respect to NLO \QCD, showing subleading Born contributions (lower panels). 
Discontinuities indicate sign changes of the LO mix contribution.
The bands correspond to scale variations, 
and in the case of ratios only the numerator is varied. 
The absolute predictions for $pp\to\lmn+2$\,jets and $pp\to\lplm+2$\,jets
are rescaled by a factor $10^{-3}$.
}
\label{fig:Vjj_sublBorn_pTV}
\end{figure*}

In this section we quantify the numerical impact of subleading Born and
photon-induced ($p\gamma$ and $\gamma\gamma$ initial states) contributions to
$\PV+2$\,jet production with leptonic decays, \ie tree-level contributions
of $\ord(\alphaS\alpha^3)$ and $\ord(\alpha^4)$.\footnote{The subleading Born
contributions of $\ord(\alpha^4)$ are dominated by diboson production with
semi-leptonic decays. In order to avoid a double counting between diboson and
$\PV+\,$jets processes we do not include those contributions in any of our
predictions in the following sections.}

\begin{figure*}[t]
\centering
   \includegraphics[width=\relplotwidth\textwidth]{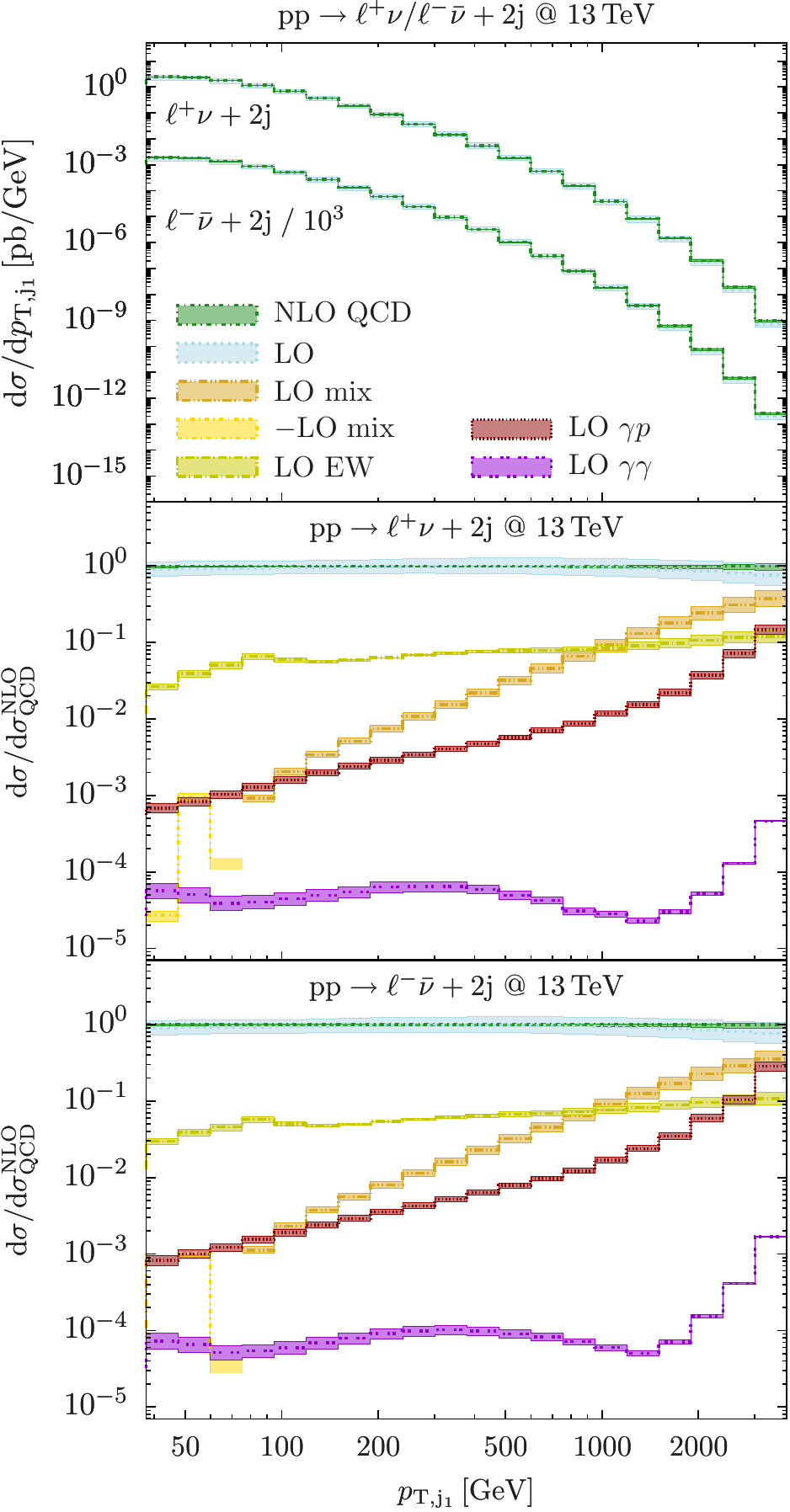}
         \qquad 
   \includegraphics[width=\relplotwidth\textwidth]{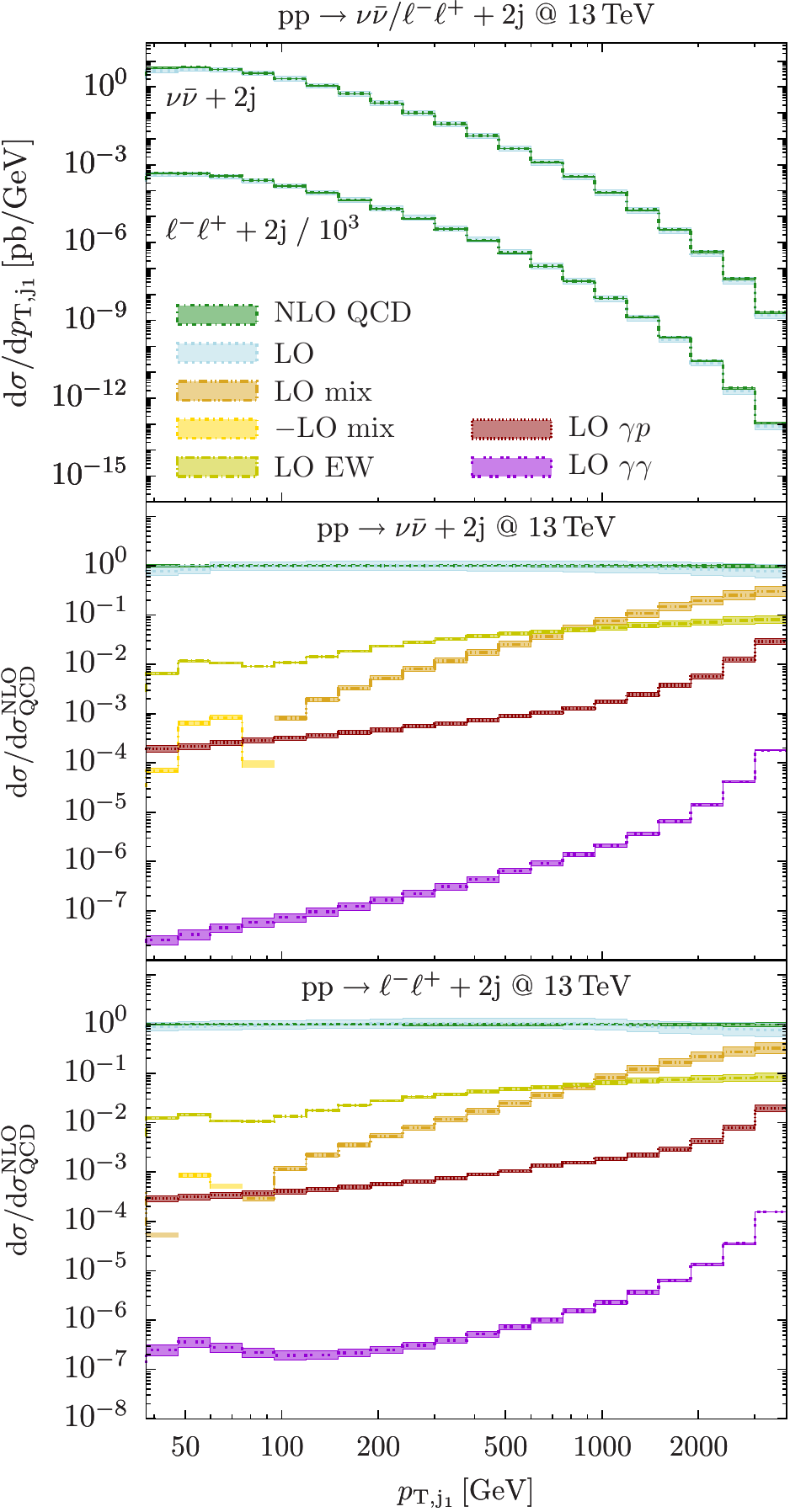}
\caption{
Distributions in the transverse momenta of the hardest jet $p_{\rT,j_1}$ for
off-shell $\PWpm+2\jet$ (left) and $\PZ+2\jet$ (right) production and decay.  Curves
and bands as in Fig.  \ref{fig:Vjj_sublBorn_pTV}.
}
\label{fig:Vjj_sublBorn_pTj1}
\end{figure*}

Figures~\ref{fig:Vjj_sublBorn_pTV} and \ref{fig:Vjj_sublBorn_pTj1} illustrate the
subleading contributions for the distributions in the transverse momenta of the reconstructed vector-boson 
and of the hardest jet, respectively. Although mostly suppressed by several
orders of magnitude at small energies, at large energies $p\gamma$-initiated
production can have a sizable impact on the \pT spectrum of the
vector boson, whereas the LO mix contribution grows up to
several tens of percent in the multi-TeV region of the jet-\pT spectrum.%
These effects can both be understood as induced by PDFs: in current PDF fits including
\QED corrections~\cite{Ball:2013hta} the photon density at high Bjoerken $x$
strongly increases, while at the same time a relative increase of quark PDFs
over the gluon PDF induces an enhancement of the four-quark channel (which involves LO mix terms) over the two-quark channel. 
Although strongly suppressed in the full \pT range, 
it is interesting to note that the LO mix contributions to the \pT spectrum of the
reconstructed vector bosons feature a different
behaviour in the case of $\ell^+\nu jj$ vs. $\ell^-\bar \nu jj$ and $\ell^+\ell^- jj$ vs. $\nu \bar \nu jj$ production (see~\reffi{fig:Vjj_sublBorn_pTV}).
In all cases we observe a sign flip that results from the 
interference of resonant EW diagrams with non-resonant QCD amplitudes (see the discussion of 
``pseudo resonances'' in~\cite{Kallweit:2014xda}).
However, the location of the sign flip and the subsequent onset of a sizable 
negative contribution is significantly displaced in the different related processes.
This can be attributed to the fact that the position of the sign flip 
is very sensitive to phase-space boundaries and the relative yields of the 
various contributing partonic channels, which in turn is sensitive to differences in the PDF luminosities that enter the various processes.

With respect to the large impact of the $p\gamma$-initiated  
production at large vector-boson \pT, one should, however, keep
in mind that the photon PDF is still very poorly constrained in this
regime~\cite{Ball:2013hta}. Therefore, we do not include these contributions in
any of the predictions for $\PV+\,$jets production in the rest of the paper.

Having a merging of different jet multiplicities in mind, we want to note that
the LO mix contributions to $pp\to\PV+2$\,jet production discussed here are
in fact identical with the \QCDmEW mixed bremsstrahlung contributions to $pp\to
\PV+1$\,jet production. The multijet merging approach introduced in 
the next section guarantees a consistent inclusion of such effects without double 
counting.

\section{Multijet merged predictions for \texorpdfstring{$\boldsymbol{\PV+}\,$}{V+}jets at NLO \protect\QCDpEW}
\label{se:meps}

In order to address the need of NLO \QCDpEW accuracy for observables that
receive sizable contributions from multijet radiation, in this section we
introduce an approach that allows one to readily implement NLO \QCDpEW effects
in the context of multijet merging. The benefits of multijet merging are first
illustrated through a na\"ive combination of fixed-order calculations for
$\PV+1\,$jet and $\PV+2\,$jet production based on exclusive sums. Subsequently,
we introduce an approximate treatment of \EW corrections, based on
infrared-subtracted virtual contributions, which allows us to include \EW
corrections in the \MEPSatNLO multi-jet merging
framework~\cite{Hoeche:2009rj,Hoeche:2012yf} in a rather straightforward way.
Finally, based on a fully automated implementation of this approach in
\SherpaOpenLoops, we present an inclusive simulation of vector-boson plus
multijet production that provides NLO \QCDpEW accuracy for $\PV+0,1,2\,$jet
final states.

\subsection[Combining \texorpdfstring{${pp\to \PV+1,2}$\,}{pp->V+1,2}jets with exclusive sums]{Combining \texorpdfstring{$\boldsymbol{pp\to \PV+1,2}$\,}{pp->V+1,2}jets with exclusive sums}
\label{se:exclusive_sums}

\begin{figure}
\centering
  \includegraphics[width=\relplotwidth\textwidth]{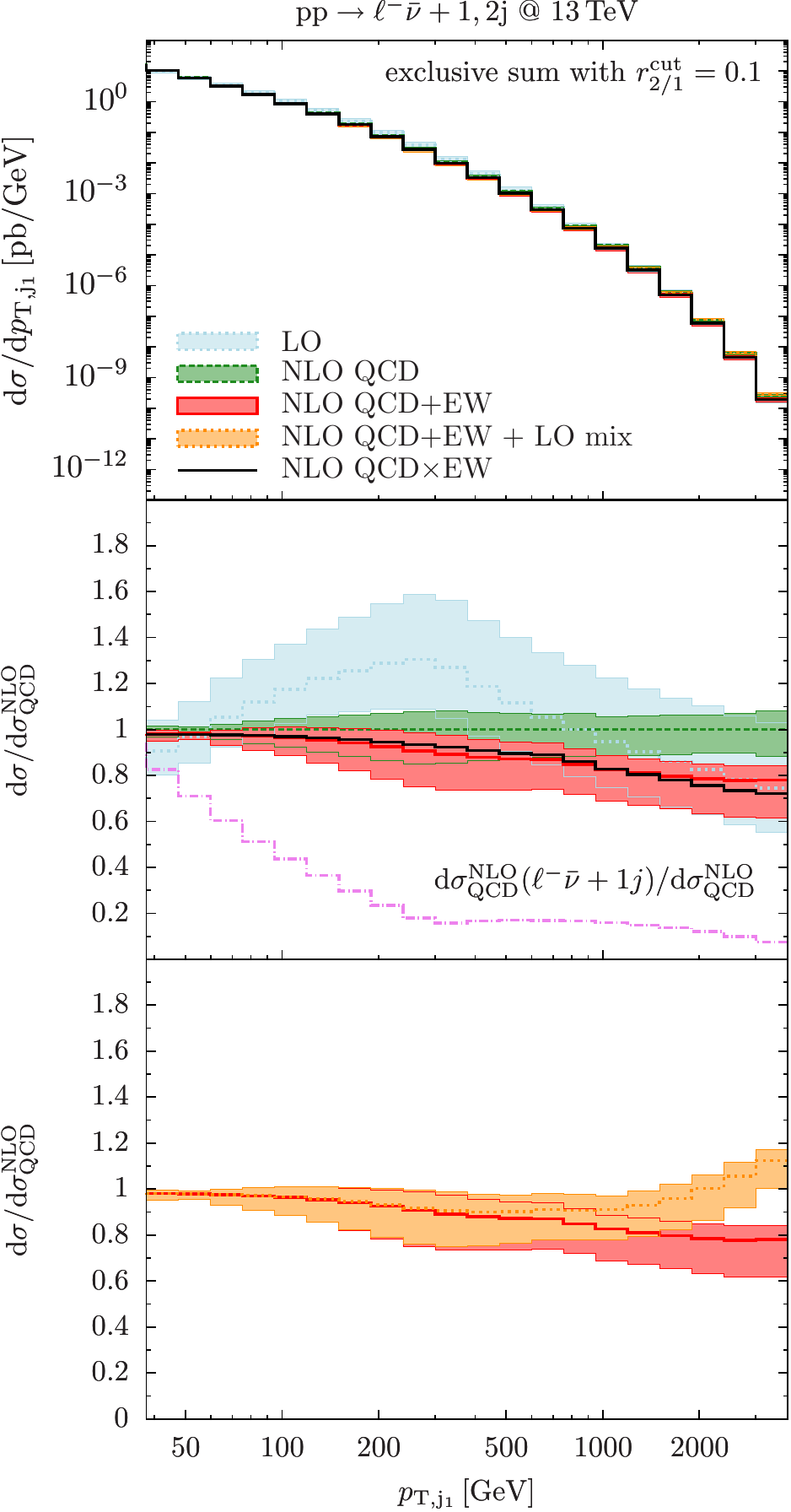}
         \qquad 
    \includegraphics[width=\relplotwidth\textwidth]{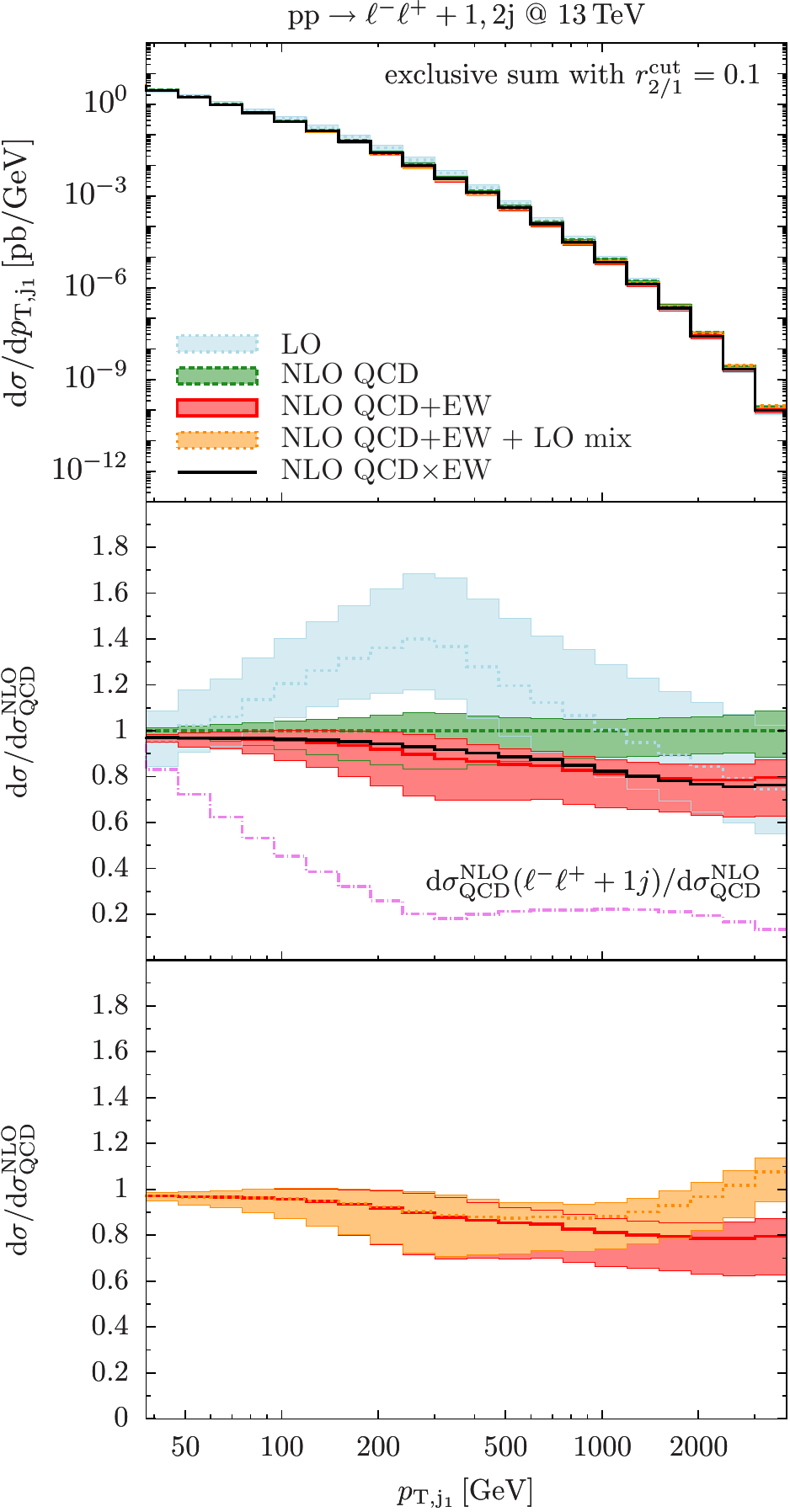}
\caption{
	  Differential distribution in the transverse momentum of the hardest jet, 
	  $p_{\rT,j_1}$, for $\lmn\,+$\,jets (left) and $\lplm+$\,jets (right). 
	  Shown are predictions merged with exclusive sums using 
	  $r_{2/1}^{\mathrm{cut}}=0.1$. 
	  The upper panels display absolute LO (light blue), 
	  NLO \QCD (green), NLO \QCDpEW (red), NLO \QCDtEW (black) and 
	  NLO \QCDpEWpLOmix (orange) predictions, where ``LO mix''
	  denotes \QCDmEW mixed Born contributions of $\ord(\alphaS\alpha^3)$ 
	  in the two-jet sample.
	  Relative corrections with respect to NLO \QCD are displayed in the 
	  lower panels. The bands correspond to scale variations, and in the 
	  case of ratios only the numerator is varied. The dashed magenta 
	  curves illustrate the relative importance of one-jet contributions 
	  ($\rratio<\rcut$) with respect to the combined one-
	  and two-jet sub-samples at NLO \QCD.
	}
\label{fig:Vjj_pTj1_exclsum}
\end{figure}

\begin{figure}
\centering
  \includegraphics[width=\relplotwidth\textwidth]{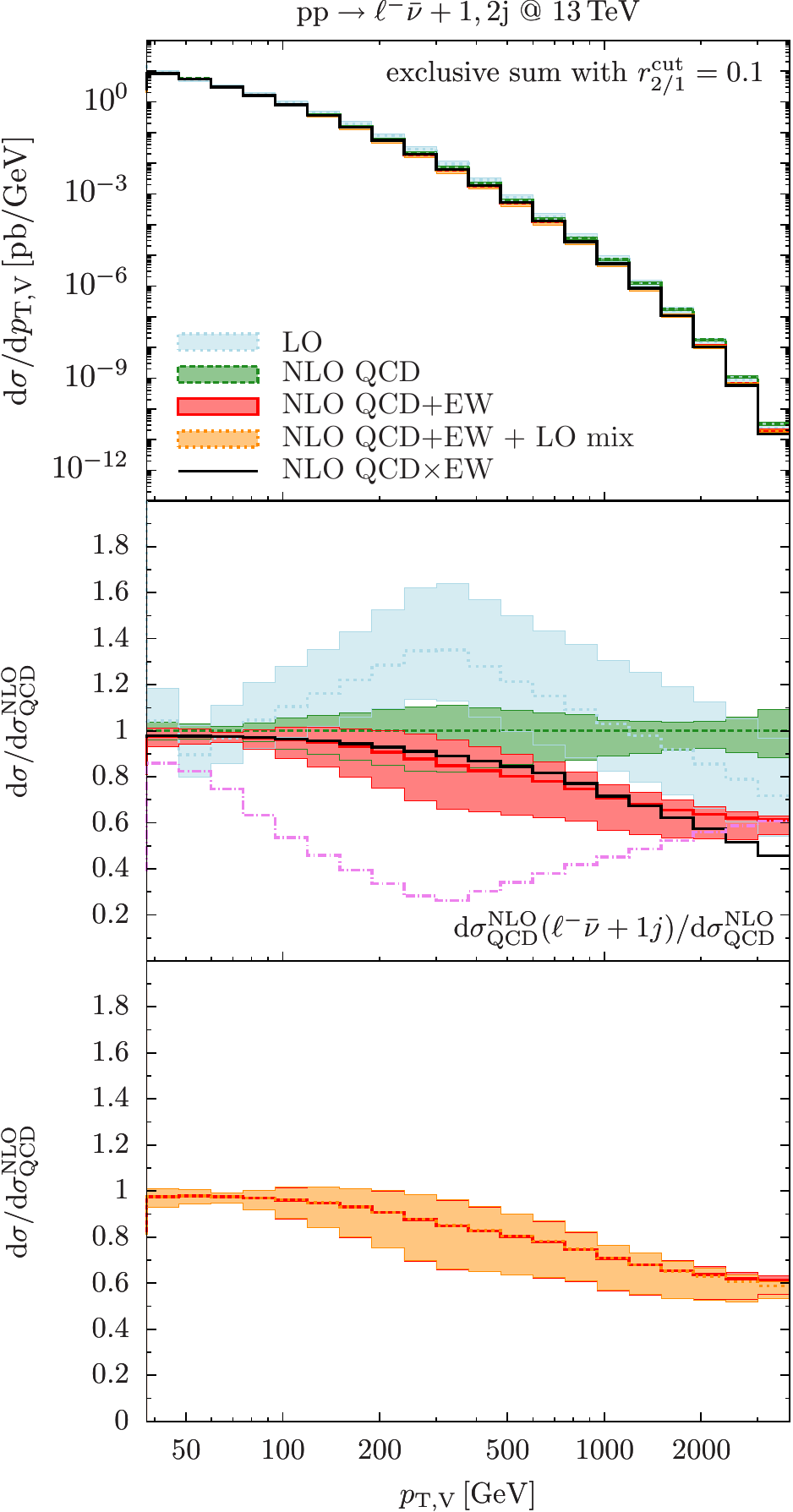}
         \qquad 
    \includegraphics[width=\relplotwidth\textwidth]{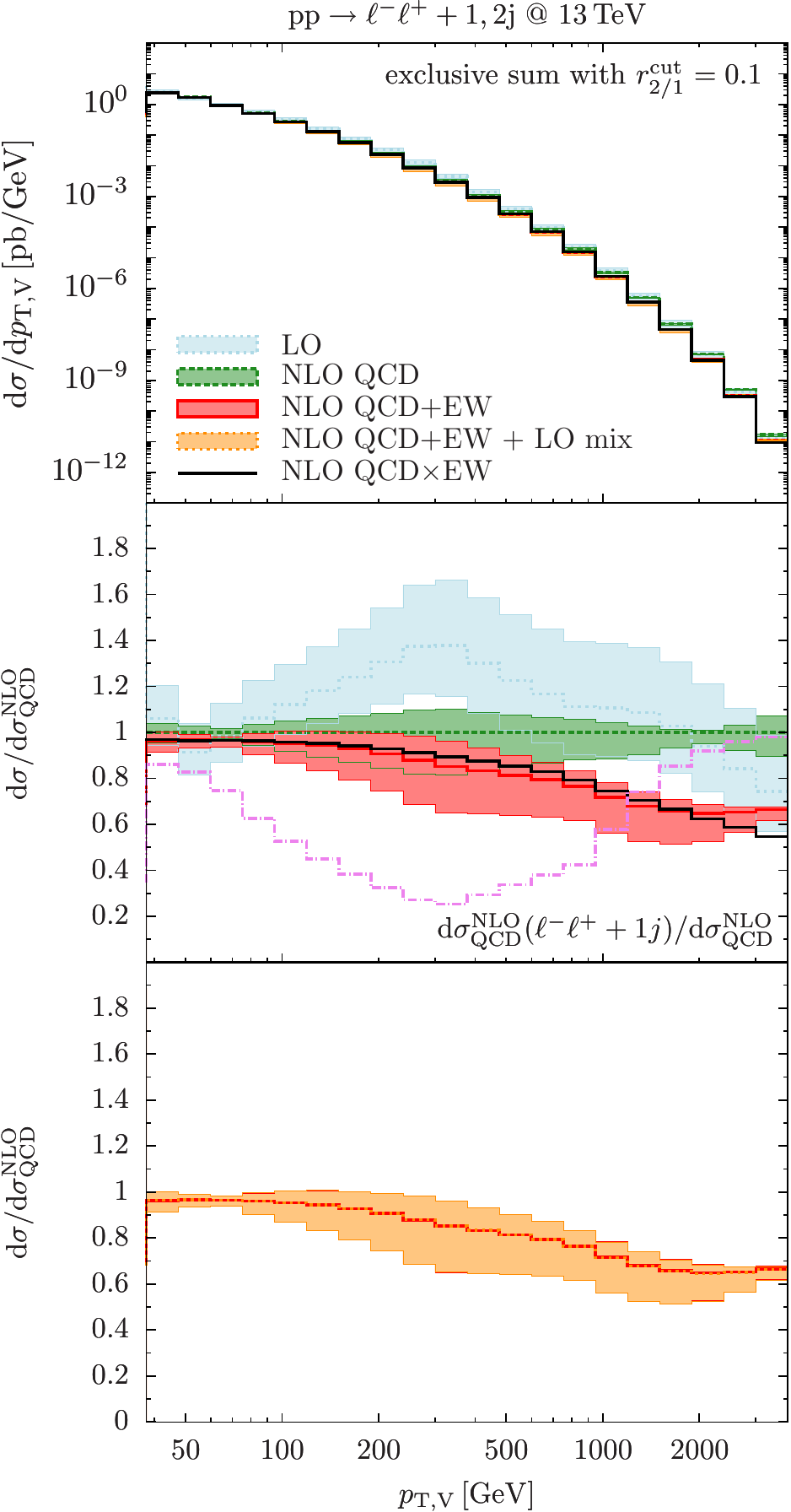}
\caption{
	  Differential distribution in the transverse momentum of the reconstructed vector 
	  boson, $\pTV$, for $\lmn\,+$\,jets (left) and $\lplm+$\,jets 
	  (right). Shown are predictions merged with exclusive sums using 
	  $r_{2/1}^{\mathrm{cut}}=0.1$. Curves and bands as in 
	  \reffi{fig:Vjj_pTj1_exclsum}.
	}
\label{fig:Vjj_pTV_exclsum}
\end{figure}

From the discussion of giant $K$-factors in~\refse{se:vj}
it should be clear that a
theoretically well behaved and phenomenologically sensible prediction of inclusive
$\PV+$\,jets cross sections can only be achieved combining 
NLO \QCDpEW cross sections for $\PV+1$\,jet and $\PV+$\,multijet production.
In this section, using a na\"ive merging approach based on exclusive sums~\cite{AlcarazMaestre:2012vp} we
illustrate
how the combination of one- and two-jet NLO samples can stabilise the perturbative \QCD convergence
of one-jet inclusive observables and guarantee a consistent behaviour of \EW corrections. 
Exclusive sums consist of combinations of fixed-order NLO calculations 
with variable jet multiplicity, where double counting 
is avoided by imposing appropriate cuts on the jet transverse momenta~\cite{AlcarazMaestre:2012vp}.
To combine $\PV+1\,$jet and $\PV+2\,$jet samples, we use the dimensionless variable
\beqar
\rratio&=&\frac{p_{\rT,j_2}}{p_{\rT,j_1}},
\eeqar
where $p_{\rT,j_1}$ and $p_{\rT,j_2}$ are the transverse momenta of the first two jets in the acceptance
region \refeq{eq:JETcuts}, and $p_{\rT,j_2}=0$ if there is only one jet within the acceptance.
The exclusive sum is built by imposing a 
$\rratio$ cut that separates the phase space into complementary regions,
$\rratio<\rcut$ and $\rratio>\rcut$. In order to avoid a double counting of 
topologies with two hard jets, the $\PV+1\,$jet sample is restricted to the 
region $\rratio<\rcut$, which corresponds to one-jet topologies, whereas the
$\PV+2\,$jet sample is used to populate the $\rratio>\rcut$ region, characterised by 
the presence of
two hard jets.

In Figures~\ref{fig:Vjj_pTj1_exclsum} and \ref{fig:Vjj_pTV_exclsum}
we present leading-jet and vector-boson \pT distributions for
inclusive $\lmn\,+$\,jets and $\lplm+$\,jets production, where 
the one- and two-jet contributions are combined using a separation cut
$\rratio=0.1$.
In the \pT distribution of the hardest jet we observe, as expected,
that above a few hundred GeV the impact of two-jet topologies is
overwhelming. In contrast, for $\pTjone<300$\,GeV their contribution tends to be suppressed by the
the acceptance cut on the second jet, $\pTjtwo>30$\,GeV, 
which effectively corresponds to $\rcut = 30\,\GeV/\pTjone>0.1$.
Thanks to the fact that the huge contributions
from two-jet topologies are included starting from Born level and 
supplemented by NLO \QCDpEW corrections, the exclusive-sums approach  
leads to a drastic improvement of the perturbative convergence 
as compared to 
fixed-order predictions for inclusive $\PV+$\,jet production in~\reffi{fig:Vj_pTV_pTj1} (left).
In fact, in the full \pT range considered we observe moderate NLO \QCD 
corrections and scale uncertainties.
Moreover, NLO \EW effects in~\reffi{fig:Vjj_pTj1_exclsum}
feature a consistent Sudakov behaviour, 
with $-20\%$ corrections around 2\,TeV.
Including also \QCDmEW mixed Born terms of $\ord(\alphaS\alpha^3)$ (LO mix) in
the two-jet sample, we observe that at the TeV scale their contribution becomes
sizable and can even overcompensate the negative effects of \EW Sudakov type.
Apart from these quantitative considerations, it is important to realise that
mixed Born contributions in the two-jet region ($\rratio>\rcut$) represent the
natural continuation of NLO mixed bremsstrahlung in the one-jet region
($\rratio>\rcut$). Their inclusion is thus crucial for a consistent combination
of different jet multiplicities.

In the vector-boson \pT distribution (\reffi{fig:Vjj_pTV_exclsum}) we observe
that, similarly as in~\reffi{fig:Vjj_pTj1_exclsum}, the relative weight of
$\PV+2\,$jet topologies grows with \pT up to about 300\,GeV as a result of the
acceptance cut on the second jet. However, in contrast to the case of the jet
\pT, in the region of high vector-boson \pT, where the separation cut $\rcut=0.1$ comes into
play, we see that one-jet contributions become increasingly important again.
This indicates that the higher a boost of the \PW boson is required by the
observable, the less likely it is to have two jets of comparable \pT, leading to
a hierarchical pattern of \QCD radiation. In this situation NLO calculations for
$\PV+1\,$jet prodution are expected to be reliable, and in fact we find that
inclusive $\PV+1\,$jet predictions and exclusive sums provide similarly well
behaved results. In both cases the quality of the perturbative \QCD expansion
turns out to be good, and in the multi-TeV regime we observe the usual negative
NLO \EW effects, which can become as large as $-40\%$. We also note that, as
compared to fixed-order $\PV+1\,$jet inclusive results in
\reffi{fig:Vj_pTV_pTj1} (left), exclusive sums lead to a smaller difference
between the \QCDpEW and \QCDtEW prescriptions. Finally, at high vector-boson \pT
we find that, consistent with the subleading role of two-jet topologies, mixed
Born contributions to $\PV+2$\,jets are irrelevant.

\subsection{Virtual approximation of NLO \EW corrections}
\label{sec:virtapp}

As discussed in the following, virtual \EW corrections with an appropriate
infrared subtraction can provide a fairly accurate approximation of exact
NLO \EW effects.  The fact that such an approximation does not require the
explicit integration of subtracted real-emission matrix elements represents
an important technical simplification.  In particular, since Born
and infrared-subtracted \EW virtual contributions live on the same $n$-parton
phase space, the combination of contributions with variable jet multiplicity 
can be realised with a multijet merging approach of LO complexity.
The main physical motivation for a virtual \EW approximation is given by the fact that Sudakov \EW
logarithms---the main source of large NLO \EW effects at high energy---arise
only from virtual corrections.  Moreover, in various cases, such as for
vector-boson production in association with one~\cite{Kuhn:2007cv} or two
jets~\cite{Denner:2014ina}, it turns out that a virtual \EW approximation can
provide percent-level accuracy for a wide range of observables and energy
scales, also well beyond the kinematic regions where 
Sudakov \EW logarithms become large.

Motivated by these observations, we adopt the following virtual approximation for the
NLO \EW corrections to $\PV+n$\,jet production,
\begin{equation}\label{eq:virtapprox}
\done\sigma_{n,\NLO\,\EWv}
    \,=\; \Big[\mr{B}_{n}(\Phi_n)
               +\mr{V}_{n,\EW}(\Phi_n)
               +\mr{I}_{n,\EW}(\Phi_n)
          \Big]\,\done\Phi_{n}.
\end{equation}
Here, $\mr{B}_{n}(\Phi_n)$ stands for the Born contribution of
$\ord(\alphaS^{n}\alpha^2)$, and $\mr{V}_{n,\EW}(\Phi_n)$ denotes the exact
one-loop \EW corrections of $\ord(\alphaS^{n}\alpha^3)$. The cancellation
of virtual infrared singularities is implemented through the
$\mr{I}_{n,\EW}(\Phi_n)$ term, which represents the NLO \EW generalisation of
the Catani--Seymour $\mathbf{I}$ operator~\cite{Dittmaier:1999mb,Dittmaier:2008md,Kallweit:2014xda}. This latter
term does not contain the \EW $\mathbf{K}$ and $\mathbf{P}$ operators.
It results from the endpoint term of the analytic integration over 
all dipole subtraction terms of $\ord(\alphaS^{n}\alpha^3)$, which
arise from the insertion of \QED and \QCD dipole kernels in
$\ord(\alphaS^{n}\alpha^2)$ squared Born matrix elements and
$\ord(\alphaS^{n-1}\alpha^3)$ \QCDmEW mixed Born terms, respectively.

\begin{figure}
\centering
   \includegraphics[width=\relplotwidthsmall\textwidth]{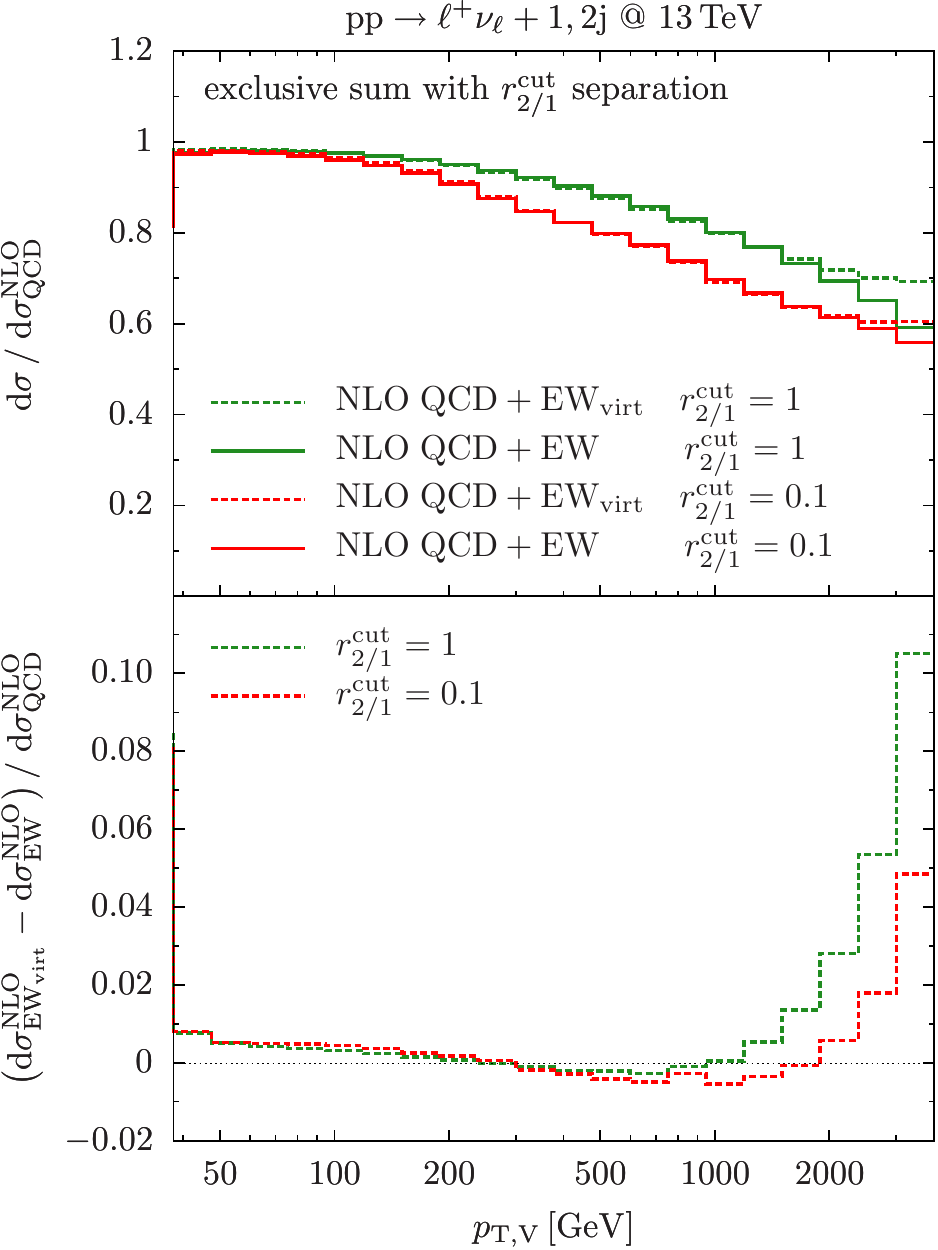}
         \qquad 
   \includegraphics[width=\relplotwidthsmall\textwidth]{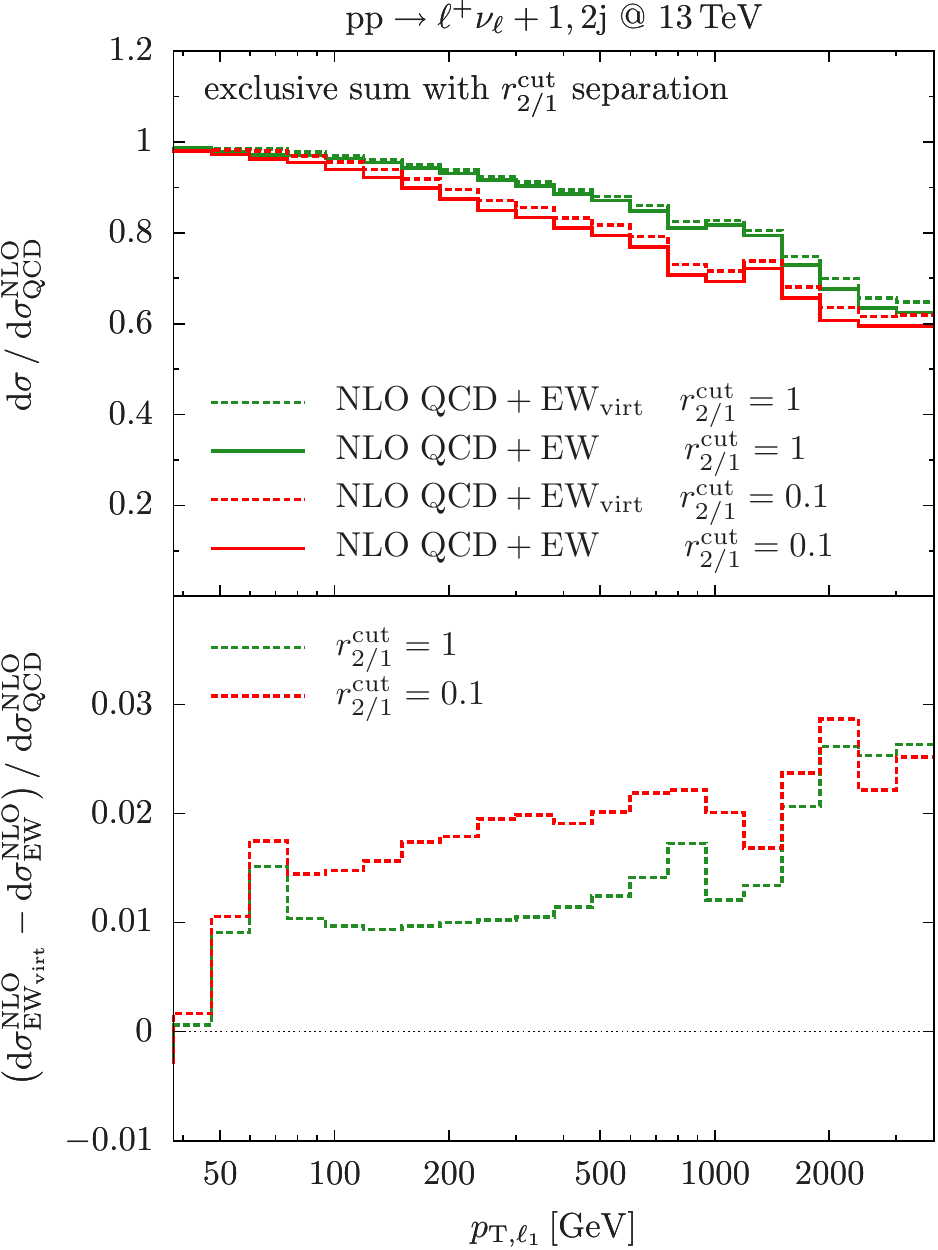}\\[4ex]
   \includegraphics[width=\relplotwidthsmall\textwidth]{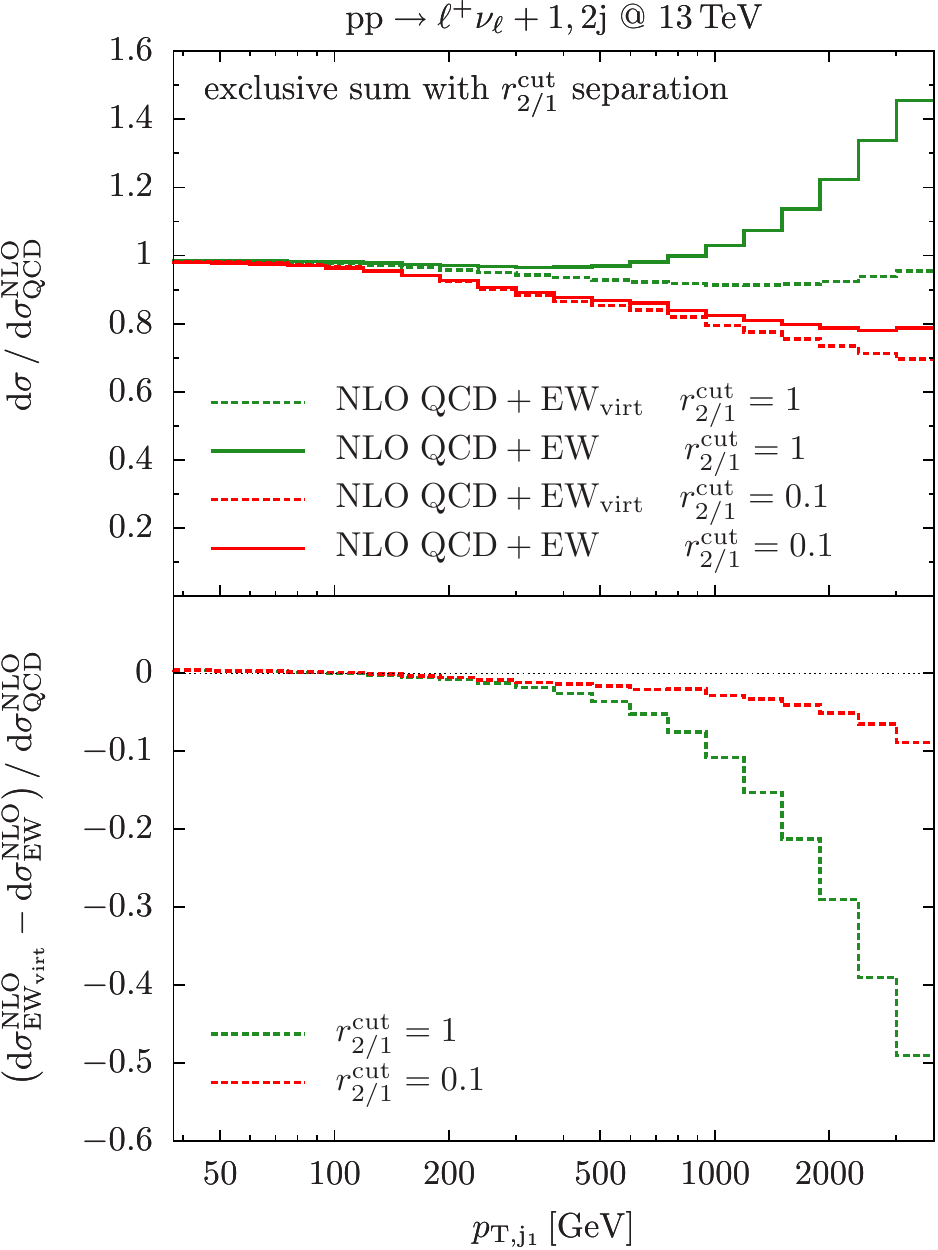}
         \qquad 
   \includegraphics[width=\relplotwidthsmall\textwidth]{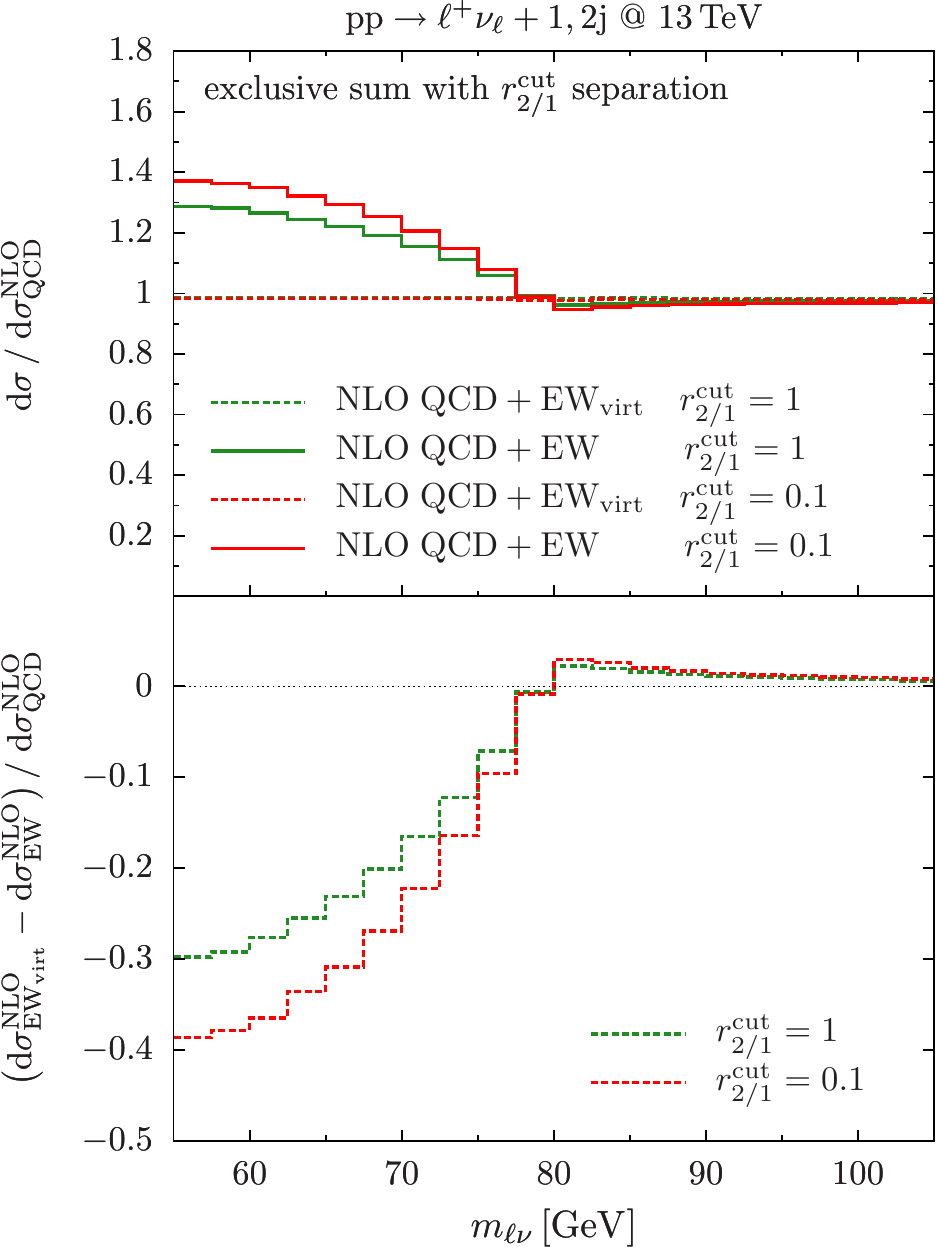}\\
\caption{
	  Exact NLO \EW predictions for $\lpn\,+$\,jets production at 13\,TeV 
	  are compared to the virtual approximation (NLO \EWvirt) of 
	  \refeq{eq:virtapprox}. All results are normalised to NLO \QCD 
	  predictions. The red curve represent parton-level predictions 
	  for $\lpn\,+1,2$\,jet combined in the exclusive-sums 
	  approach with a separation cut $\rcut=0.1$, while conventional 
	  predictions for $\lpn\,+1\,$jet ($\rcut=1$) are shown 
	  in green.
	}
\label{fig:exnejj_exclusive_sum_approx}
\end{figure}

In the following the shorthand \EWvirt will be used to denote the virtual \EW
approximation of~\refeq{eq:virtapprox}. 
The accuracy of this approximation is illustrated in
\reffis{fig:exnejj_exclusive_sum_approx}{fig:nenexjj_exclusive_sum_approx}
by comparing it to exact NLO \EW results for various (physical and unphysical) differential observables in $pp\to\ell\ell/\ell\nu/\nu\nu+1,2$\,jet 
production.\footnote{Process-dependent correction factors are introduced in \reffis{fig:exnejj_exclusive_sum_approx}{fig:nenexjj_exclusive_sum_approx}
such that the integrated NLO \QCDpEWvirt predictions match the complete
NLO \QCDpEW results. These factors are $k_{\nn}\approx 1.00$ for $\nn$\,+jets, 
$k_{\ell\nu}\approx 0.99$ for $\ell\nu$\,+jets and $k_{\ell\ell}\approx 0.98$ for $\lplm$\,+jets.}
Exact and approximate results are compared both for
the case of a conventional  NLO calculation for $\PV+1$\,jet ($\rcut=1$) and
combining NLO predictions for $\PV+1,2$\,jets with exclusive sums ($\rcut=0.1$). 
Exclusive sums provide a quantitative indication of the accuracy of the \EWvirt
approximation in a framework that mimics, although in a rough way,
the multijet merging approach that will be adopted in
\refses{se:mepsatnlo}{se:results_merged}.

\begin{figure}
\centering
   \includegraphics[width=\relplotwidthsmall\textwidth]{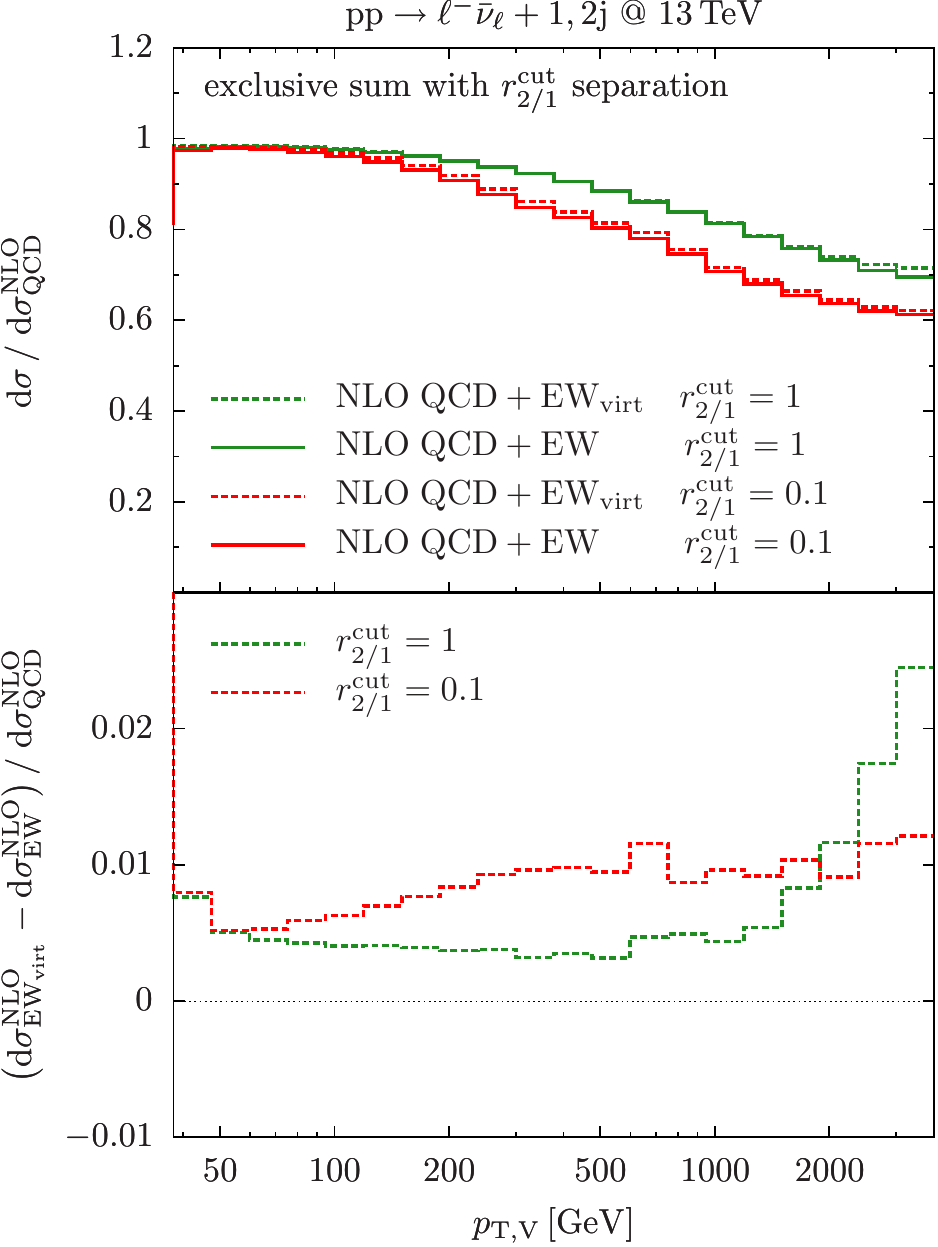}
         \qquad 
   \includegraphics[width=\relplotwidthsmall\textwidth]{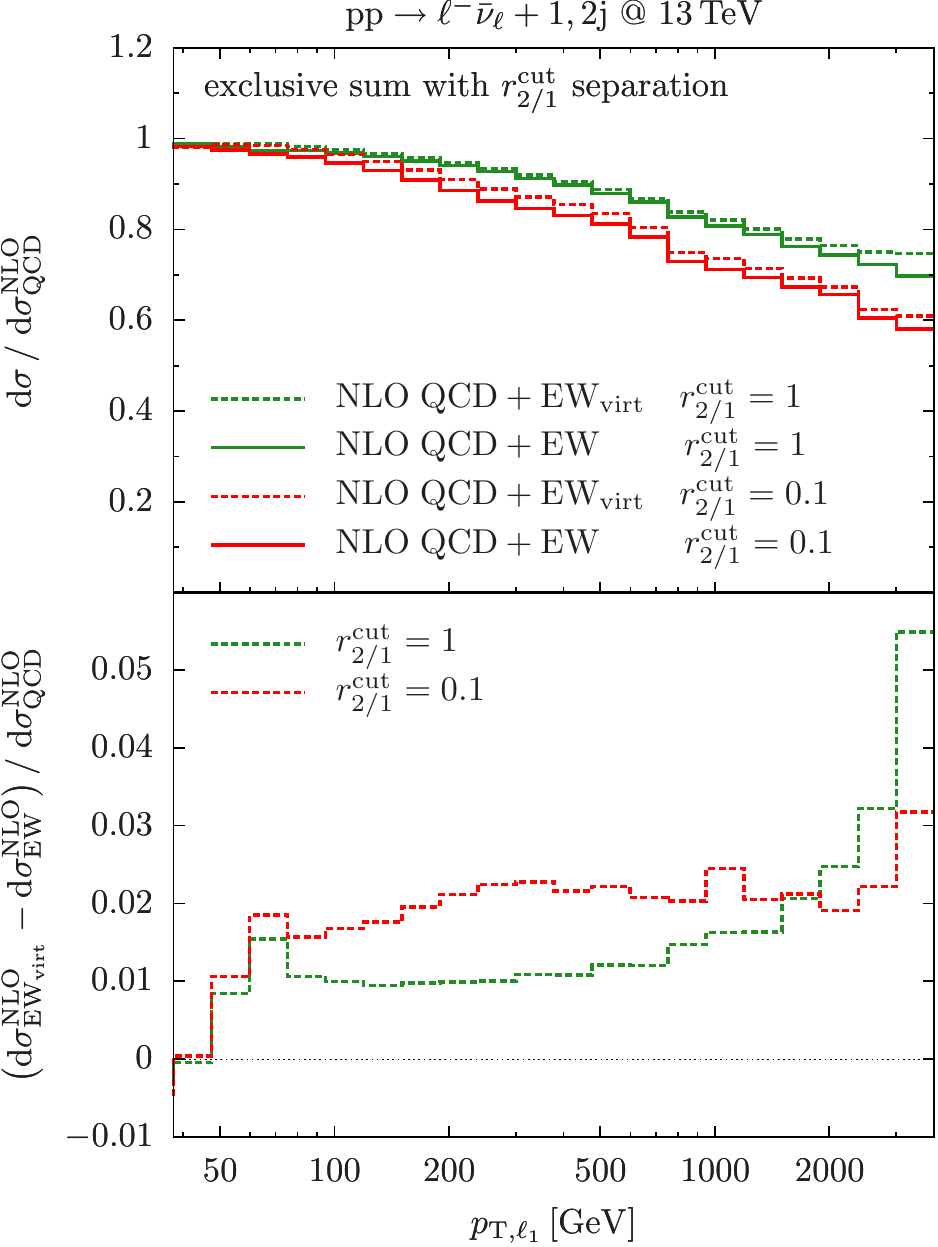}\\[4ex]
   \includegraphics[width=\relplotwidthsmall\textwidth]{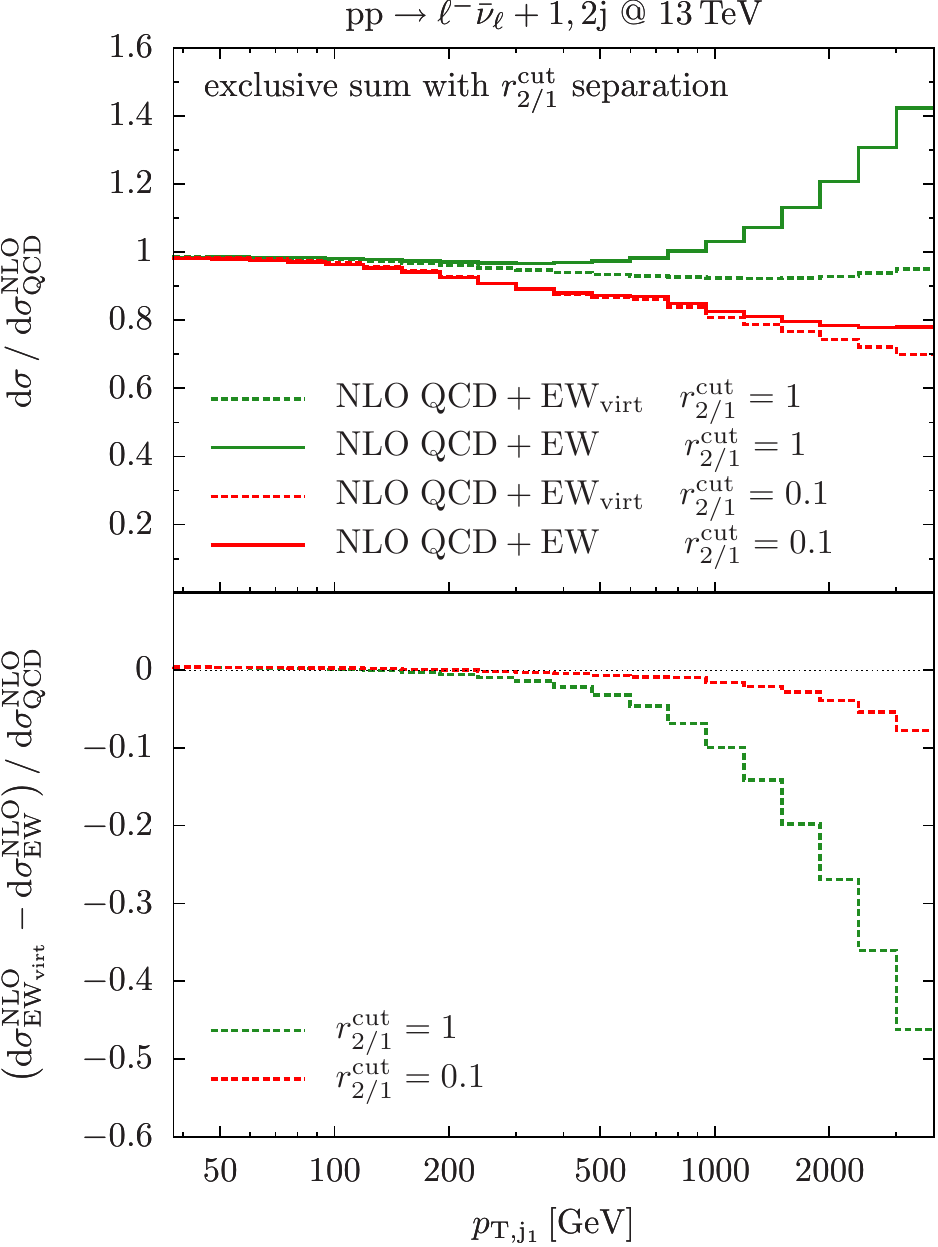}
         \qquad 
   \includegraphics[width=\relplotwidthsmall\textwidth]{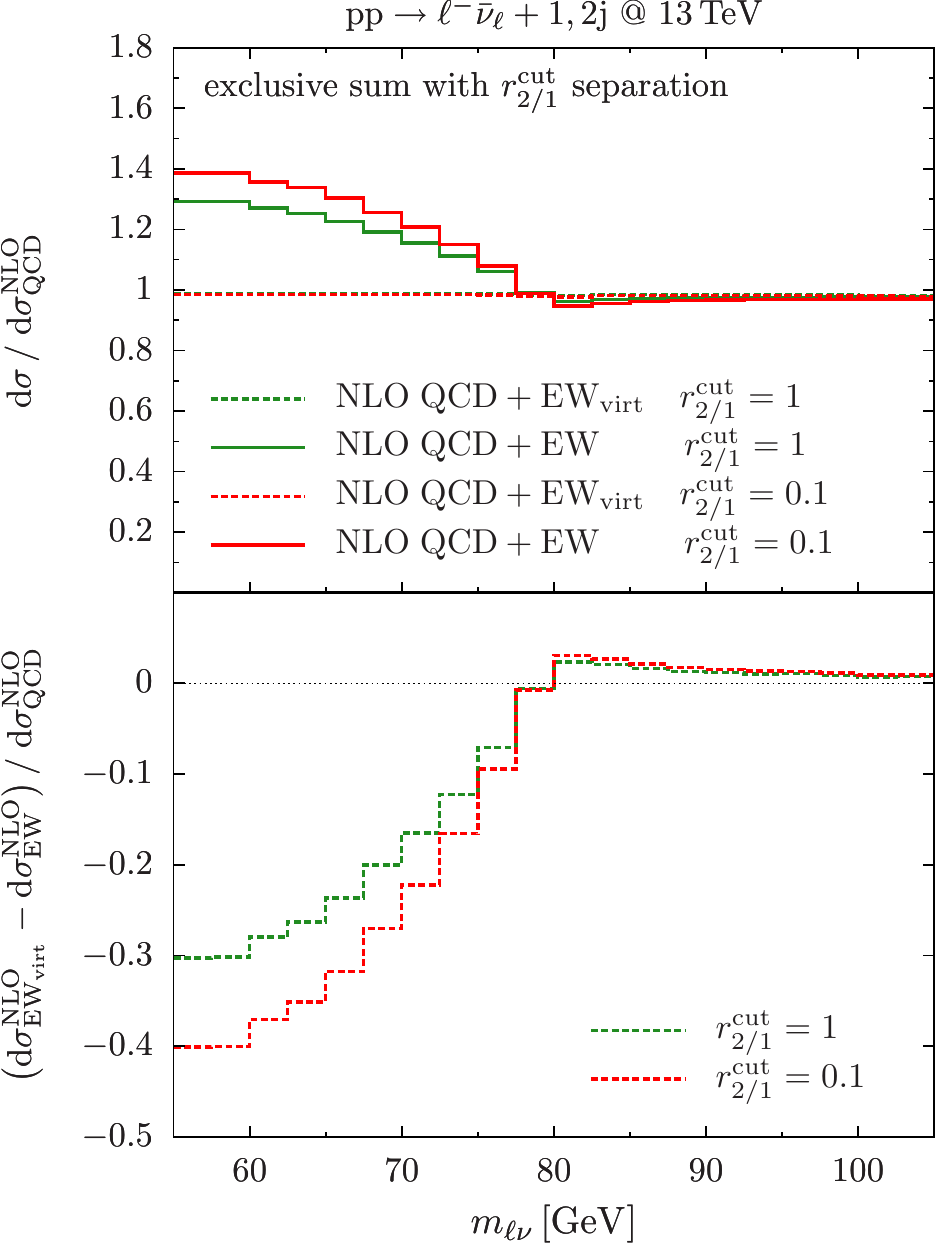}\\
\caption{
	  Exact NLO \EW predictions for $\ell^-\bar\nu_\ell+$\,jets production at 
	  13\,TeV are compared to the virtual approximation (NLO \EWvirt) 
	  of \refeq{eq:virtapprox}. Normalisation and exclusive-sum 
	  separation cuts are as in \reffi{fig:exnejj_exclusive_sum_approx}.
	}
\vspace*{4ex}
\label{fig:enexjj_exclusive_sum_approx}
\end{figure}

\begin{figure}
\centering
   \includegraphics[width=\relplotwidthsmall\textwidth]{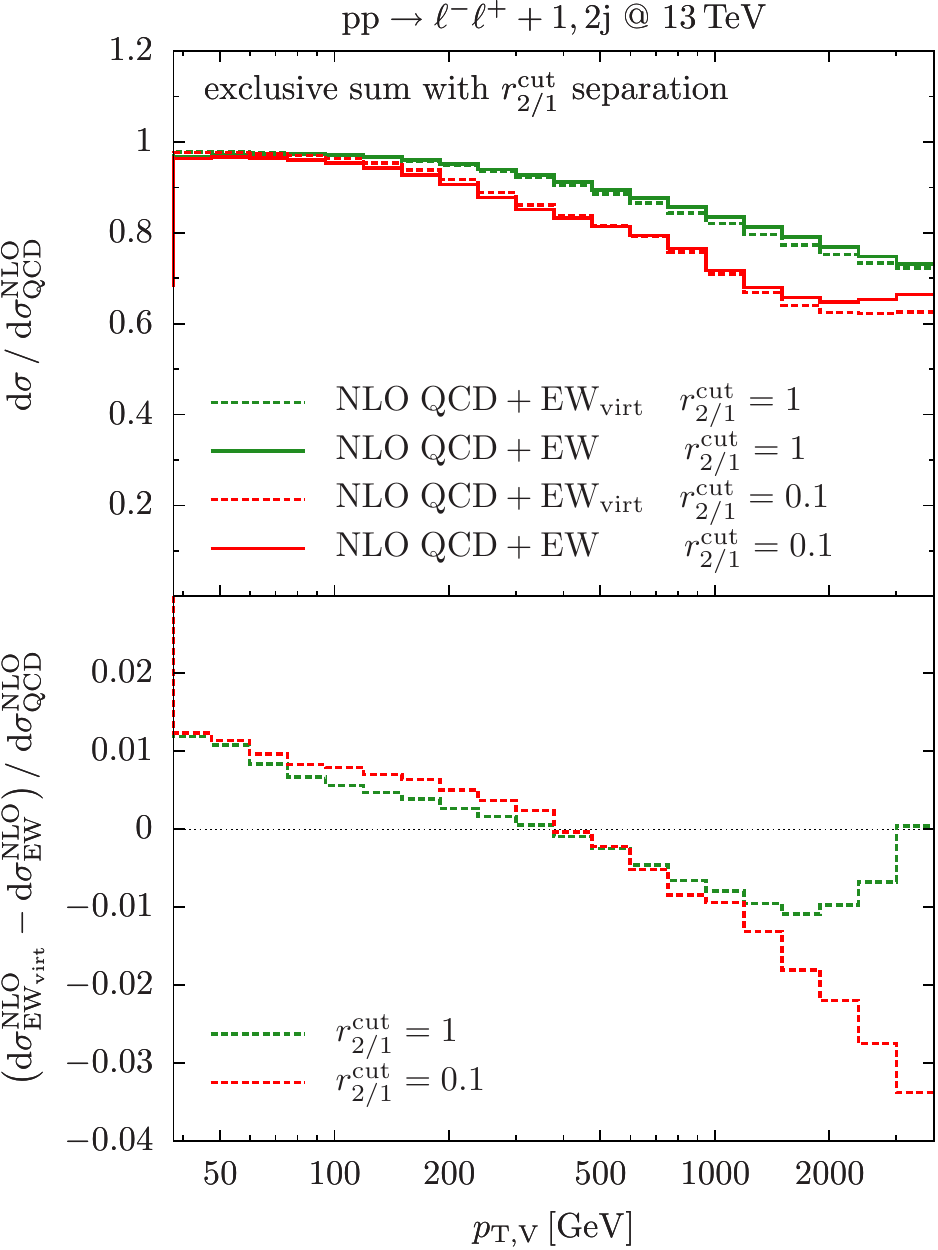}
         \qquad 
   \includegraphics[width=\relplotwidthsmall\textwidth]{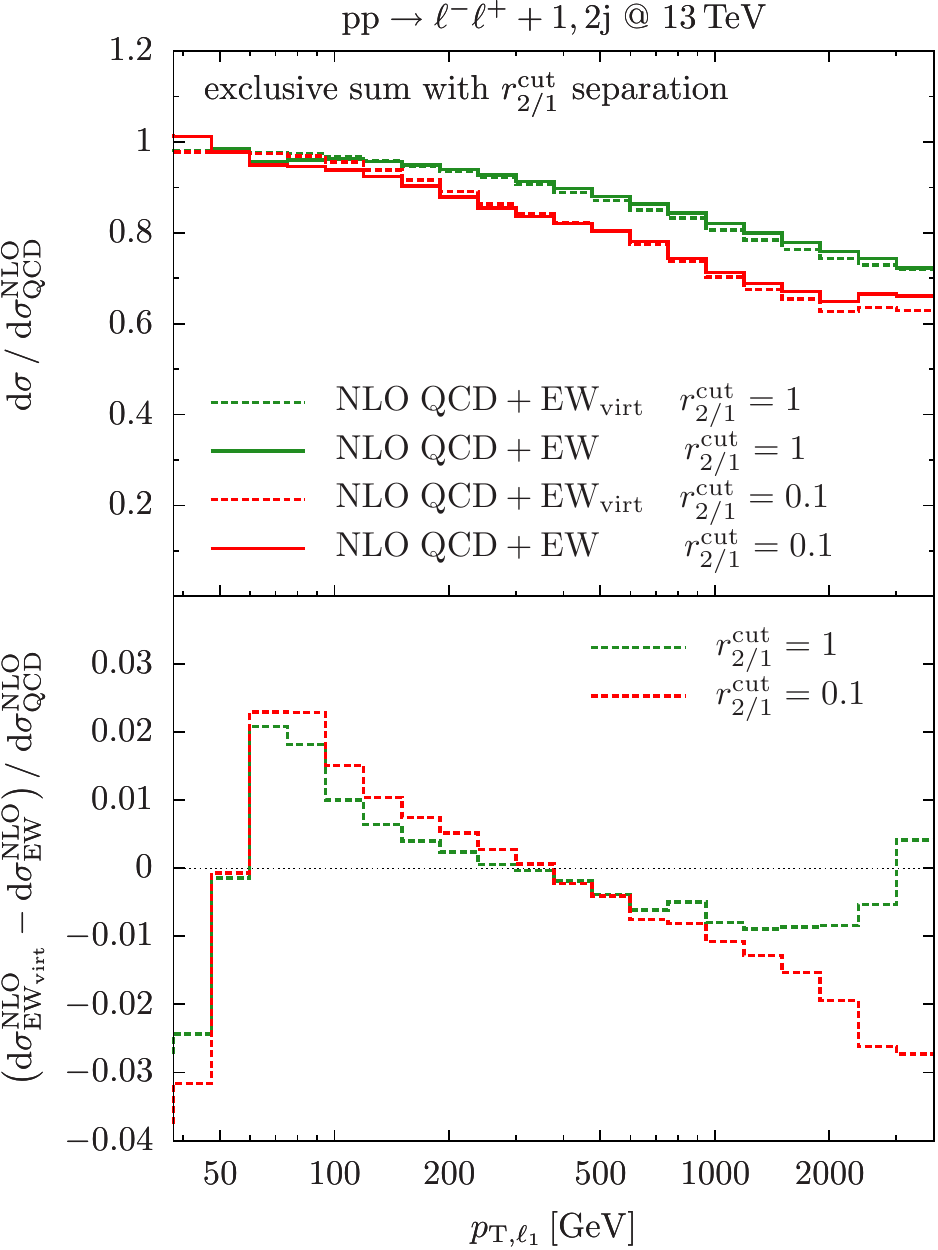}\\[4ex]
   \includegraphics[width=\relplotwidthsmall\textwidth]{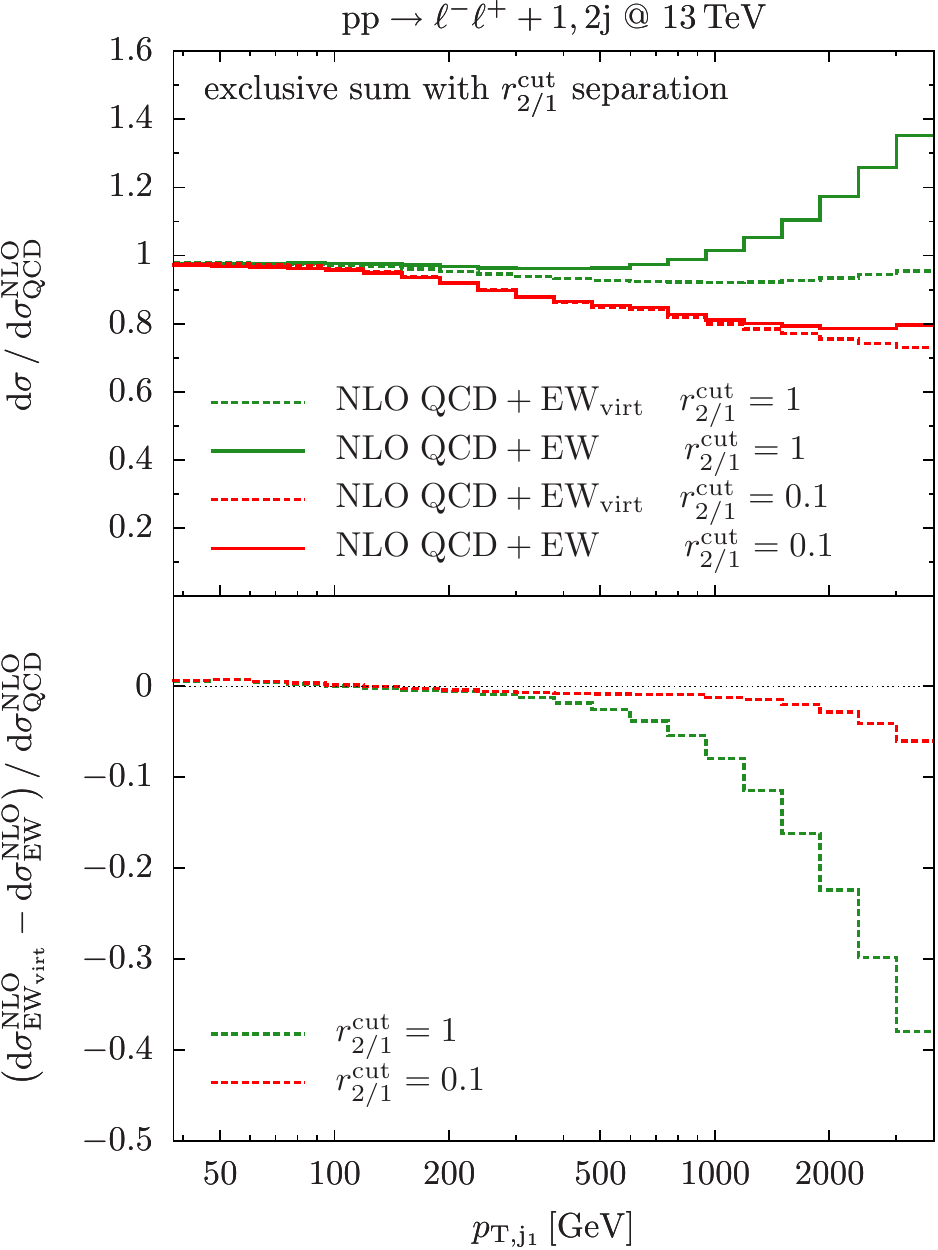}
         \qquad 
   \includegraphics[width=\relplotwidthsmall\textwidth]{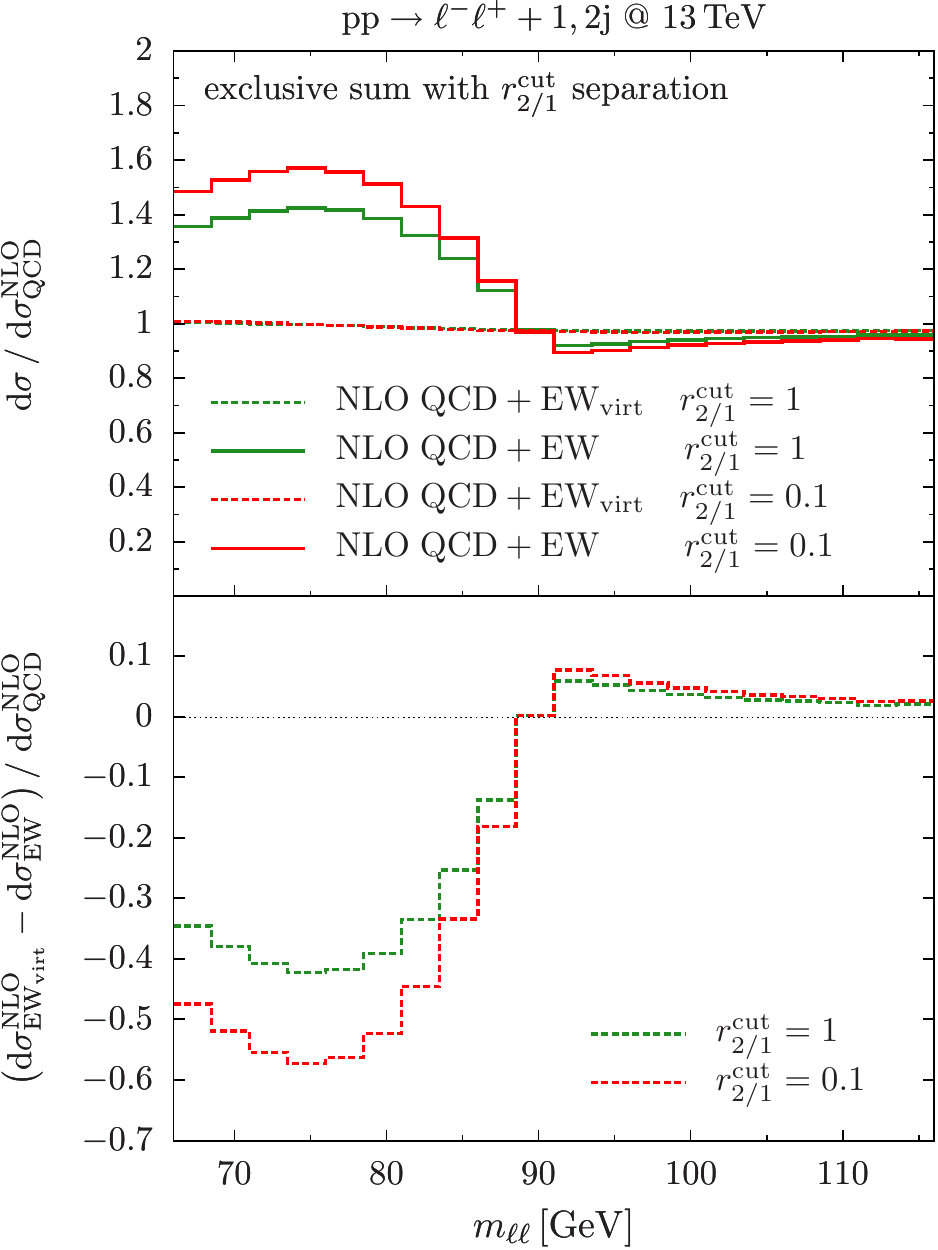}\\
\caption{
	  Exact NLO \EW predictions for $\ell^+\ell^-+$\,jets production at 13\,TeV 
	  are compared to the virtual approximation (NLO \EWvirt) of 
	  \refeq{eq:virtapprox}. Normalisation and exclusive-sum separation 
	  cuts are as in \reffi{fig:exnejj_exclusive_sum_approx}.
	}
\vspace*{4ex}
\label{fig:eexjj_exclusive_sum_approx}
\end{figure}

\begin{figure}
\centering
   \includegraphics[width=\relplotwidthsmall\textwidth]{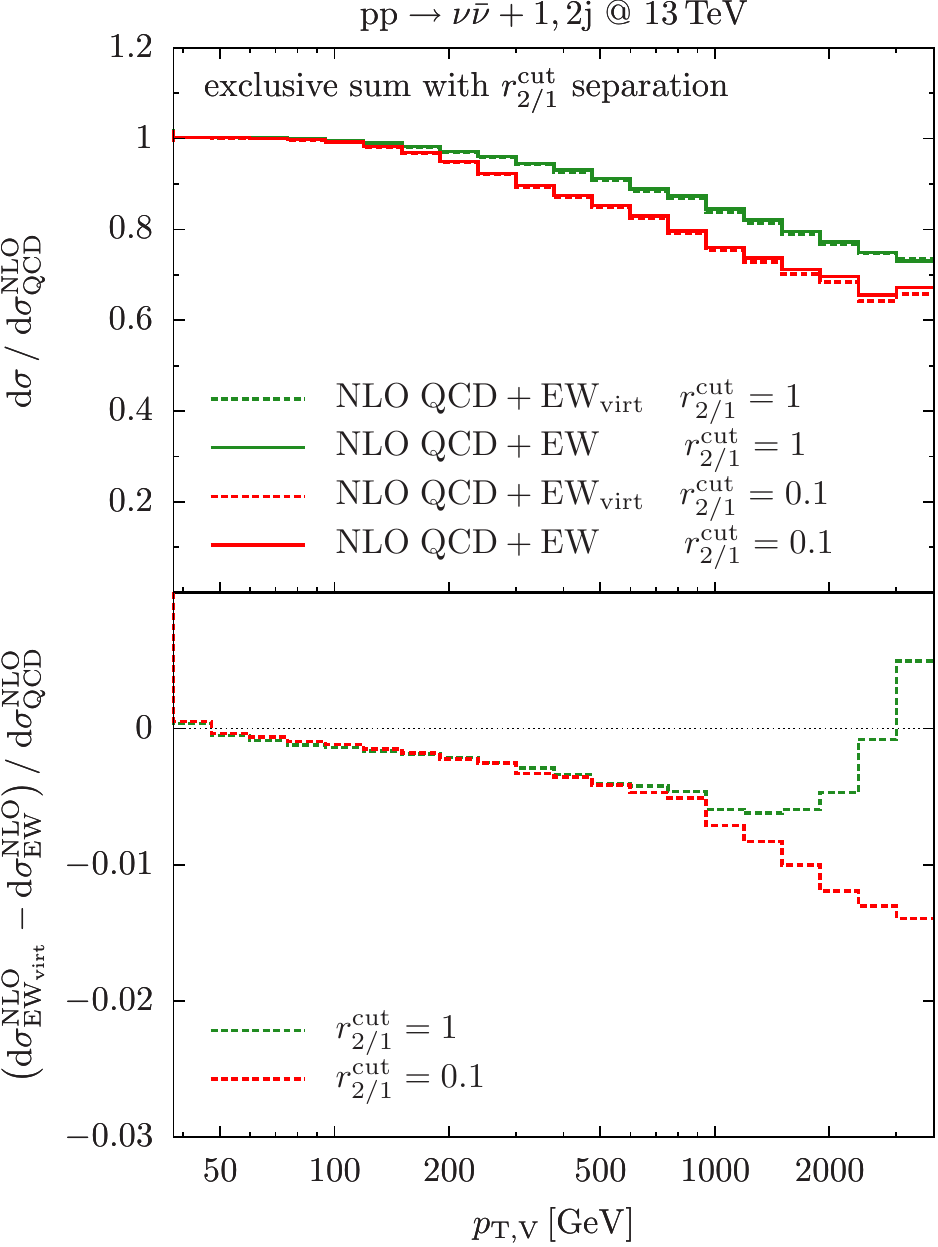}
   \qquad
   \hspace*{\relplotwidthsmall\textwidth}   \\[4ex]
   \includegraphics[width=\relplotwidthsmall\textwidth]{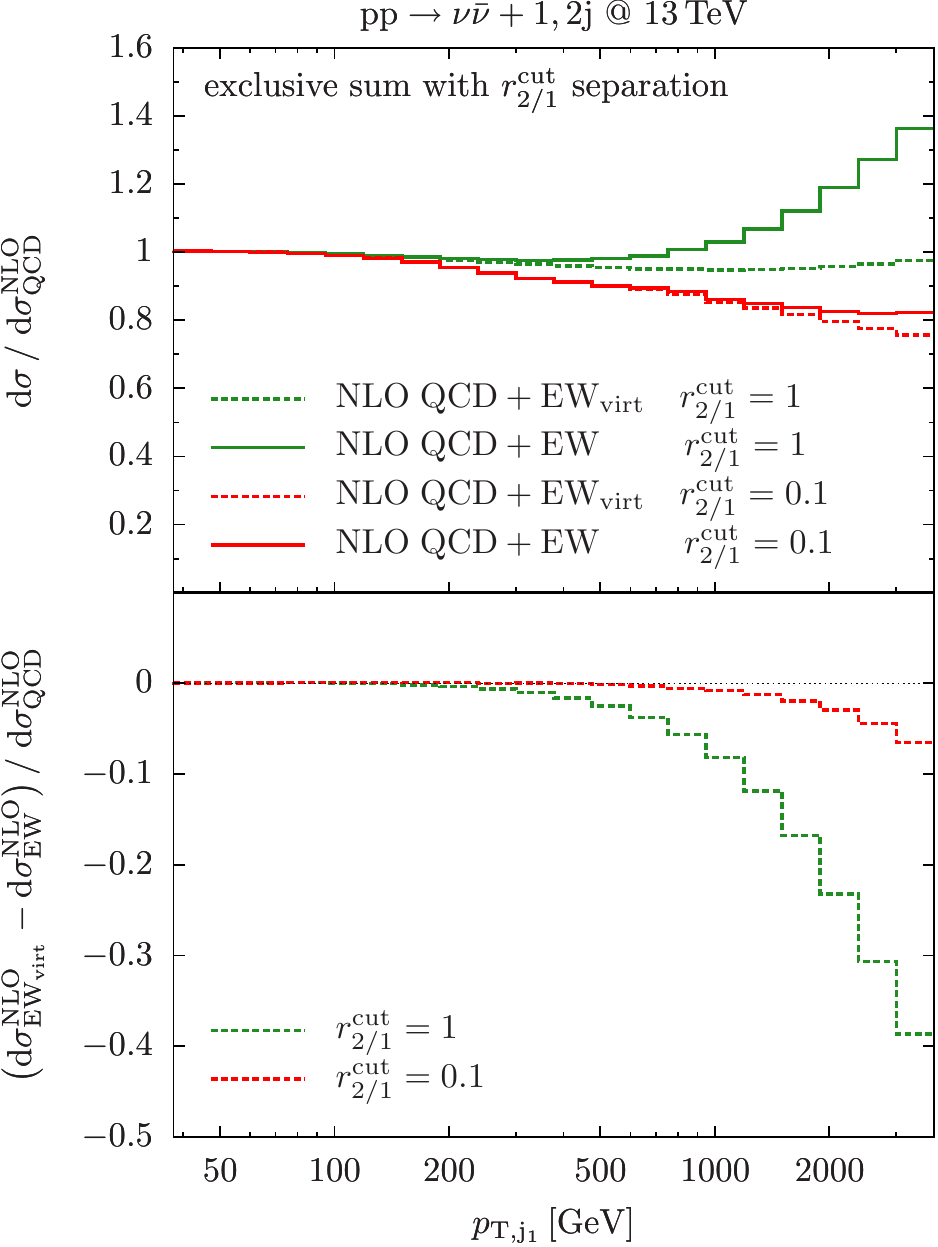}
   \qquad
      \includegraphics[width=\relplotwidthsmall\textwidth]{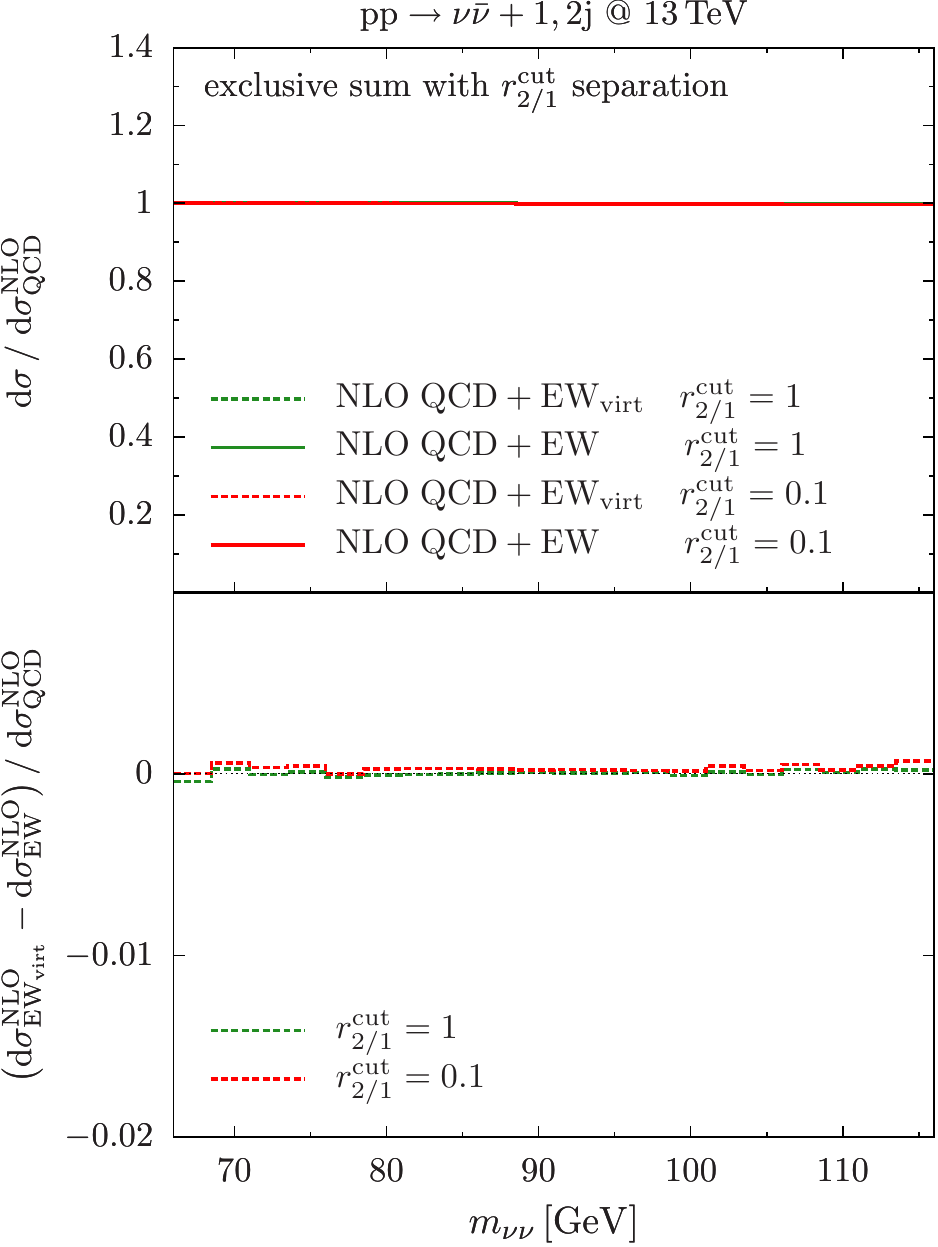}\\
\caption{
	  Exact NLO \EW predictions for $\nu_\ell\bar{\nu}_\ell+$\,jets production at 
	  13\,TeV are compared to the virtual approximation (NLO \EWvirt) 
	  of \refeq{eq:virtapprox}. Normalisation and exclusive-sum 
	  separation cuts are as in \reffi{fig:exnejj_exclusive_sum_approx}.
	}
\vspace*{4ex}
\label{fig:nenexjj_exclusive_sum_approx}
\end{figure}

For the various processes and distributions in 
\reffis{fig:exnejj_exclusive_sum_approx}{fig:nenexjj_exclusive_sum_approx}
the \EWvirt approximation turns out to be in
generally good agreement with exact NLO predictions.  The most striking
exception is given by the $m_{\ell\ell}$ and $m_{\ell\nu}$  invariant-mass distributions
in the off-shell region below the Breit--Wigner peak. In this case, real \QED radiation off
the charged leptons leads to corrections of a few tens of percent, which can not be reproduced by the 
\EWvirt approximation as exclusive real photon emission is not included. In contrast, for distributions in the transverse
momentum of the vector bosons or of the charged leptons that arise from their
decays, we observe very good agreement, typically at the 1--2\% level, from
low \pT up to the multi-TeV region.  

The leading-jet \pT distribution represents a special case. Here, the \EWvirt
approximation performs quite well up to about 500\,GeV, but at the TeV scale it 
is plagued by sizable inaccuracies.  We have checked that this is
largely due to the contribution of mixed bremsstrahlung, i.e.~to the \QCDmEW
interference between matrix elements that describe the real emission of \QCD partons at
$\ord(\alphaS^n\alpha^3)$.  Such contributions are not covered by the
\EWvirt approximation, while in a standard NLO \EW calculation for 
$\PV+1$\,jet ($\rcut=1$) they can reach 30--50\% in the multi-TeV region.  In
contrast, in the exclusive-sums approach mixed bremsstrahlung is suppressed
by the separation cut between 1-jet and 2-jet regions ($\rcut=0.1$), and
the discrepancy between exact \EW corrections and \EWvirt approximation is reduced to 
less than 10\% at 3\,TeV.
On the one hand, this level of agreement can be further improved
by lowering the value of the separation cut. Thus in our implementation of 
multijet merging we will adopt a merging cut that corresponds to $\rcut\ll 0.1$
in the multi-TeV region.
On the other hand, for a realistic description of \EW effects, it is clear that
the sizable contribution from mixed bremsstrahlung should be included 
also above the merging cut.
In the \MEPS framework described in \refse{se:ewmepsextension},
this will be achieved by complementing any $n$-jet Born contribution
of $\ord(\alphaS^n\alpha^2)$ by mixed Born contributions of 
$\ord(\alphaS^{n-1}\alpha^3)$ for any jet multiplicity $n\ge 2$ that is 
included in the merging procedure.\footnote{Note that mixed Born contributions
do not exist for $n\le 1$.} Such mixed Born contributions 
will provide an effective description of mixed bremsstrahlung that arises from regions with
$n-1$ jets. Moreover, their accuracy will be (approximately) increased by one order in $\alphaS$
through the implementation of NLO \EWvirt corrections of $\ord(\alphaS^{n}\alpha^3)$.

In summary, in absence of kinematic constraints that confine vector bosons
in the off-shell regime below the resonance region, combining the \EWvirt
approximation of~\refeq{eq:virtapprox} with mixed bremsstrahlung
contributions can reproduce NLO \EW predictions with an accuracy of 1--2\%
up to transverse momenta of the order of 1\,TeV or more.

\subsection{\texorpdfstring{\protect\MEPS}{MEPS} merging at NLO \QCD}
\label{se:mepsatnlo}

As a basis to combine NLO \EW corrections with multijet merging,
in this section we recapitulate the essential features of the \MEPS merging method~\cite{Hoeche:2009rj,Hoeche:2010kg,Hoeche:2012yf,Gehrmann:2012yg}. 
This technique allows one to generate inclusive event samples with
variable jet multiplicity in such a way that events with $n=0,1,\dots,\nmax$ jets are
described in terms of corresponding $n$-jet matrix elements at LO or NLO accuracy. 
To this end, resolved jets are separated from unresolved emissions by means of a 
so-called merging scale, $\qcut$, and the phase space is split into different
regions according to the number of resolved jets.
More precisely, the  phase-space partitioning is formulated in terms of the $\kT$-type jet-resolution parameters $Q_{1}>Q_{2}>\dots >Q_{\nmax}$,
which represent the resolution scales of the first, second, and subsequent emissions. The 
$n$-jet regions for $0\le n < \nmax$ are thus defined through 
\begin{equation}
Q_{1}>\dots > Q_{n}>\qcut>Q_{n+1}>\dots\;.
\end{equation} 
In the leading-order formulation of the \MEPS method~\cite{Hoeche:2009rj}, called \MEPSatLO, 
the exclusive cross sections with $0\le n < \nmax$ resolved jets are 
generated according to\footnote{Here we employ the notation of~\cite{Hoeche:2014rya} 
in a slightly simplified form. For a more detailed discussion of technical aspects 
we refer to the original publications \cite{Hoeche:2009rj,Hoeche:2010kg,
  Hoeche:2012yf,Gehrmann:2012yg}.}
\begin{equation}\label{eq:mepsatlo}
    \done\sigma_{n}^\text{(\MEPSatLO)}
    \,=\;\done\Phi_{n}\,
          \mr{B}_{n}(\Phi_n)\,\Theta(Q_{n}-\qcut)\,
          \mc{F}_{n}(\mu_Q^2\,;<\!\qcut),
\end{equation}
where $\mr{B}_{n}(\Phi_n)$ summarises the relevant squared Born matrix
elements convoluted with PDFs and summed/averaged over all partonic degrees
of freedom.  The theta function ensures that all partons in the matrix
elements correspond to resolved jets, while $\mc{F}_{n}(\mu_Q^2\,;<\!\qcut)$
denotes a truncated vetoed parton shower that is restricted to the unresolved
regions, $Q<\qcut$, as explained in more detail below.
For the highest matrix-element multiplicity, $n=\nmax$, the region 
$Q_{\nmax}>Q>\qcut$ is inclusive with respect to higher-order radiation. Thus, the 
$\qcut$-veto is relaxed to a $Q_{\nmax}$-veto, and the 
truncated parton shower can fill the whole phase space below.

The truncated vetoed shower supplements multijet matrix elements with
Sudakov suppression factors that render resolved jet emissions equally
exclusive as shower emissions. In combination with the CKKW scale
choice~\cite{Catani:2001cc,Krauss:2002up}, this guarantees a 
smooth transition from matrix-element to parton-shower predictions across
$\qcut$ and ensures the restauration of the parton shower's resummation 
properties in the matrix-element region. As a result, the $\qcut$ dependence of physical observables is kept
at a formally subleading level with respect to the logarithmic accuracy of
the parton shower. The implementation of the above aspects of the merging procedure 
requires, for each multijet event, the determination of a 
would-be shower history consisting of a 
$2\to 2$ core process, characterised by a certain core scale $\mu_\core$, 
and a series of subsequent branchings at scales 
$t_1,t_2,\dots,t_M$. In the \MEPS approach, shower histories are determined by
probabilistic clustering of multijet final states 
based on the inversion of the \Sherpa parton shower. 

The truncated parton shower $\mc{F}_{n}(\mu_Q^2\,;<\!\qcut)$ in
\refeq{eq:mepsatlo} starts at the resummation scale $\mu_Q^2=t_0=\mu_\core^2$ and is stopped and
restarted at each reconstructed branching scale $t_1,\dots,t_M$. At each stage 
a kernel corresponding to the actual partially clustered configuration 
is used. Finally, the shower terminates 
at the infrared cutoff, $t_c$.  The Sudakov form factor 
that guarantees the exclusiveness of $n$-jet contributions
is generated by vetoing the
entire event in case of any resolved emission ($Q>\qcut$) of the truncated shower
for $t_0>t>t_c$. 
Since the role of the Sudakov suppression 
is to avoid double counting
between contributions with different numbers of resolved 
jets, unresolved emissions ($Q<\qcut$) are not vetoed.\footnote{Note that, 
for $n$-jet configurations, in spite of
$Q_n>\qcut$, also truncated shower emissions with $t>t_n$ can give rise to
unresolved jets with $Q<\qcut$ due to the different nature of the shower
evolution variable $t$ and the $\kT$-measure $Q$.}

The factorisation scale is set equal to the core scale,
$\mu_F=\mu_\core$, while the strong coupling $\alpha_S$ in multijet Born matrix elements
is evaluated at the renormalisation scale $\mu_R=\mu_\CKKW$, defined through 
\beq\label{eq:ckkwscale}
\alpha_S^N(\mu^2_\CKKW)=\alpha_S^{N-M}(\mu^2_\core)\;\alpha_S(t_1)\dots\alpha_S(t_M),
\eeq
where $\alpha_S^N$ and $\alphaS^{N-M}$ are the overall $\alpha_S$ factors for the 
LO cross section of the actual multijet process and for the related $2\to 2$ core process, 
respectively. 

In the case of $\PV+\,$jets, the shower history is determined by stepwise 
clustering of $\PV+\,$multijet events 
based on the relative probability of all possible \QCD and \EW splitting 
processes, using matrix-element information to select allowed 
states only.\footnote{For example, in a $gq\to\lplm q$ configuration identifying 
a $q\to qg$ splitting would be allowed by the parton shower and preferred in 
many regions of phase space over the alternatives. However, this would lead to a $gg\to\lplm$ 
configuration and, thus, identifying such a splitting needs to be prevented.} 
In particular, also the creation of vector bosons and their (off-shell) decays are 
treated as possible splitting processes.
Thus the clustering of $\PV+$multijet events 
terminates with three possible $2\to 2$ core processes: 
$pp\to 2\ell$, $pp\to \PV j$ and $pp\to jj$.   
The corresponding  default core scales in \Sherpa read\footnote{The 
core scale $\mu_{\core,jj}$ is driven by the smallest Mandelstam invariant, i.e.~by the scale 
associated with the dominant topology in the $pp\to jj$ core process. 
In practice $\mu_{\core,jj}$  is fairly close to the jet transverse momentum after clustering.}
\beq\label{eq:corescale}
\mu_{\core, \ell\ell}=m_{\ell\ell},\qquad
\mu_{\core, V\!j}=\frac{1}{2} E_{T,V} =\frac{1}{2}\sqrt{M^2_V+p^2_{T,V}},\qquad
\mu_{\core, jj}=\frac{1}{2} \left(\frac{1}{\hat s} -\frac{1}{\hat t} -\frac{1}{\hat u} \right)^{-\frac{1}{2}}.
\eeq 
Note that excluding \EW splittings from the clustering procedure would always lead to a Drell--Yan 
core process and a core scale $\mu_\core=m_{\ell\ell}=\ord{(M_{\PZ,\PW})}$,
which is clearly inappropriate at high transverse momenta.
Including all \QCD and \EW splittings in the clustering algorithm
is thus crucial for the consistent determination of the 
hard core processes and the related scale. 
In particular, it allows for shower histories where $\PV+\,$multijet production 
proceeds via hard dijet production and subsequent soft vector-boson 
emission, which corresponds to the dominant mechanism of $\PV+\,$jets production 
at high jet \pT.

The \MEPSatNLO merging method~\cite{Hoeche:2012yf,Gehrmann:2012yg}
upgrades LO merging to NLO \QCD in the
\MCatNLO framework~\cite{Frixione:2002ik,Frixione:2003ei,Hoeche:2011fd,Hoeche:2012fm}.
It can be summarised through the following formula for exclusive $n$-jet cross sections,
\begin{equation}\label{eq:mepsatnlo}
  \begin{split}
    \done\sigma_{n}^\text{(\MEPSatNLO)}
    \,=&\;\biggl[\done\Phi_{n}\,
          \tilde{\mr{B}}_{n}(\Phi_n)\,
          \bar{\mc{F}}_{n}(\mu_Q^2\,;<\!\qcut)
\\
    &{}\;\;
    +\done\Phi_{n+1}\,
         \tilde{\mr{H}}_{n}(\Phi_{n+1})\,
         \Theta(\qcut-Q_{n+1})\,
         \mc{F}_{n+1}(\mu_Q^2\,;<\!\qcut)\biggr]\,\Theta(Q_{n}-\qcut)\;.
  \end{split}
\end{equation}
As discussed in more detail below, 
the $\tilde{\mr{B}}_{n}(\Phi_n)$ term 
corresponds to so-called soft events in \MCatNLO
and describes
$n$ resolved partons ($Q_{n}>\qcut$)
at matrix-element level including virtual corrections.
The $\tilde{\mr{H}}_{n}(\Phi_{n+1})$ term corresponds to so-called hard events
in \MCatNLO. It involves subtracted matrix elements with $n$ resolved partons, plus an additional parton whose emission
is constrained in the unresolved region $(Q_{n+1}<\qcut)$ in order to avoid double counting with 
matrix elements of higher jet multiplicity. Of course, for $n=\nmax$ 
this constraint on the real emission is not required, and the corresponding 
theta function in \refeq{eq:mepsatnlo} is omitted.

Similarly as in the LO case, soft and hard events in~\refeq{eq:mepsatnlo}
are used as seeds of truncated vetoed parton showers with
starting scale $\mu_Q=\mu_\core$ and a veto against emissions with
$Q>\qcut$. The veto is relaxed when the maximum jet multiplicity $n=\nmax$ is reached.
In \MEPSatNLO, the truncated shower  that is applied to soft events,
$\bar{\mc{F}}_{n}(\mu_Q^2\,;<\!\qcut)$, is matched to the first matrix-element emission.
To this end, the first emission is generated by the kernel\footnote{Here the veto against emissions with 
$Q>\qcut$ is implicitly understood.}~\cite{Schumann:2007mg,Hoeche:2010kg}
\begin{equation}\label{eq:compound_kernel}
   \tilde{\mr{D}}_{n}(\Phi_{n+1})\,=\;
   \mr{D}_{n}(\Phi_{n+1})\,\Theta(t_{n}-t_{n+1})\;
   +\sum_{j=0}^{n-1}\mr{B}_{n}(\Phi_n)\,\mr{K}_{j}(\Phi_{1,n+1})\,
      \Theta(t_j-t_{n+1})\,\Theta(t_{n+1}-t_{j+1})\,\Big|_{\,t_0=\mu_Q^2}\;.
\end{equation}
Here, $\mr{D}_{n}(\Phi_{n+1})$ denotes exact Catani--Seymour subtraction terms.
They are used to generate emissions with hardness $t_{n+1}<t_n$, which arise from $n$-parton configurations,
and they match the full-colour infrared singularity structure of real-emission matrix elements.
The remaining terms in \refeq{eq:compound_kernel}
describe intermediate emissions with hardness $t_{n+1}\in [t_j,t_{j+1}]$
that arise from partially clustered configurations with $0\le j<n$ partons
and corresponding Catani--Seymour kernels $\mr{K}_j$ in the usual leading-colour approximation of the parton shower. 
The matching of the truncated vetoed parton shower to the first NLO emission results in the following expression for
hard events,   
\begin{equation}\label{eq:mepsatnlo_h}
  \begin{split}
    \tilde{\mr{H}}_{n}(\Phi_{n+1})=&\;
    \mr{R}_{n}(\Phi_{n+1})-
      \tilde{\mr{D}}_{n}(\Phi_{n+1})\,\Theta(\mu_Q^2-t_{n+1})\;,
  \end{split}
\end{equation}
where $\mr{R}_{n}(\Phi_{n+1})$ stands for real-emission matrix elements.
The soft term in~\refeq{eq:mepsatnlo} reads 
\begin{equation}\label{eq:mepsatnlo_bbar}
  \begin{split}
    \tilde{\mr{B}}_{n}(\Phi_{n})=&\;
      \mr{B}_{n}(\Phi_{n})+\tilde{\mr{V}}_{n}(\Phi_{n})
      +\int\done\Phi_1\,\tilde{\mr{D}}_{n}(\Phi_{n},\Phi_1)\,\Theta(\mu_Q^2-t_{n+1})\;.
  \end{split}
\end{equation}
It comprises a Born contribution, $\mr{B}_{n}(\Phi_{n})$, a term 
$\tilde{\mr{V}}_{n}(\Phi_{n})$ consisting of virtual \QCD corrections and 
initial-state collinear counterterms,\footnote{Such contributions correspond 
to the $\mu_F$ dependent part of the integrated $\mathbf{P}$ operator 
in the Catani--Seymour approach.}
and the integrated subtraction terms~\refeq{eq:compound_kernel}
associated with the truncated parton shower.
Similarly as for LO merging, we set $\mu_F=\mu_\core$, and
the renormalisation scale is chosen according to~\refeq{eq:ckkwscale}.

\subsection{Extension of \texorpdfstring{\protect\MEPS}{MEPS} merging to NLO \texorpdfstring{\QCDpEW}{QCD+EW}}
\label{se:ewmepsextension}

Let us now turn to the extension of the \MEPSatNLO formalism to 
also include NLO \EW effects.
While the method that we are going to introduce is entirely general, 
for the convenience of the discussion, in the following 
we will adopt a counting of $\alphaS$ and $\alpha$ couplings
that corresponds to the specific case of $\PV+\,$multijet production
with off-shell vector-boson decays. In this case,
in phase-space regions with $n$ resolved jets,
LO and NLO \QCD contributions of $\ord(\alphaS^{n}\alpha^2)$ and 
$\ord(\alphaS^{n+1}\alpha^2)$ will be supplemented by NLO \EW corrections
of $\ord(\alphaS^{n}\alpha^3)$ and mixed \QCDmEW Born terms of 
$\ord(\alphaS^{n-1}\alpha^3)$.

Besides all relevant tree plus virtual amplitudes and 
Catani--Seymour counterterms---which are already available in \SherpaOpenLoops 
in the framework of fixed-order NLO \QCDpEW automation---a complete
implementation of \MEPS merging at NLO \QCDpEW requires additional
technical ingredients that are still missing to date.
In particular, the \Sherpa parton shower,
extended to \QCDpQED, should be 
matched to the real emission of photons and
\QCD partons at $\ord(\alphaS^{n}\alpha^3)$ in the \SMCatNLO framework.
Moreover, a consistent
showering and clustering approach for events associated with mixed \QCDmEW
matrix elements is needed.
While we expect that such technical prerequisites will be fulfilled in the near future,
based on the good quality of the NLO \EWvirt
approximation of~\refse{sec:virtapp} and the fact that 
it does not require resolved emissions of photons or QCD partons at NLO EW,
in the following we present a first approximate, but reliable, 
extension of NLO multijet merging to also include NLO \EW effects.
This approach is based on the implementation of the NLO \EWvirt approximation
in the $\tilde{\mr{B}}_{n}(\Phi_{n})$ soft term of \refeq{eq:mepsatnlo}. 
While all other aspects of
\MEPSatNLO, including the truncated vetoed \QCD parton shower, are kept unchanged,
the NLO EW improved $n$-jet  soft term takes the form
\begin{equation}\label{eq:mepsatnloewa}
    \tilde{\mr{B}}_{n,\QCDpEW}(\Phi_{n})=
     \tilde{\mr{B}}_{n}(\Phi_{n})
        +{\mr{V}}_{n,\EW}(\Phi_{n})
        +\mr{I}_{n,\EW}(\Phi_{n})
        +\mr{B}_{n,\mix}(\Phi_{n})\;.
\end{equation}
Here $\tilde{\mr{B}}_{n}(\Phi_{n})$ is the usual NLO \QCD soft term~\refeq{eq:mepsatnlo_bbar}, and
 $\mr{B}_{n,\mix}(\Phi_{n})$ denotes \QCDmEW mixed Born contributions of $\ord(\alphaS^{n-1}\alpha^3)$.
The terms ${\mr{V}}_{n,\EW}(\Phi_{n})$ and $\mr{I}_{n,\EW}(\Phi_{n})$ represent 
the renormalised virtual corrections of $\ord(\alphaS^{n}\alpha^3)$
and the NLO \EW generalisation of the Catani--Seymour $\mathbf{I}$ operator, respectively,  as discussed in~\refse{sec:virtapp}.

The $\mr{I}_{n,\EW}$ term cancels all $\ord(\alphaS^{n}\alpha^3)$
infrared divergences in the virtual \EW corrections.  This corresponds to
an approximate and fully inclusive description of the emission of photons and \QCD
partons at $\ord(\alphaS^{n}\alpha^3)$.  More precisely, only contributions
of soft and final-state-collinear type are included, while
initial-state collinear contributions and related PDF counterterms ($\mathbf{K}$ and
$\mathbf{P}$ operators in the Catani--Seymour framework) are not taken into account. This implies
a (small) spurious 
$\ord(\alphaS^{n}\alpha^3)$ dependence associated to the 
uncancelled factorisation scale dependence of the 
$\ord(\alphaS^{n}\alpha^2)$ and $\ord(\alphaS^{n-1}\alpha^3)$
Born terms.
In contrast, all relevant ultraviolet divergences and related renormalisation scale variations of
$\ord(\alphaS^{n}\alpha^3)$ are consistently included and cancelled. 
To this end, virtual \EW corrections 
(${\mr{V}}_{n,\EW}$) and \QCDmEW mixed Born terms
($\mr{B}_{n,\mix}$) have to be kept together in \refeq{eq:mepsatnloewa},
since only their combination is free from 
renormalisation-scale logarithms at $\ord(\alphaS^{n}\alpha^3)$.
This approach will be denoted as \MEPSatNLO \QCDpEWvirt in the following.

Concerning the accuracy of the approximation~\refeq{eq:mepsatnloewa}
a few comments are in order.  First of all,
thanks to the exact treatment of virtual \EW corrections, all possible large
virtual \EW effects related to Sudakov logarithms are included by
construction.  Moreover, the merging approach guarantees that \EW correction
effects are consistently included also in phase-space regions of
higher jet multiplicity.  
Secondly, as pointed out in~\refse{sec:virtapp},
sizable NLO \EW contributions can arise also from 
the emission of \QCD partons through mixed \QCDmEW matrix elements
at NLO.  
As far as equation \refeq{eq:mepsatnloewa} is concerned,
such mixed bremsstrahlung contributions are only included in a
fully inclusive and approximate way through the $\mr{I}_{n,\EW}$ operator.
Nevertheless, the fact that mixed Born terms
($\mr{B}_{n,\mix}$) are effectively merged at LO guarantees a fairly reliable
and fully exclusive description of mixed bremsstrahlung also at high jet transverse momenta, where
the effects can be sizable.
Technically, unresolved ($Q_{n+1}<\qcut$) mixed bremsstrahlung of
$\ord(\alphaS^{n}\alpha^3)$ is generated by the interplay of the
$\ord(\alphaS^{n-1}\alpha^3)$ $\mr{B}_{n,\mix}$ terms with the \QCD parton
shower, and its resolved counterpart ($Q_{n+1}>\qcut$)  is described by the Born mixed matrix
elements with one extra jet, $\mr{B}_{n+1,\mix}$.
Finally, let us note that genuine \QED bremsstrahlung at $\ord(\alphaS^{n}\alpha^3)$ is only 
included through the na\"ive and inclusive approximation
provided by the $\mr{I}_{n,\EW}$ term.
Thus, the approximation~\refeq{eq:mepsatnloewa} cannot account for large \QED logarithms 
that can appear in differential distributions for bare leptons and similar 
exclusive observables.
Nevertheless, for a wide range of physical observables the impact of \QED bremsstrahlung 
tends to be negligible. This is the case also for many leptonic observables
if photon bremsstrahlung is treated in a rather inclusive way, \eg through
the recombination of collinear photon emissions.
In any case, leading-logarithmic \QED effects 
could be easily included in~\refeq{eq:mepsatnloewa} by a simple \QCDpQED extension of the
parton shower~\cite{Hoeche:2009xc} or a YFS-type soft photon resummation 
\cite{Schonherr:2008av}, without having to match the \QED part to NLO \QCDpQED
matrix elements. This pure shower approach could be further 
improved by including photon-emission matrix elements via LO merging 
\cite{Hoeche:2009xc}, in a similar way as discussed above for the 
case of mixed bremsstrahlung.

The NLO \QCDpEW extension of the \MEPSatNLO method based on
equation~\refeq{eq:mepsatnloewa} was implemented in \SherpaOpenLoops in a
fully automated way and applied to $\PV+\,$multijet production as described in
the following section.

\subsection[Numerical \texorpdfstring{\protect\MEPSatNLO}{MEPS@NLO} 
\texorpdfstring{\protect\QCDpEW}{QCD+EW} results for \texorpdfstring{$pp\to\protect\PV+0,1,2\,$}{pp->V+0,1,2}jets]
{Numerical \texorpdfstring{\protect\MEPSatNLO}{MEPS@NLO} \texorpdfstring{\protect\QCDpEW}{QCD+EW} results 
for \texorpdfstring{$\boldsymbol{pp\to\protect\PV+0,1,2}\,$}{pp->V+0,1,2}jets}
\label{se:results_merged}

Based on the above described multijet merging method, 
in this section we present an inclusive simulation of
$\lmn+$\,multijet production that includes NLO \QCDpEWvirt correction effects 
in phase-space regions with up to two resolved jets.
In addition to the settings summarised 
in \refeqs{eq:massesew}{eq:defsintheta} we set the renormalisation scale 
according to \refeq{eq:ckkwscale}, and both factorisation and resummation 
scales to the core scale defined in \refeq{eq:corescale}. The remaining free 
parameter, the merging scale separating the individual jet multiplicities, 
is set to $\qcut=20\,\text{GeV}$. To estimate the uncertainties of our 
calculation, we vary the renormalisation and factorisation scales by a factor 
two in a correlated way. 
The resummation and merging scales are not varied 
here as they give rise to much smaller uncertainties 
for the observables to be studied in this paper.
While this observation has already been made in various 
studies based on the \MEPSatNLO method~\cite{Cascioli:2013gfa,
Hoeche:2013mua,Hoeche:2014lxa,Hoeche:2014qda,Hoeche:2014rya,
Buschmann:2014sia,Goncalves:2015mfa}, in \refapp{app:meps} 
we show that it holds true 
also in the multi-TeV regime, where the gap between
the merging scale and the hard scattering energy can 
reach two orders of magnitude.
The presented analysis 
has been implemented in \Rivet \cite{Buckley:2010ar}. 


\begin{figure*}[t]
\centering
   \includegraphics[width=\relplotwidth\textwidth]{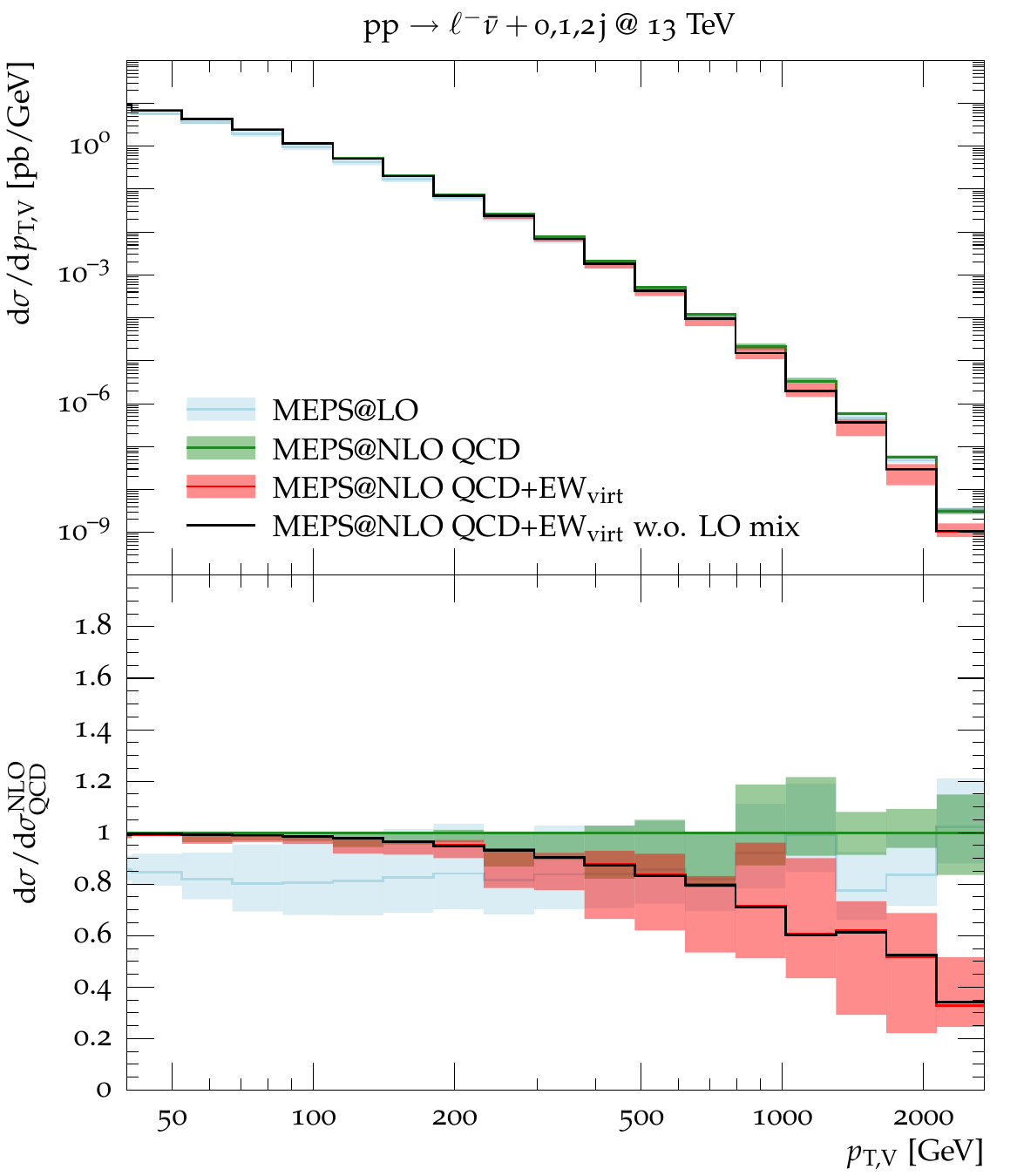}
         \qquad 
   \includegraphics[width=\relplotwidth\textwidth]{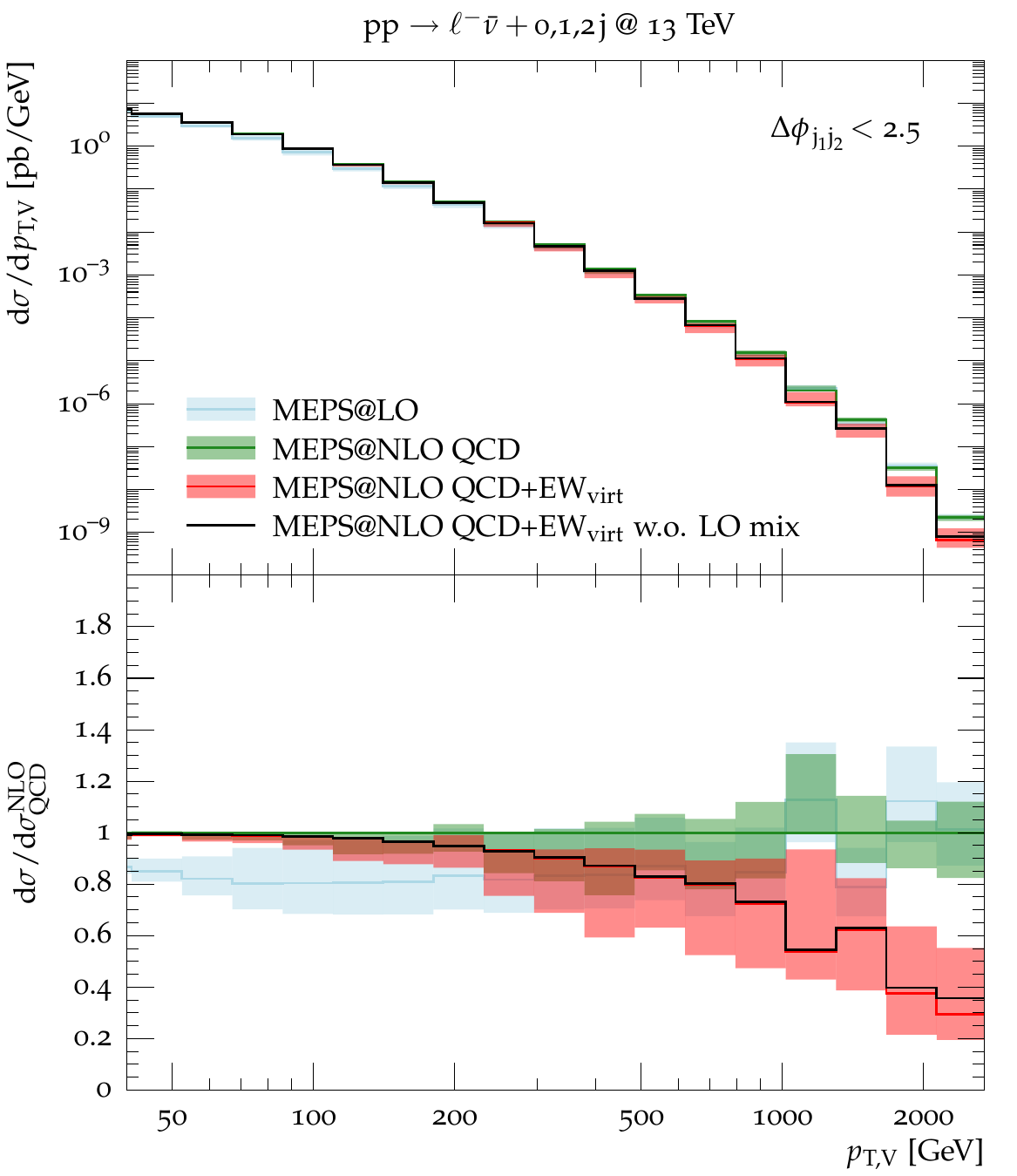}
\caption{
	  Distribution in the transverse momentum of the 
	  reconstructed vector boson in $\lmn\,+$\,jets production
	  with standard cuts (left) and in presence of an extra cut 
	  $\Delta\phi_{j_1j_2}<2.5$ (right). 
	  The upper frame displays absolute predictions obtained with \MEPSatNLO \QCD merging (green) 
	  and its extension to NLO \QCDpEW accuracy including (red) or excluding (black) 
	  mixed Born contributions to $\PV+2\,$jet topologies (LO mix).
	  Relative corrections with respect to \MEPSatNLO \QCD are shown in the lower panels. 
	  The bands correspond to scale variations, and in the case of ratios 
	  only the numerator is varied. 
	}
\label{fig:MEPS_Wen_pTV}
\end{figure*}

The first observable we study is the transverse momentum of the
reconstructed \PW boson in $\lmn\,+\,$jets production, as detailed in
\reffi{fig:MEPS_Wen_pTV}. This
observable receives significant contributions from two-jet topologies, which
are, however, typically dominated by 
a first hard jet, while the second jet tends to be much softer.  For this
reason we observe a rather similar behaviour of NLO \QCDpEW effects in fixed-order 
calculations for $\lmn\,+1\,$jet (\reffi{fig:Vj_pTV_pTj1}) and
$\lmn\,+2\,$jets (\reffi{fig:Vjj_pTV}), as well as in their
combination through exclusive sums (\reffi{fig:Vjj_pTV_exclsum}) 
and with \MEPSatNLO \QCDpEWvirt merging (\reffi{fig:MEPS_Wen_pTV}).
More precisely, apart from statistical fluctuations and minor differences due to different scale choices, \MEPSatNLO \QCDpEWvirt predictions are in good agreement with 
\mbox{$\lmn\,+2$-jet} results, both for what concerns the size of electroweak corrections and scale uncertainties.
At high \pT the impact of \EW effects in the \MEPS framework turns out to be remarkably large 
and can reach $-50\%$ or more in the multi-TeV region.
This is quantitatively consistent with the 
outcome of the factorised \QCDtEW 
prescription in inclusive $\PV+$\,jet NLO calculations,
and clearly more pronounced than what results from the
additive combination of \QCDpEW fixed-order corrections (\reffi{fig:Vj_pTV_pTj1}).
This feature can be attributed to the inclusion of 
NLO \EW effects in two-jet topologies and, to some extent, also in three-jet topologies 
via NLO matching to the parton shower.
For the vector-boson \pT distribution, 
as already observed at fixed-order NLO,
mixed Born contributions are almost negligible, and
the exclusion of back-to-back
dijet configurations  through a $\deltajj<2.5$ cut
have little impact on the behaviour of NLO \EW effects.


\begin{figure*}[t!]
\centering
   \includegraphics[width=\relplotwidth\textwidth]{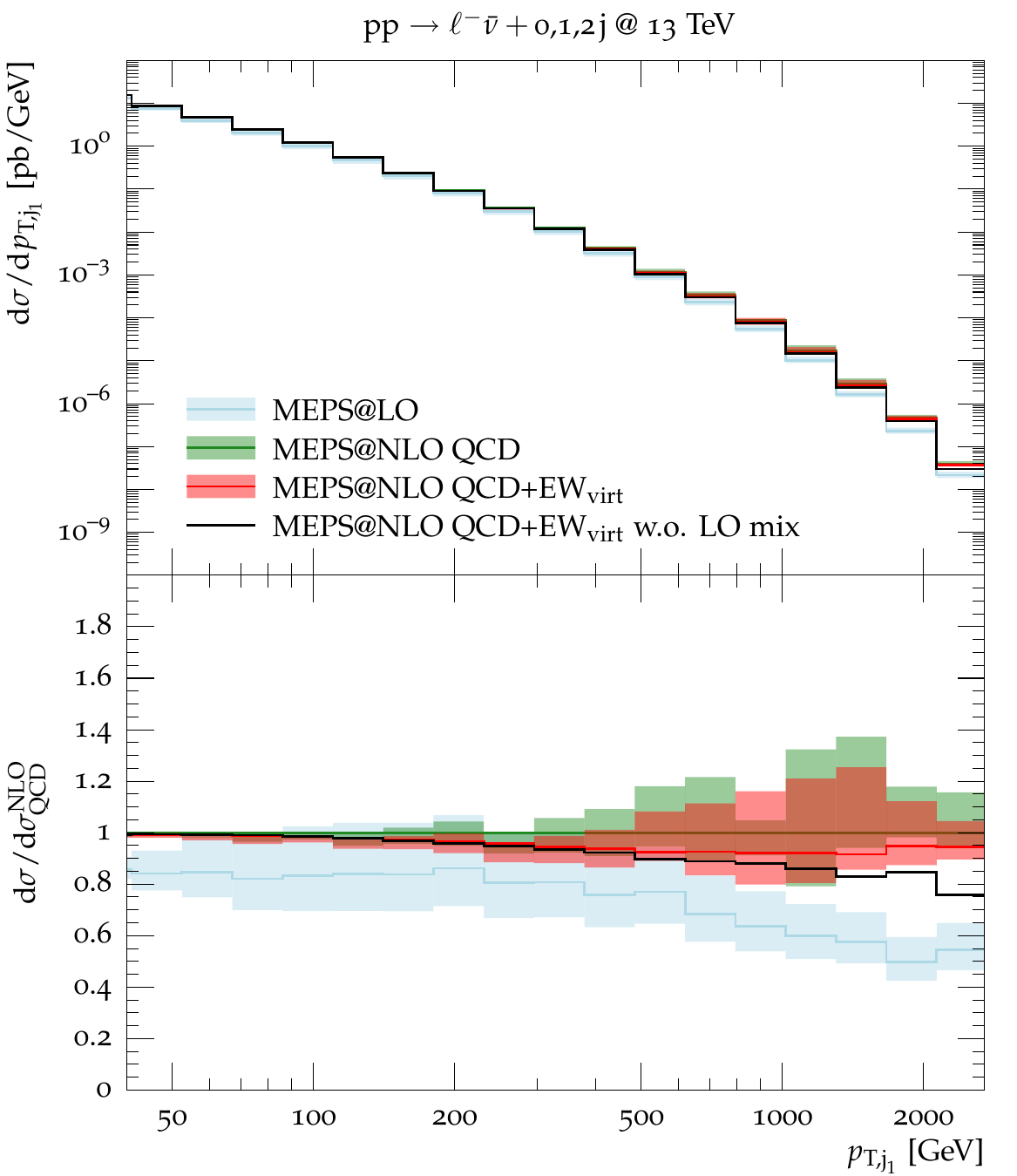}
         \qquad 
   \includegraphics[width=\relplotwidth\textwidth]{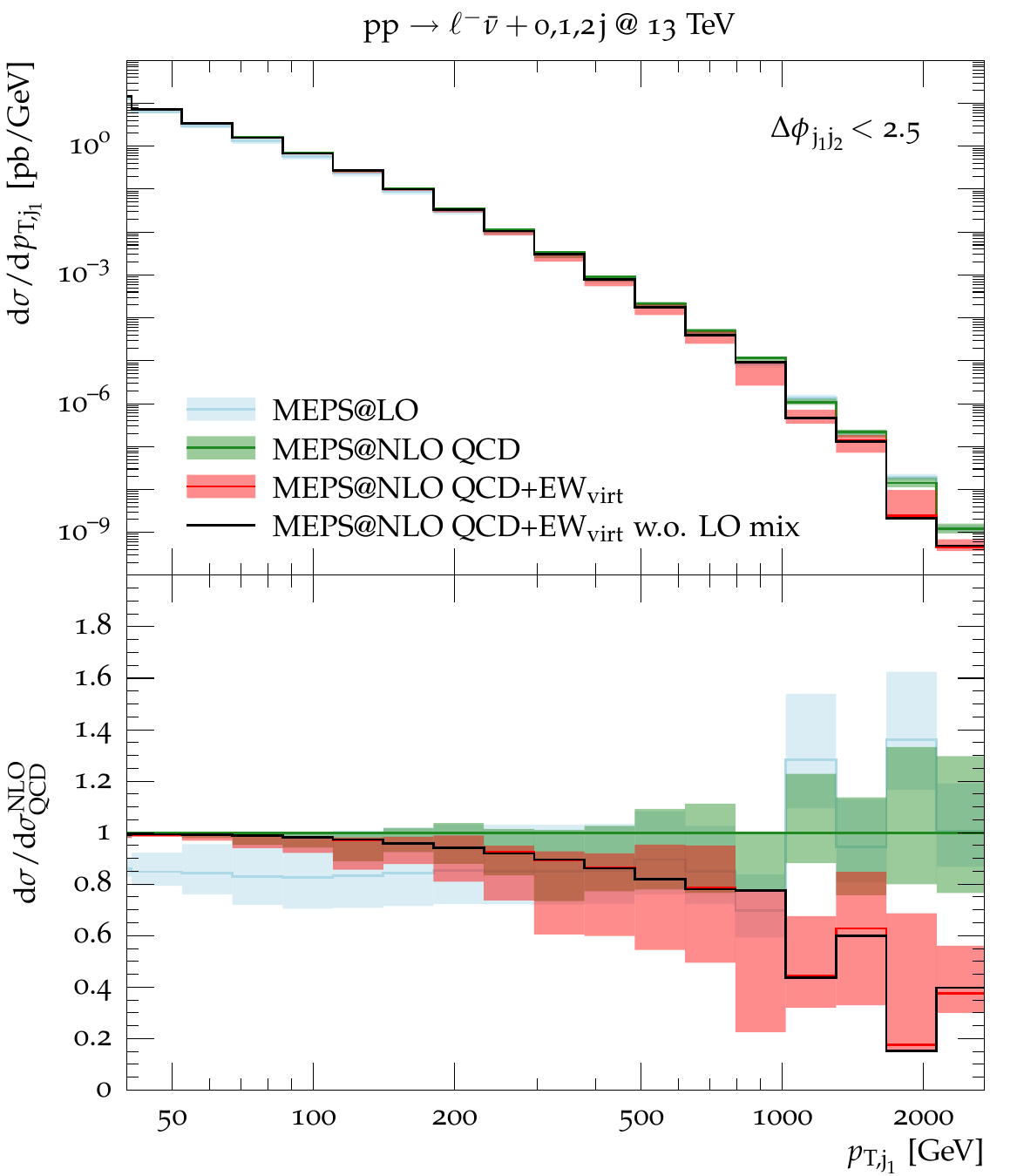}
\caption{
	  Distributions in the transverse momentum of the hardest jet 
	  with standard cuts (left) 
	  and in presence of an extra cut 
	  $\Delta\phi_{j_1j_2}<2.5$ (right). 
	  Curves and bands as in Fig. \ref{fig:MEPS_Wen_pTV}.
	}
\label{fig:MEPS_Wen_pTj1}
\end{figure*}

In \reffi{fig:MEPS_Wen_pTj1} we examine the transverse momentum of the leading
jet. As this observable exhibits a strong sensitivity to higher jet
multiplicities---in particular to topologies with two hard back-to-back
jets---it is ideally suited to be calculated using a consistent multijet
merging. In particular, similarly as for the case of exclusive sums discussed in
\refse{se:exclusive_sums}, thanks to the inclusion of dijet topologies as
genuine $\lmn+2\,$jet production processes at NLO, the \MEPSatNLO methodology
allows one to avoid giant $K$-factors and cures the pathological behaviour of
\EW corrections observed in fixed-order NLO \QCDpEW calculations for
$\PV+1$\,jet. Moreover, at sufficient hardness of both jets, the $\PV+2\,$jet
configurations are treated as a \PV~boson radiated from a dijet core process,
and the scales are set accordingly, further helping to achieve a more physical
description of such states.
This is confirmed by the decent behaviour of NLO \QCD scale uncertainties and
NLO \EW corrections in~\reffi{fig:MEPS_Wen_pTj1}. It is worth noting the sizable
impact of mixed Born contributions, qualitatively similar to that seen in
\reffi{fig:Vjj_pTj1_exclsum}, in the tail of this distribution. The absolute
size of this effect differs, however, due to the different scale choices. Once
back-to-back dijet topologies are removed through a $\deltajj$ cut, as shown on
the right hand side of \reffi{fig:MEPS_Wen_pTj1}, mixed Born contributions are
suppressed, and the pure Sudakov-type behaviour of the NLO \EW corrections is
recovered.


\begin{figure*}[t!]
\centering
   \includegraphics[width=\relplotwidth\textwidth]{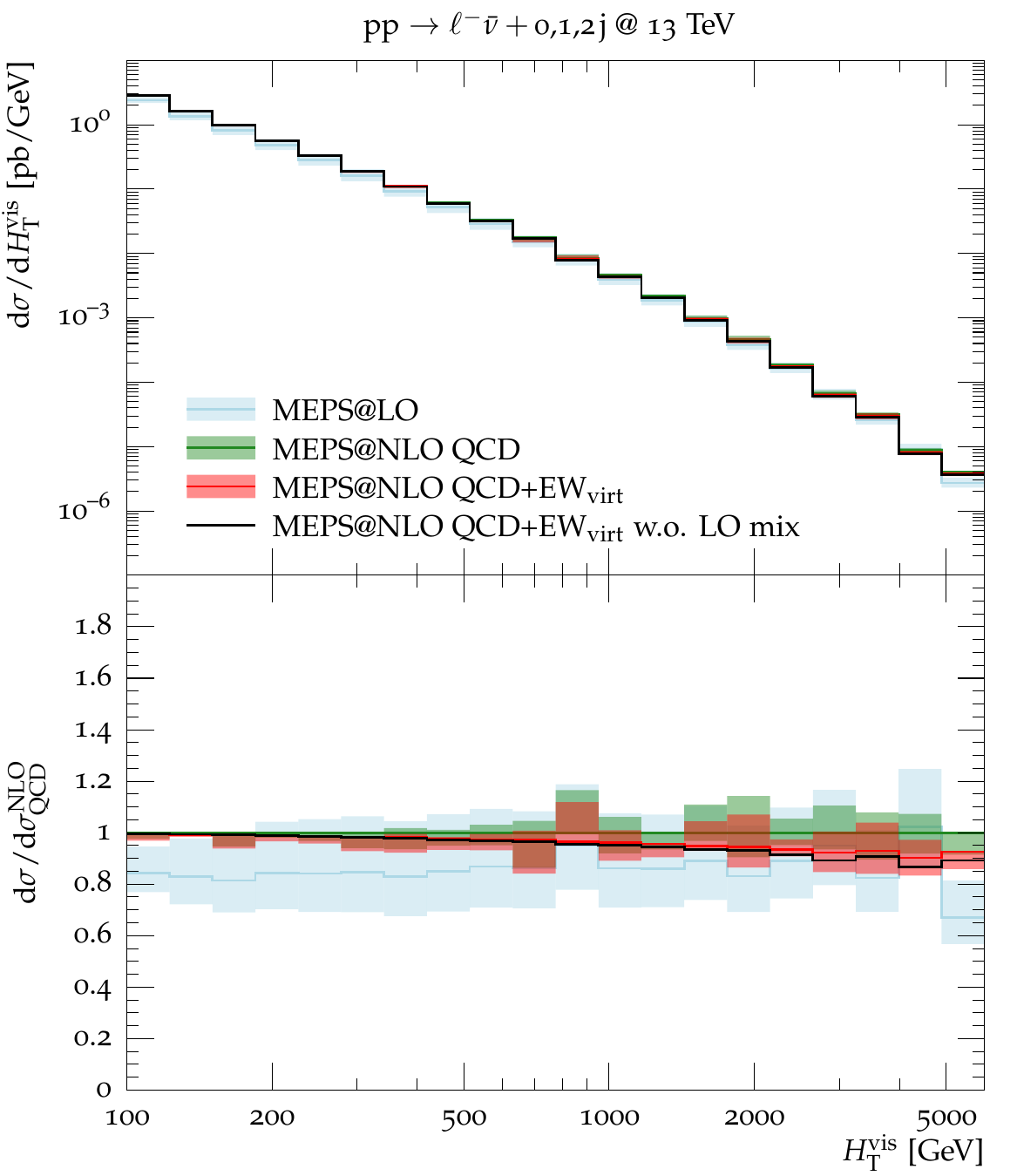}
         \qquad 
   \includegraphics[width=\relplotwidth\textwidth]{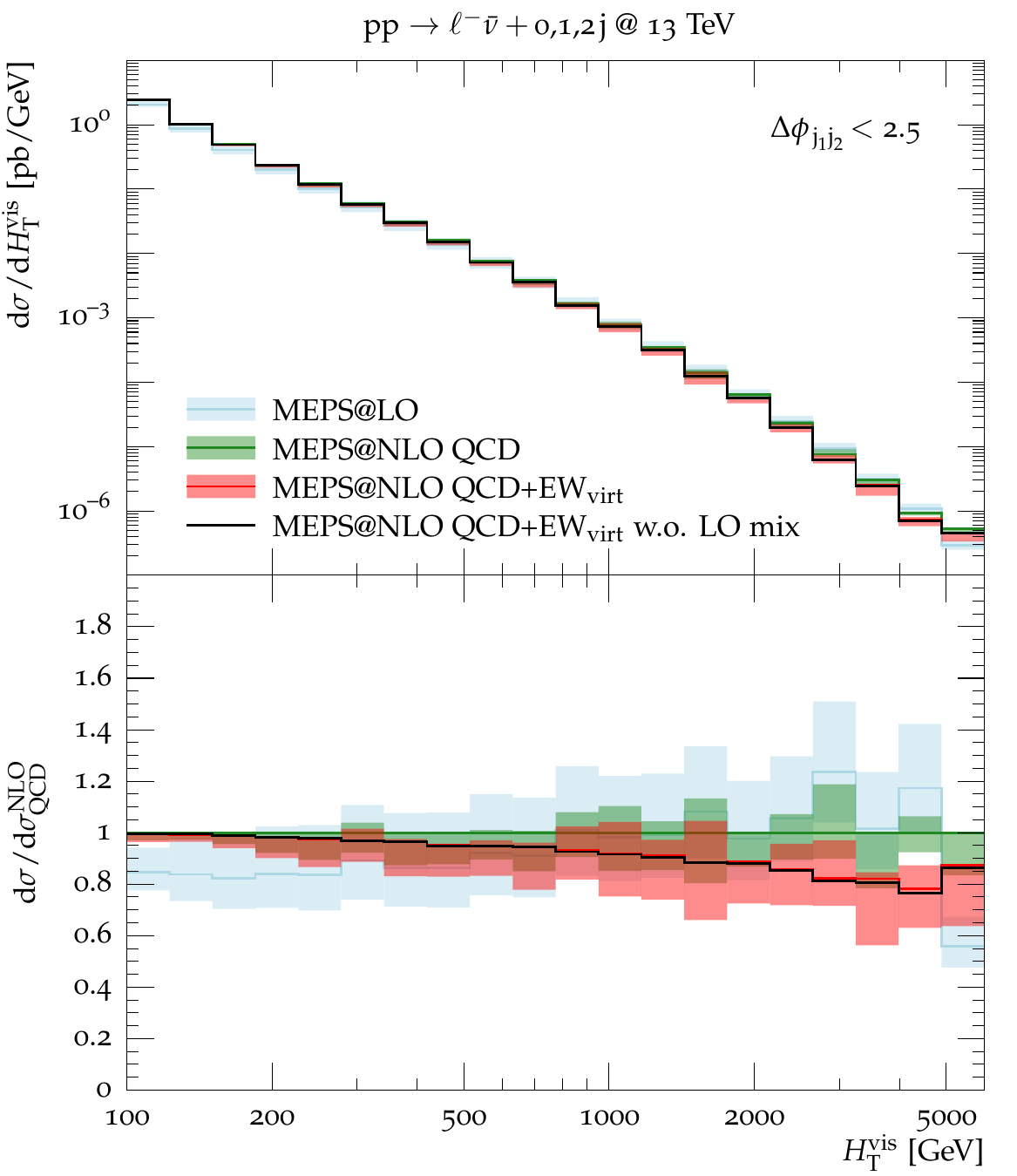}
\caption{
          Differential distribution in the total visible transverse energy 
          $\HTvis=\sum_{i} \pTji+\sum_\ell p_{\rT,\ell}$  
	  with standard cuts (left) 
	  and in presence of an extra cut 
	  $\Delta\phi_{j_1j_2}<2.5$ (right). 
	  Curves and bands as in Fig. \ref{fig:MEPS_Wen_pTV}.
	}
\label{fig:MEPS_Wen_HTtot}
\end{figure*}

Finally, in \reffi{fig:MEPS_Wen_HTtot} we examine the distribution in the scalar
sum of the transverse momenta of all visible objects, \ie jets and leptons,
$\HTvis$. Again, this is a typical example of an observable receiving
contributions from various jet multiplicities simultaneously, and is thus
expected to benefit from a multijet merging approach. As the $\pT$ of the
leading jet is a major contributor to $\HTvis$, this observable exhibits many of
the characteristics of $\pTjone$, albeit in reduced severity. Nonetheless, due
to the \MEPSatNLO approach the troublesome configurations dominated by a hard
dijet system are again rendered benign, and the NLO \QCD and NLO \EW corrections
behave as expected. Subleading Born contributions have a visible, but much
smaller impact than for the transverse momentum of only the leading jet. As
before, the inclusion of a $\deltajj$ cut tends to enhance negative \EW
correction effects in the tail.

In all investigated observables \MEPSatLO predictions are in fairly good
agreement with corresponding \MEPSatNLO predictions at the NLO \QCD level. In
particular, due to the scale choice of \eqref{eq:ckkwscale} shapes of all
differential distributions receive moderate \QCD corrections, even in the
inclusive $\pT$ distribution of the hardest jet.

\section{Summary and conclusions}
\label{se:conclusions}

The inclusion of \QCD and \EW higher-order corrections in theoretical simulations
is a central prerequisite for precision tests of the Standard Model and
for new-physics searches at the energy frontier during Run-II of the LHC. In the TeV
range, electroweak Sudakov logarithms change the shape of important kinematic
distributions significantly and often yield corrections 
that largely exceed the intrinsic uncertainties of NLO QCD predictions.
The recently achieved automation of NLO \EW corrections within the \SherpaOpenLoops
and \MunichOpenLoops Monte Carlo frameworks opens the door to access 
high precision at the energy frontier for a multitude of processes.

One example where both \QCD and \EW radiative corrections are large is the
experimentally very important process class of vector-boson production in
association with jets. Here, it is well known that the inclusive production in
conjunction with at least one hard jet is highly sensitive to multijet
radiation. In particular, in the regime of high jet \pT, fixed-order NLO
calculations for inclusive $V+1$\,jet production are plagued by giant
QCD $K$-factors, and also EW corrections behave in a pathological way. Precise
theoretical predictions can only be achieved beyond NLO or via a merging of
higher jet multiplicities. In this paper we have developed an approximate
framework for multijet merging at NLO including \QCD and \EW corrections. As a
first application, we have presented an inclusive simulation of $\PV+\,$jets
production that guarantees NLO \QCDpEW accuracy for final states involving zero,
one and two jets.
 
As a prerequisite for the described merging we have calculated NLO \QCDpEW fixed-order
results for $pp\to\PV+2$\,jets presenting, for the first time,
predictions that describe the off-shell production and decay of all electroweak
vector bosons, $V=\PW^\pm,Z/\gamma^*$, including all possible final states with 
charged leptons and neutrinos.
Off-shell and $Z/\gamma^*$ interference effects were included throughout by
means of an automated implementation of the complex-mass scheme at NLO. Detailed
NLO QCD+EW predictions are provided for various important kinematic observables,
including the \pT spectra of the leptons, the reconstructed vector bosons and
the accompanying jets.
The tails of such distributions receive large EW corrections of Sudakov
type, which can reach $-40\%$ at 2\,TeV and are maximally pronounced in $V+2$\,jet
configurations where the leading transverse momentum is carried by the vector boson.
As expected, such large Sudakov corrections  are hardly affected by the 
leptonic decays, whereas less inclusive observables, in particular the invariant mass
of the lepton pair in $Z/\gamma^*\to \ell^+ \ell^-$, exhibit a strong dependence 
on genuine \QED bremsstrahlung, with corrections of up to $50\%$. Besides 
NLO \EW corrections, we have also studied all subleading Born and 
photon-induced $\PV+2$\, jet processes which can give sizeable
contributions in the TeV range.
 
Towards a merging including \EW corrections we first combined $\PV+1$\,jet and
$\PV+2$\,jet NLO \QCDpEW predictions by means of an implementation of na\"ive
exclusive sums. Already here the perturbative convergence was found to be
largely stabilized. Within the context of exclusive sums, we have developed an
NLO \EW approximation that combines exact \EW virtual corrections with an
inclusive treatment of bremsstrahlung effects. In the kinematic regime dominated
by large virtual Sudakov corrections, the agreement with respect to the full
calculation was found to be mostly at the few percent level. This approximation
allowed us to include NLO \EW corrections into the \MEPSatNLO multijet merging
framework of \Sherpa in a straightforward way.
Within the \MEPSatNLO framework we have provided multijet-merged predictions for
$\lmn\,+$\,jets production including \QCDpEW corrections up to two jets. Thanks
to the inclusion of dijet topologies as genuine $\lmn+2\,$jet production
processes at NLO, the \MEPSatNLO methodology allowed us to stabilize the
perturbative convergence and to cure the pathological behaviour of \EW
corrections observed in fixed-order NLO \QCDpEW calculations for $\PV+1$\,jet.

In a forthcoming paper, we plan to investigate multijet-merged
cross-section ratios for different $\PV+\,$jets processes, including a thorough study
of theoretical uncertainties. Precise predictions for such ratios including \EW
corrections can reduce important systematic uncertainties 
in monojet searches for dark matter and many other BSM searches, 
resulting in significant improvements of the experimental sensitivity.

The findings presented here motivate similar studies for a wide range of other
Standard Model processes, and at the same time further developments towards a
complete parton-shower matching and multi-jet merging at NLO \QCDpEW.

\acknowledgments
We thank A.~Denner, S.~Dittmaier and L.~Hofer for providing us
with the one-loop tensor-integral library \Collier. 
We also want to thank A.~Denner for providing us detailed numerical results
of \cite{Denner:2014ina}, and for useful discussions concerning the complex-mass scheme.
This research was supported in part by the Swiss
National Science Foundation (SNF) under contract PP00P2-128552
and by the Research Executive Agency (REA) of the European Union
under the Grant Agreements PITN--GA--2010--264564 ({\it
LHCPhenoNet}), PITN--GA--2012--315877 ({\it MCnet}) and 
PITN--GA--2012--316704 ({\it HiggsTools}). 


\begin{appendix}

\section{Further NLO predictions for \texorpdfstring{$\boldsymbol{\protect\PV+2}$}{V+2} jet production}
\label{app:vjj}

\begin{figure*}[p]
\centering
   \includegraphics[width=\relplotwidth\textwidth]{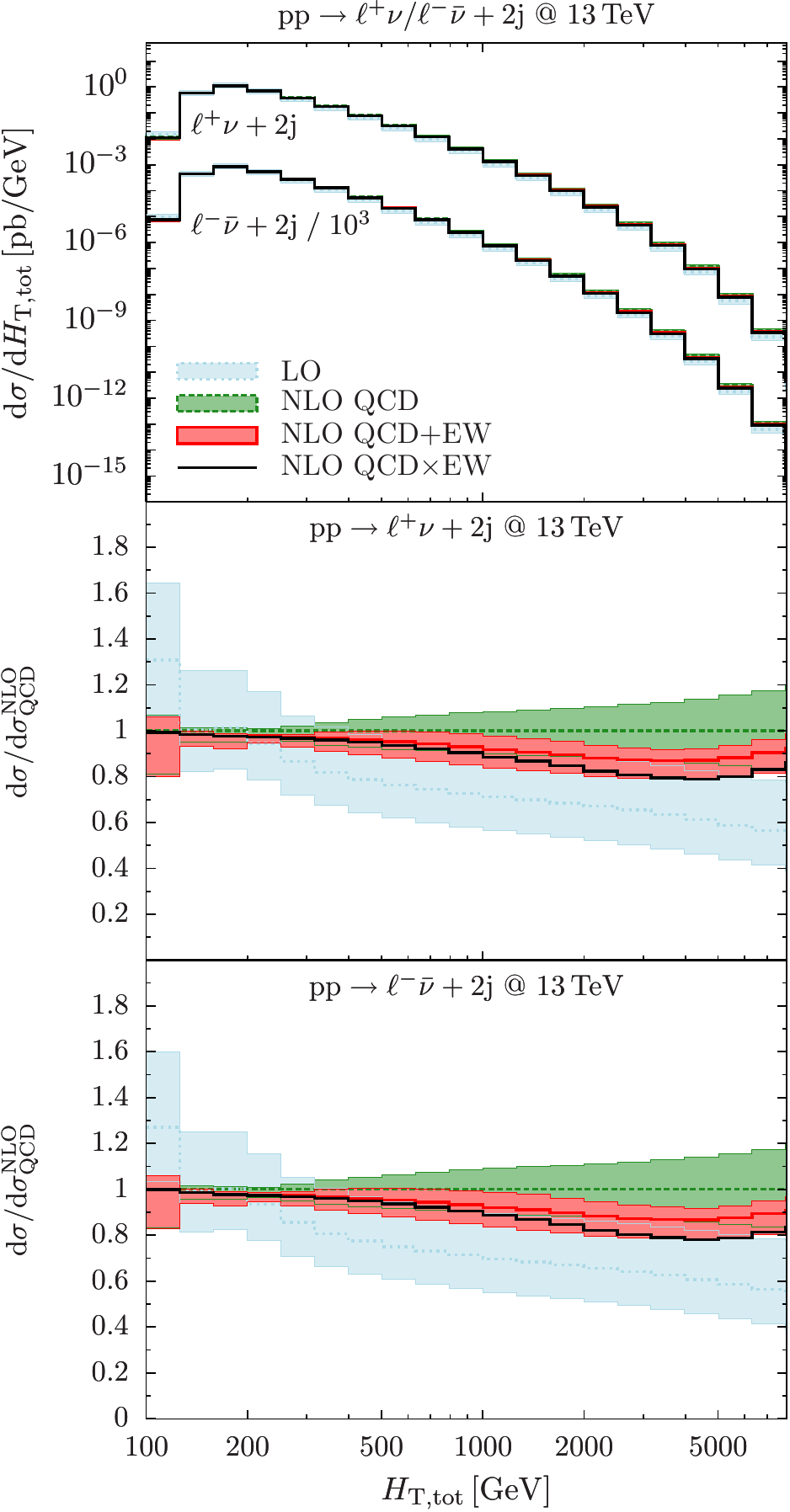}
         \qquad 
   \includegraphics[width=\relplotwidth\textwidth]{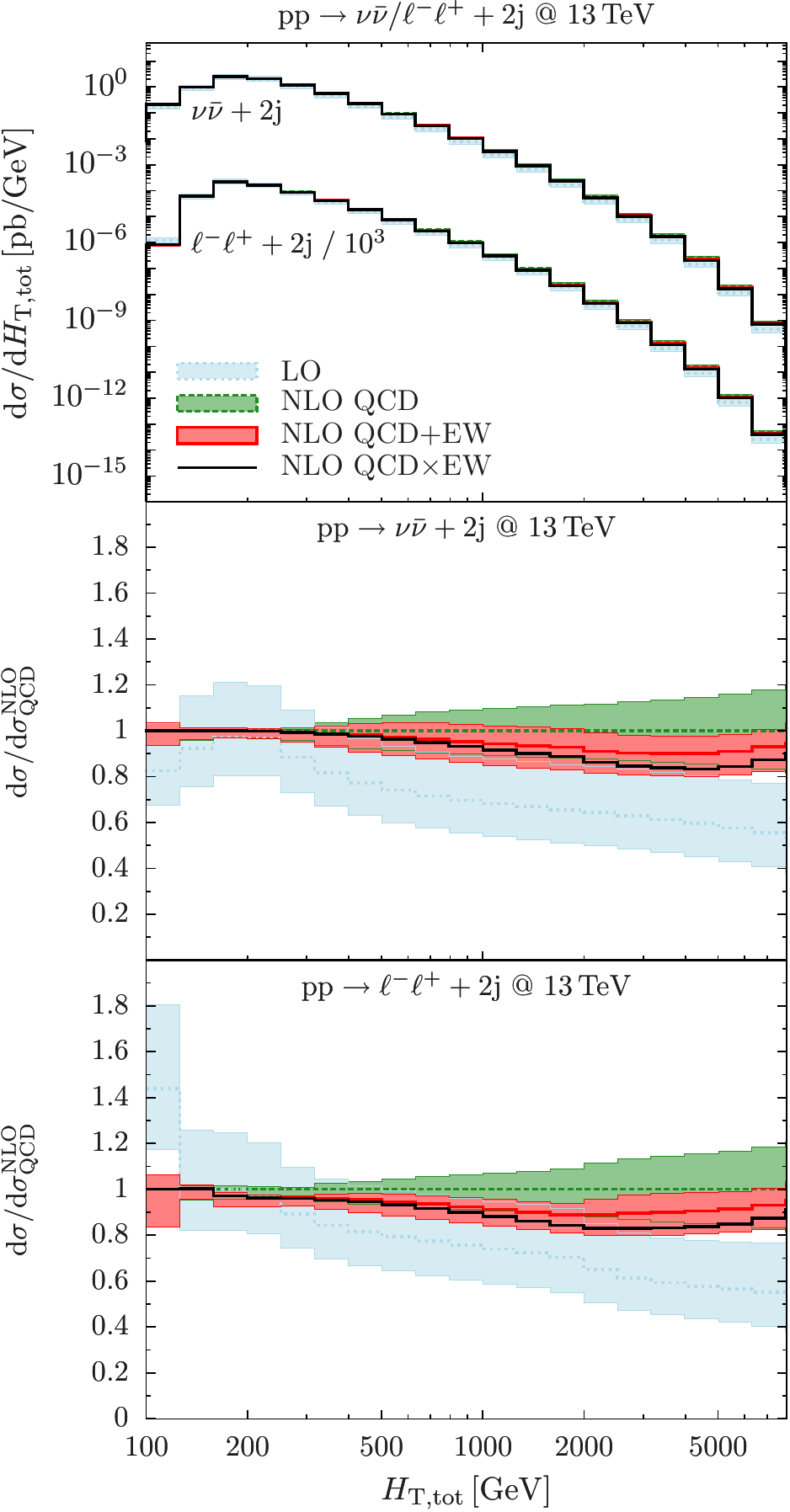}
\caption{
Differential distribution in $\HTtot=\sum_{k} p_{T,j_k}+\sum_\ell p_{T,\ell}+ \missingET$. 
Curves and bands as in Fig. \ref{fig:Vj_pTV_pTj1}.
}
\label{fig:Vjj_HTtot}
\end{figure*}

\begin{figure*}[p]
\centering
   \includegraphics[width=\relplotwidth\textwidth]{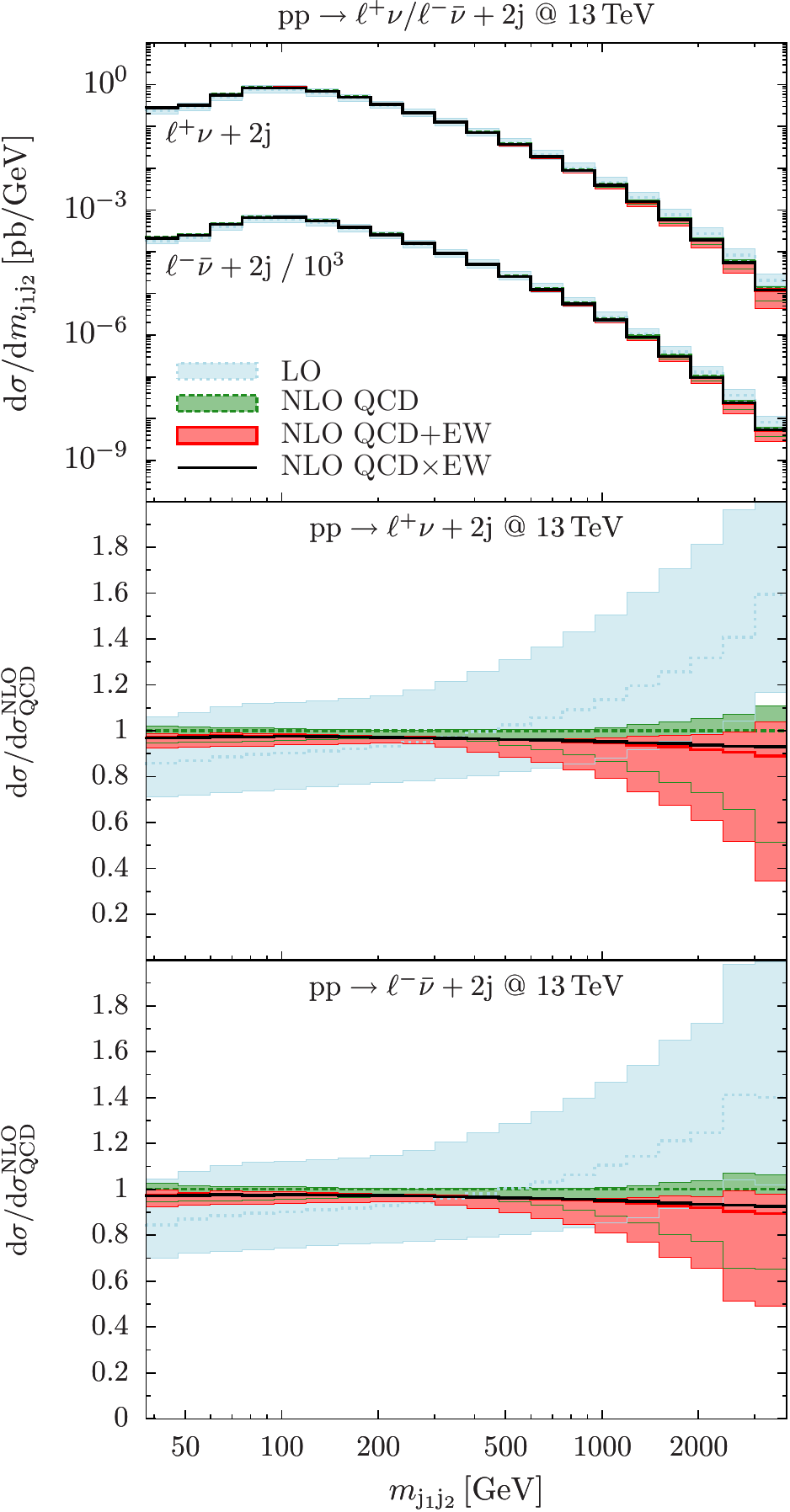}
         \qquad 
   \includegraphics[width=\relplotwidth\textwidth]{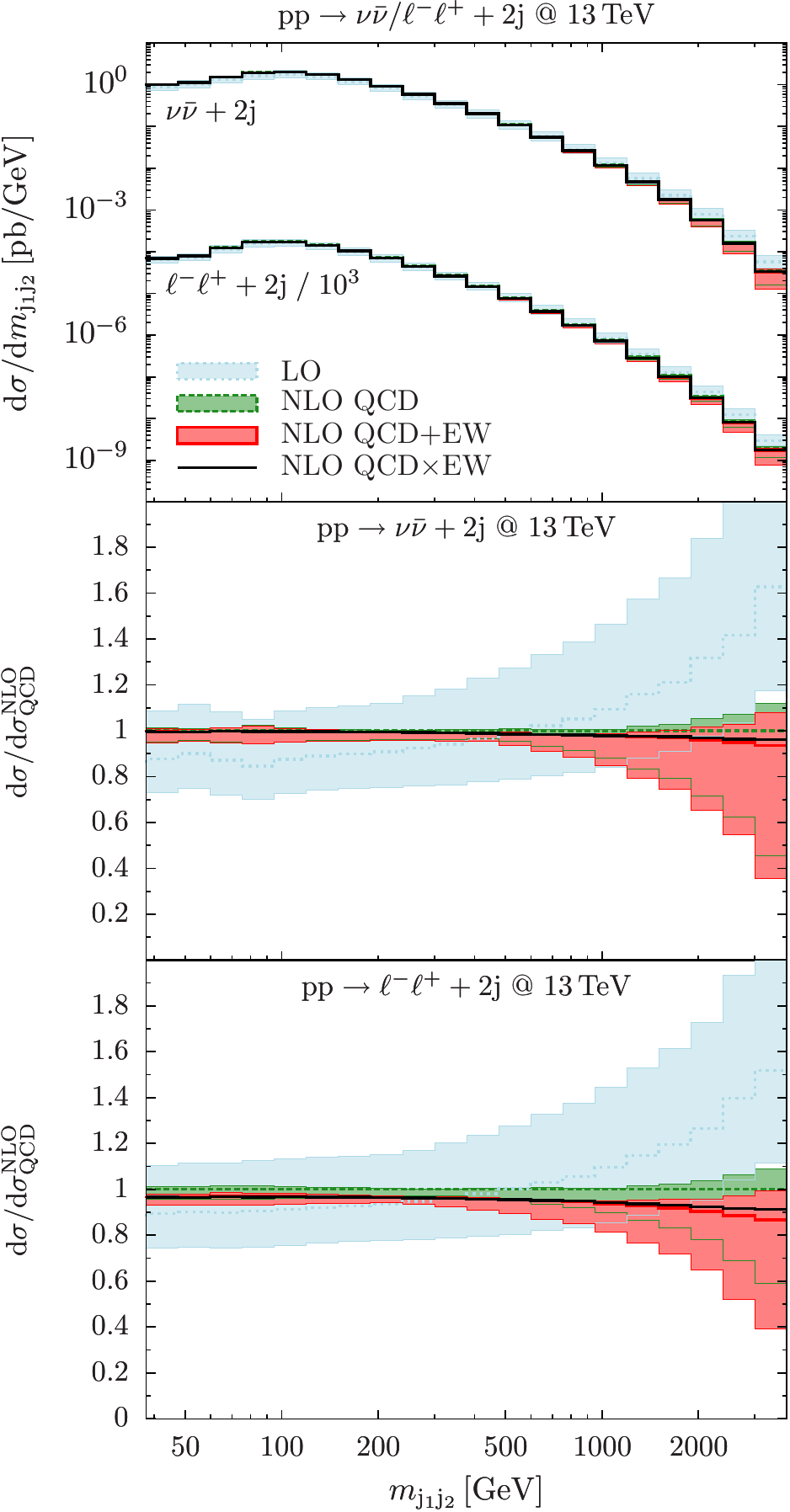}
\caption{
Differential distributions in the invariant mass $m_{\mathrm{j_1j_2}}$ of the two hardest jets.
Curves and bands as in Fig. \ref{fig:Vj_pTV_pTj1}.
}
\label{fig:Vjj_mjj}
\end{figure*}

In \reffis{fig:Vjj_HTtot}{fig:Vjj_mjj} we present further fixed-order NLO
\QCDpEW results for $pp\to\lpn+2$\,jets, $pp\to\lmn+2$\,jets,
$pp\to\lplm+2$\,jets and $pp\to\nn+2$\,jets at the LHC. In particular, we show
distributions in $\HTtot$ and the invariant mass of the two leading jets,
$\mjj$. In the tail of the $\HTtot$ distribution, NLO \QCD and \EW corrections
approach the 70\% and 10\% level, respectively, and the \QCDtEW curve suggests
that due to the sizeable \QCD corrections the importance of NLO \EW corrections
is underestimated by about a factor two in the NLO \QCDpEW prediction. The
corrections in $\mjj$ show a very different picture. Here, NLO \EW corrections
are very small and almost completely independent of the dijet mass up to the
multi-TeV range. However, in this regime, LO \EW contributions from $V+2$\,jet
production via vector-boson fusion will become sizable. Thus, a detailed study
of \EW corrections in $\mjj$ requires the inclusion of the subleading one-loop
corrections of $\ord(\alphaS\alpha^4)$ and $\ord(\alpha^5)$.

\section{Multijet merging systematic uncertainties in the TeV range}
\label{app:meps}

\begin{figure*}[t!]
\centering
   \includegraphics[width=\relplotwidth\textwidth]{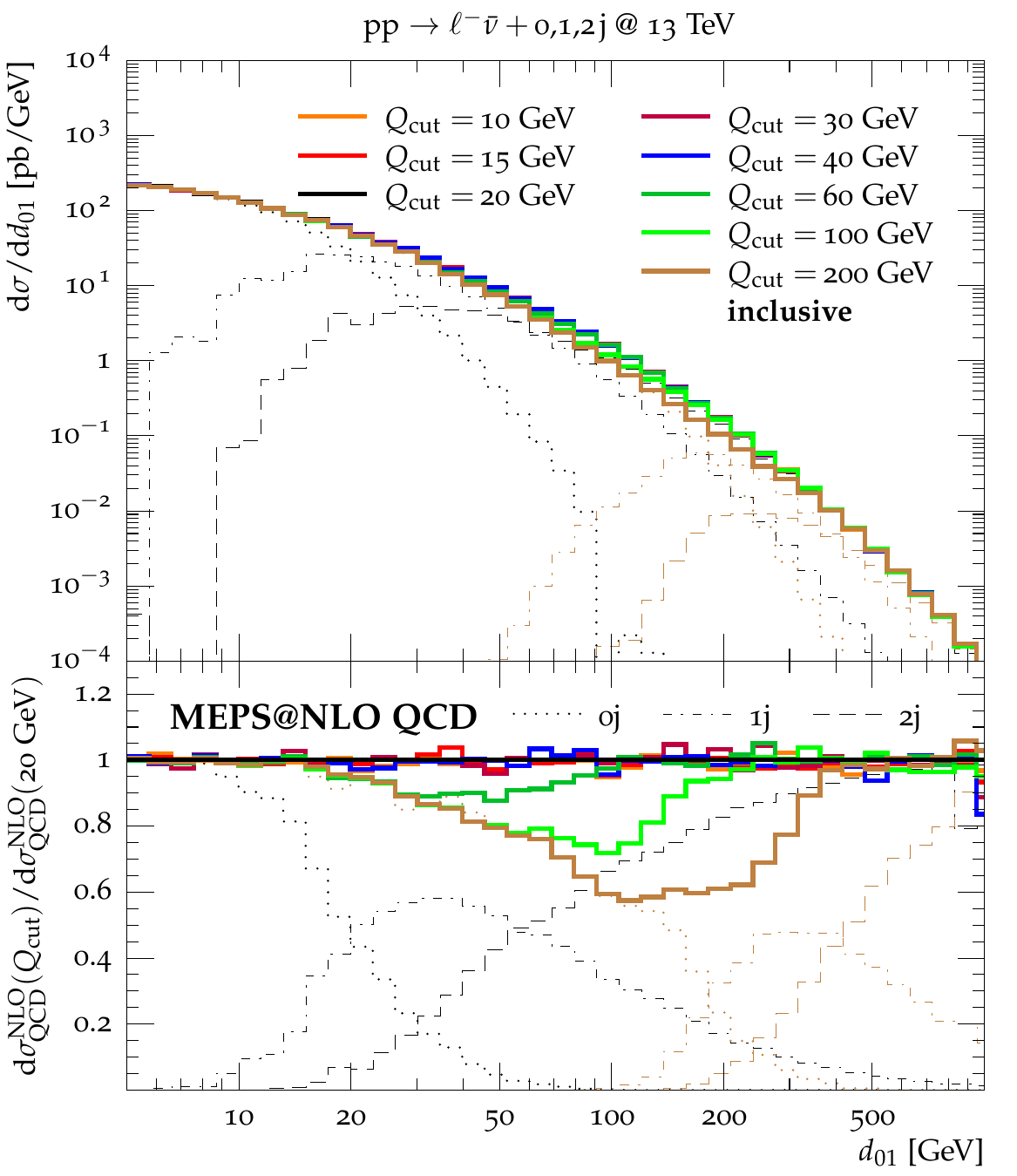}
         \qquad 
   \includegraphics[width=\relplotwidth\textwidth]{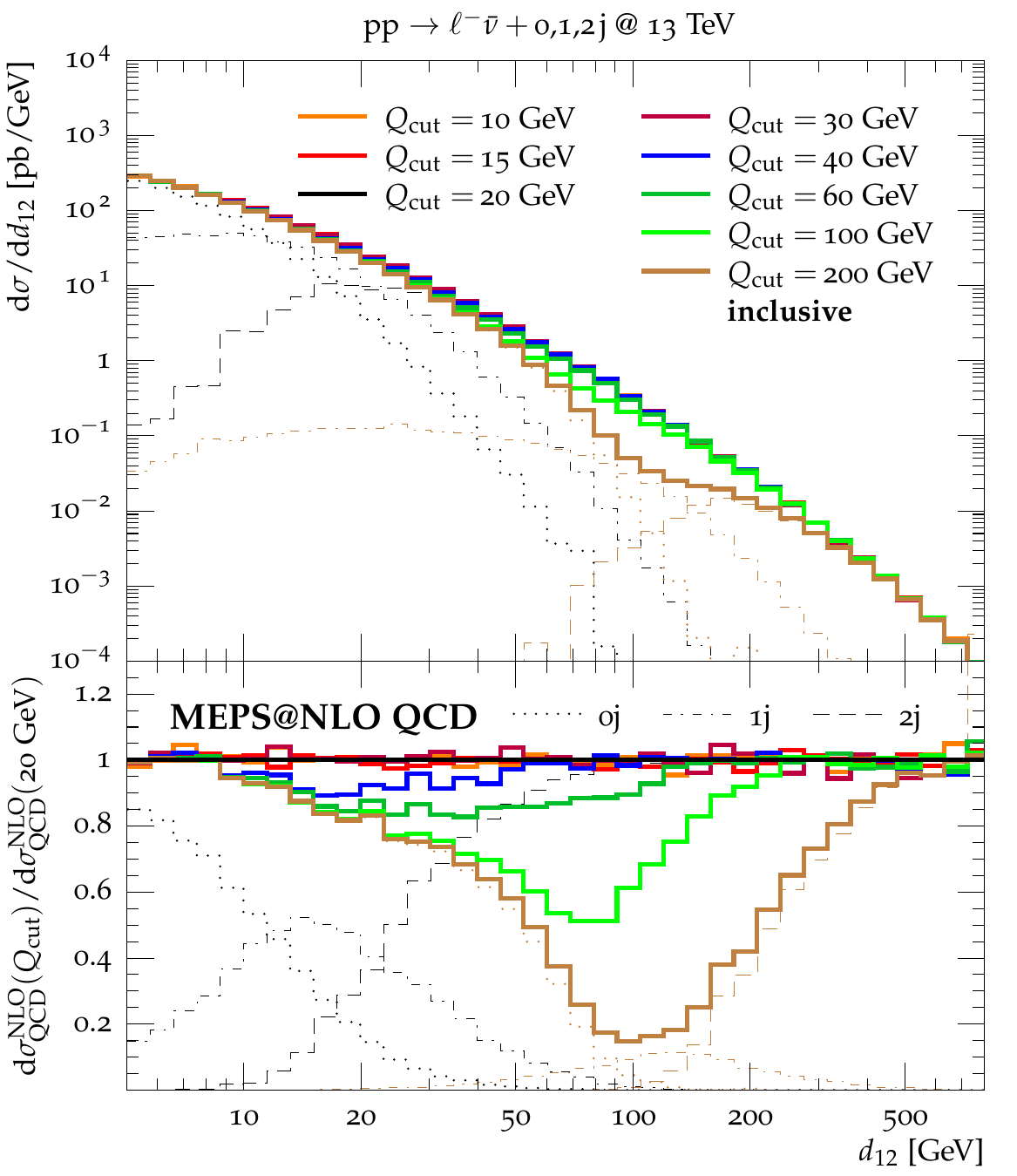}
   \\
   \includegraphics[width=\relplotwidth\textwidth]{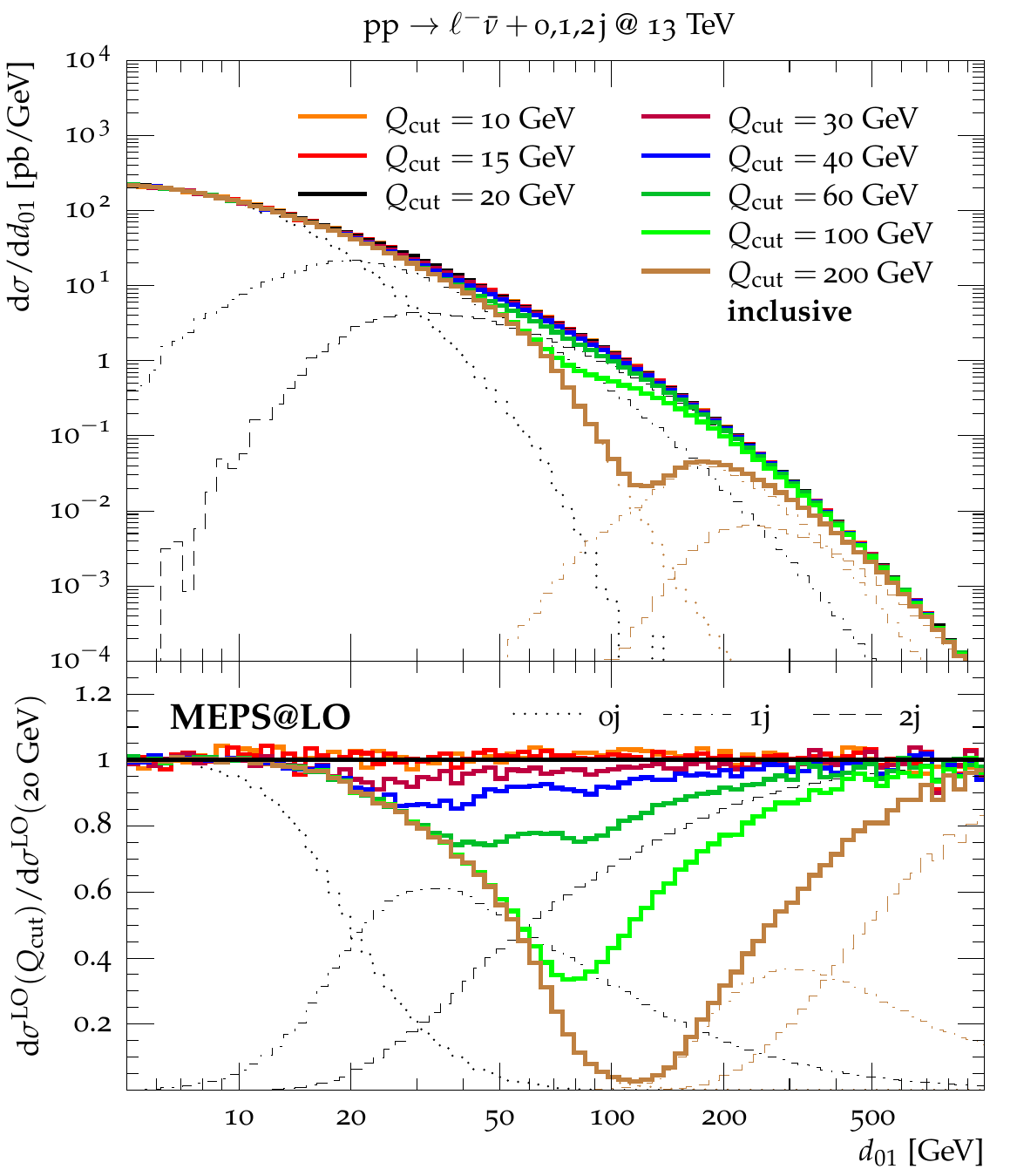}
         \qquad 
   \includegraphics[width=\relplotwidth\textwidth]{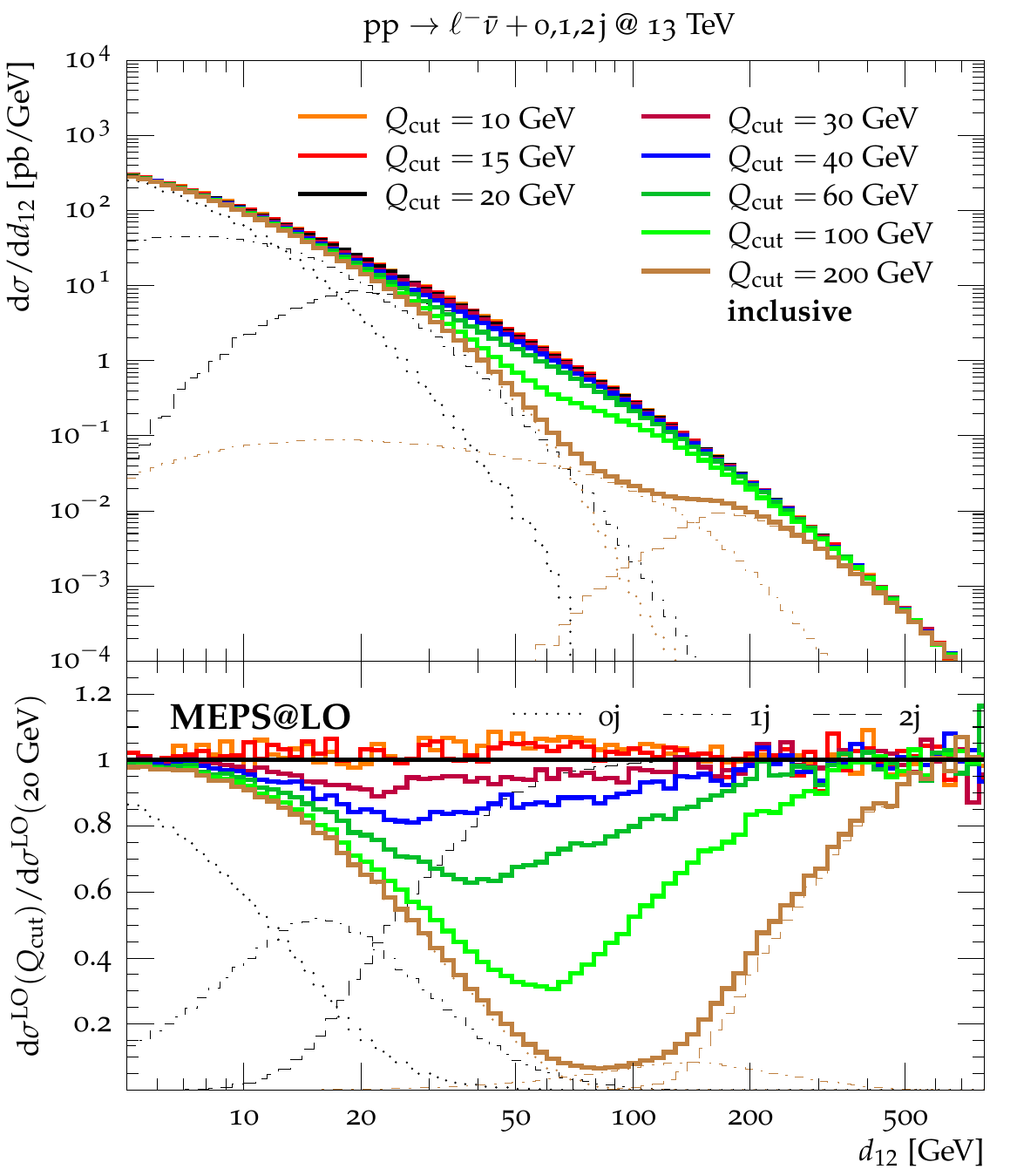}
\caption{
	  Multijet merging systematics of the $1\to 0$ (left) and $2\to 1$ 
	  (right) $k_\perp$ jet resolutions  ($R=0.6$) in $pp\to\ell^-\bar\nu+\text{jets}$ 
	  events in leading order (bottom) and next-to-leading order (top) 
	  multijet merging in the \protect\MEPS scheme. Only basic lepton 
	  acceptance cuts are applied. The contributions 
	  of the individual jet multiplicities are indicated by dotted, 
	  dashdotted and dashed lines for $\qcut=20$ and $200$ GeV.
	}
\label{fig:MEPS_Wen_systs_dij_inc}
\end{figure*}

\begin{figure*}[t!]
\centering
   \includegraphics[width=\relplotwidth\textwidth]{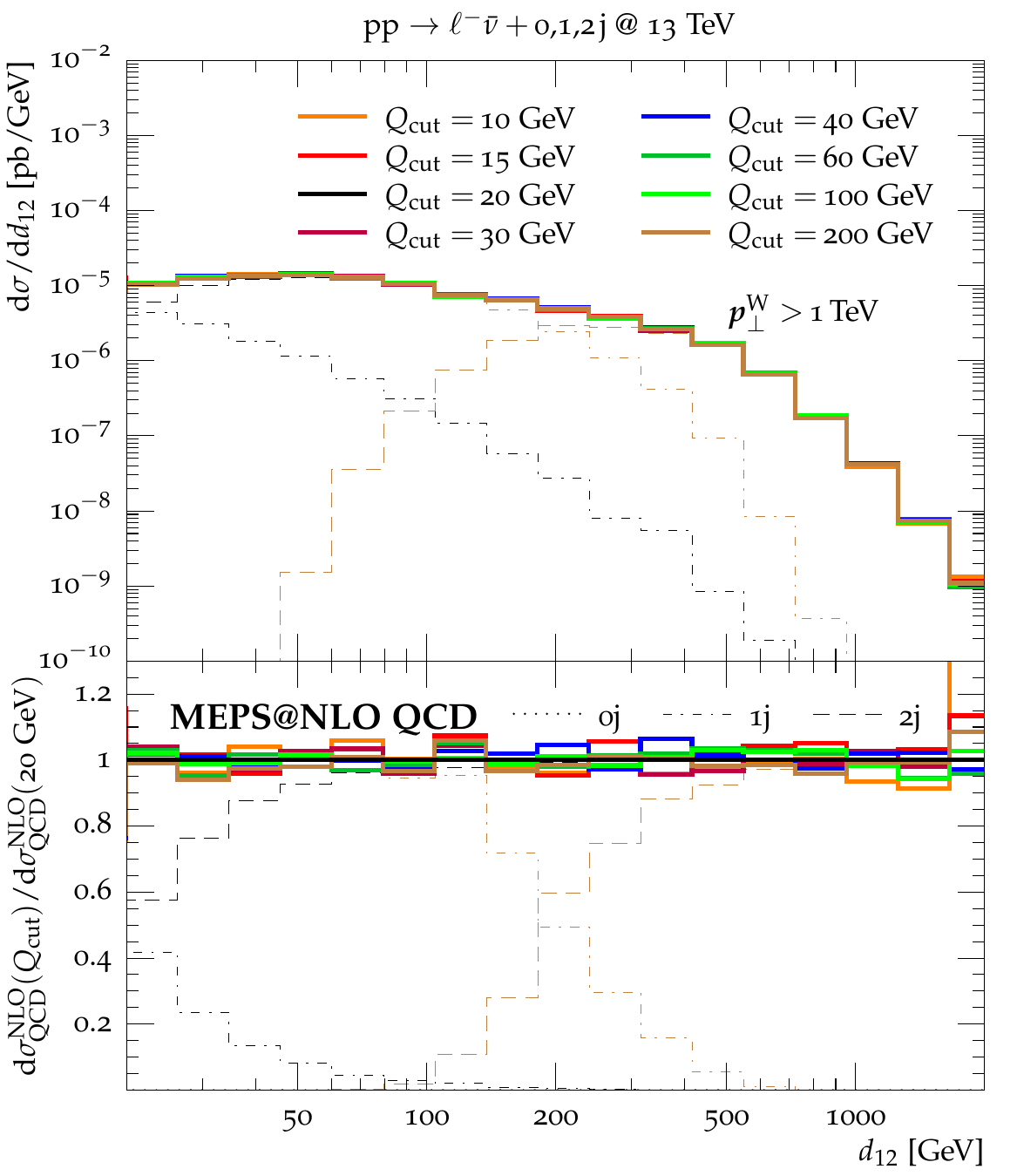}
         \qquad 
   \includegraphics[width=\relplotwidth\textwidth]{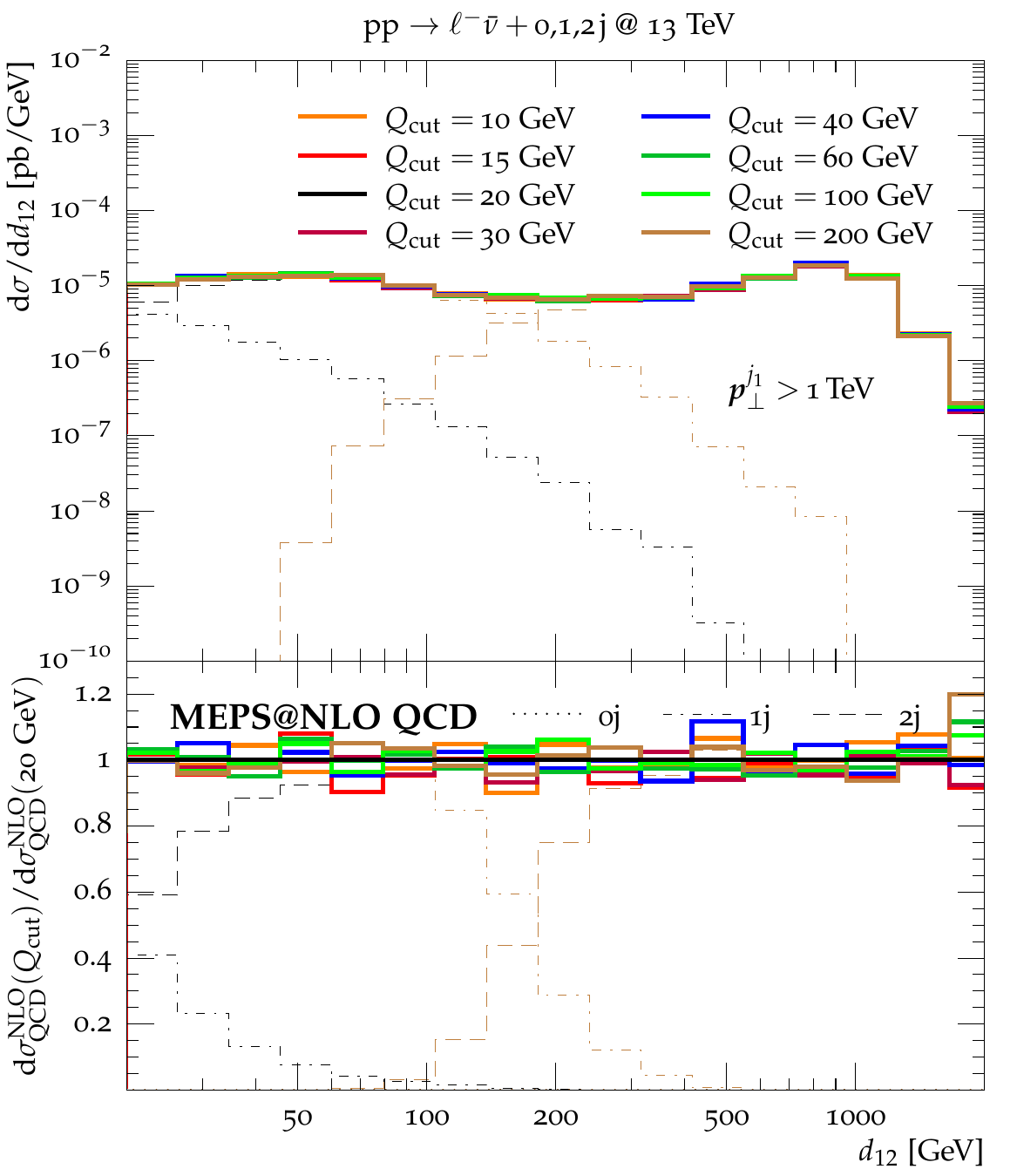}\\
   \includegraphics[width=\relplotwidth\textwidth]{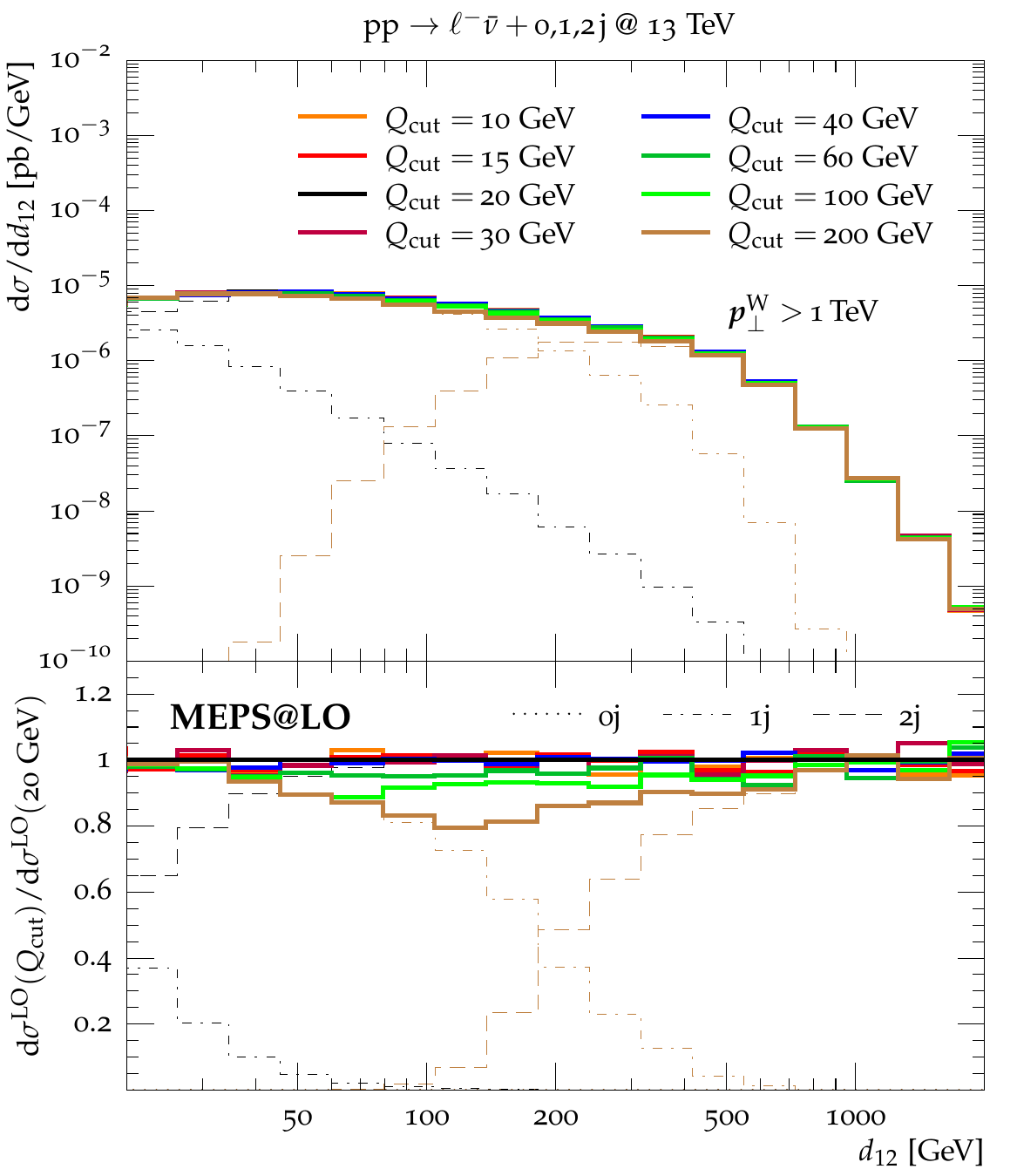}
         \qquad 
   \includegraphics[width=\relplotwidth\textwidth]{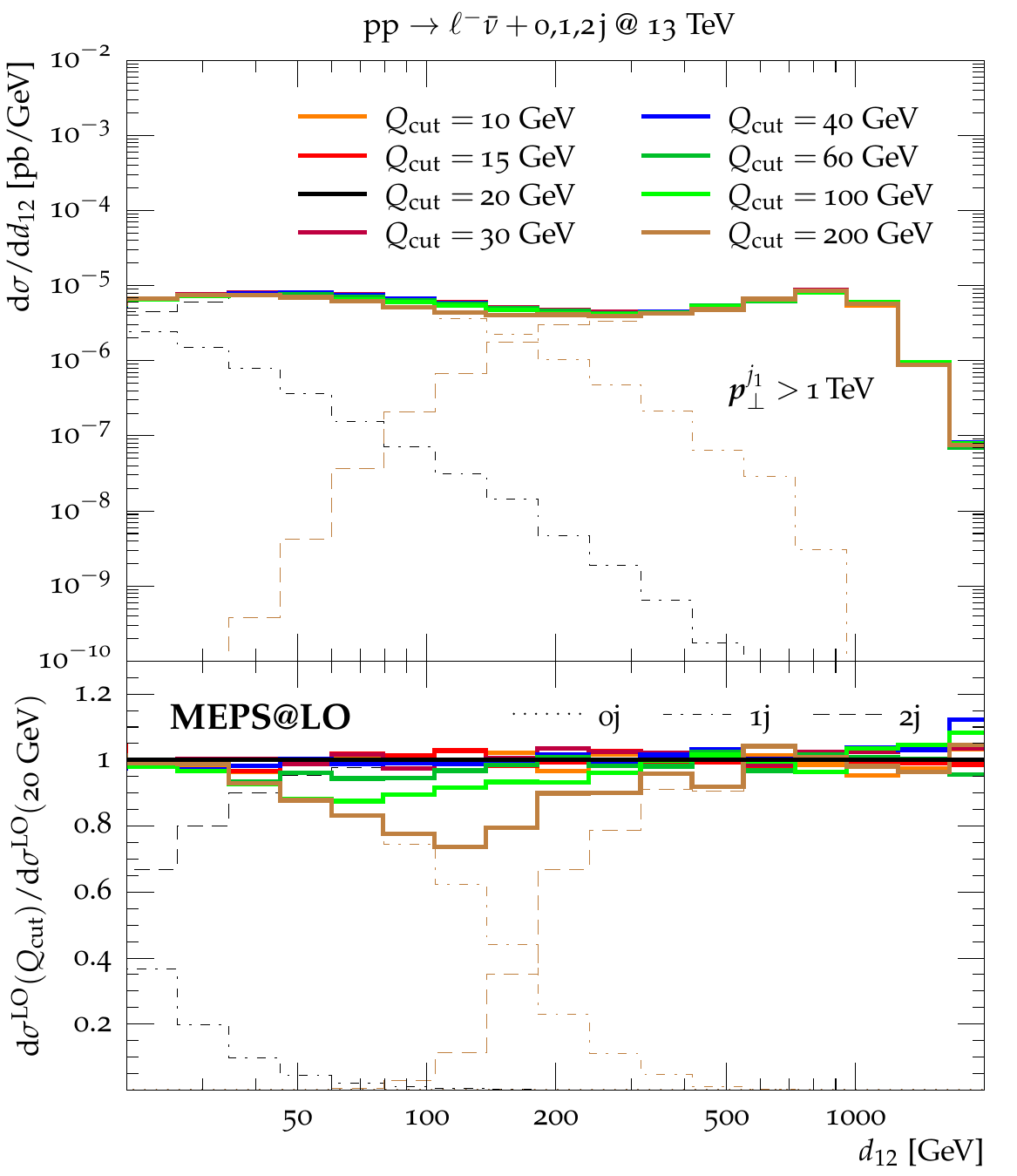}
\caption{
	  Multijet merging systematics of the $2\to 1$ $k_\perp$ jet 
	  resolution  ($R=0.6$) in $pp\to\ell^-\bar\nu+\text{jets}$ in events with a 
	  reconstructed $W$ boson (left) or leading jet (right) with 
	  $p_\perp>1\,\text{TeV}$ in leading order (bottom) and 
	  next-to-leading order (top) 
	  multijet merging in the \protect\MEPS scheme. Only basic lepton 
	  acceptance cuts are applied. The contributions 
	  of the individual jet multiplicities are indicated by dotted, 
	  dashdotted and dashed lines for $\qcut=20$ and $200$ GeV.
	}
\label{fig:MEPS_Wen_systs_dij_tev}
\end{figure*}

\begin{figure*}[t!]
\centering
   \includegraphics[width=\relplotwidth\textwidth]{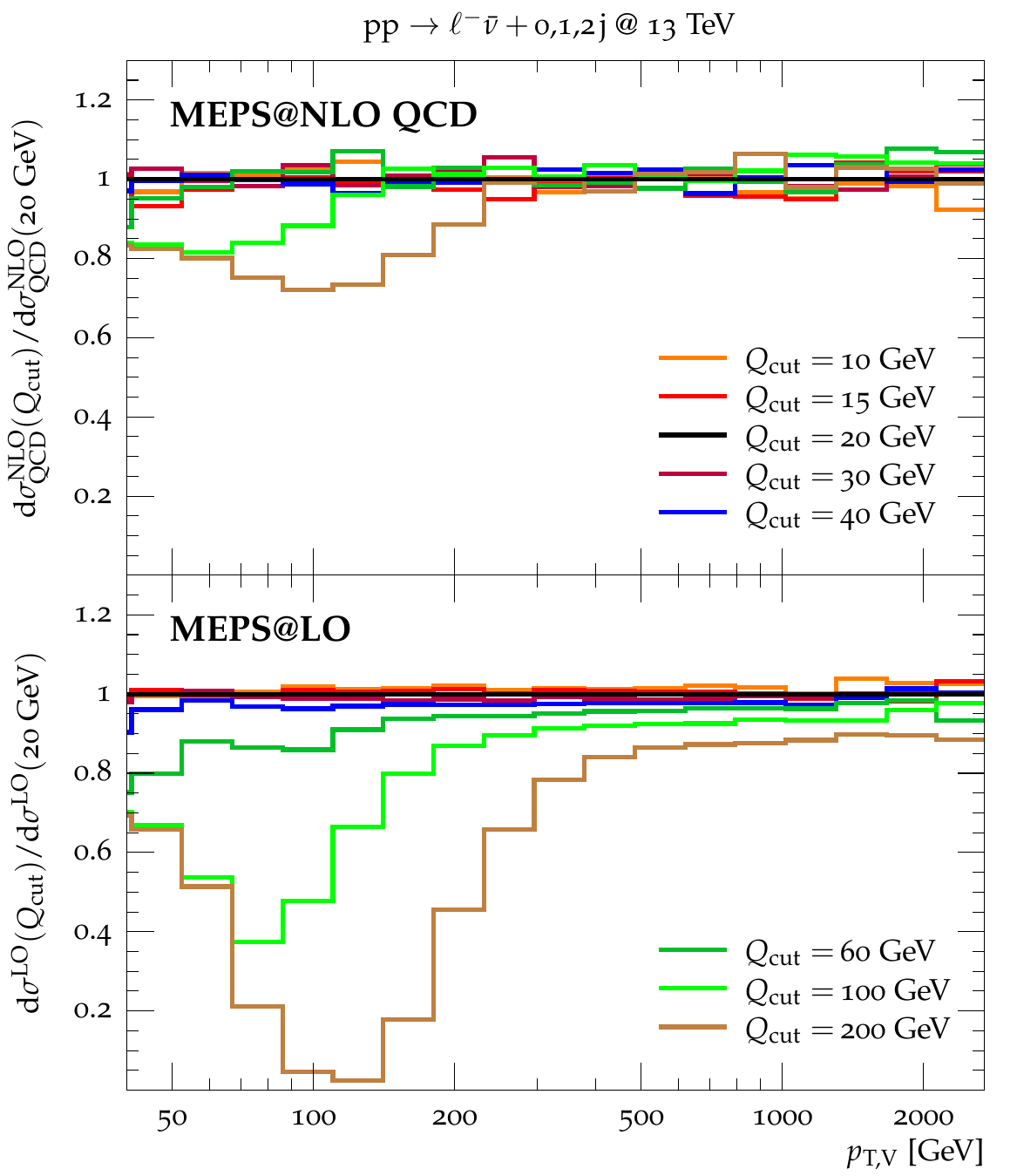}
         \qquad 
   \includegraphics[width=\relplotwidth\textwidth]{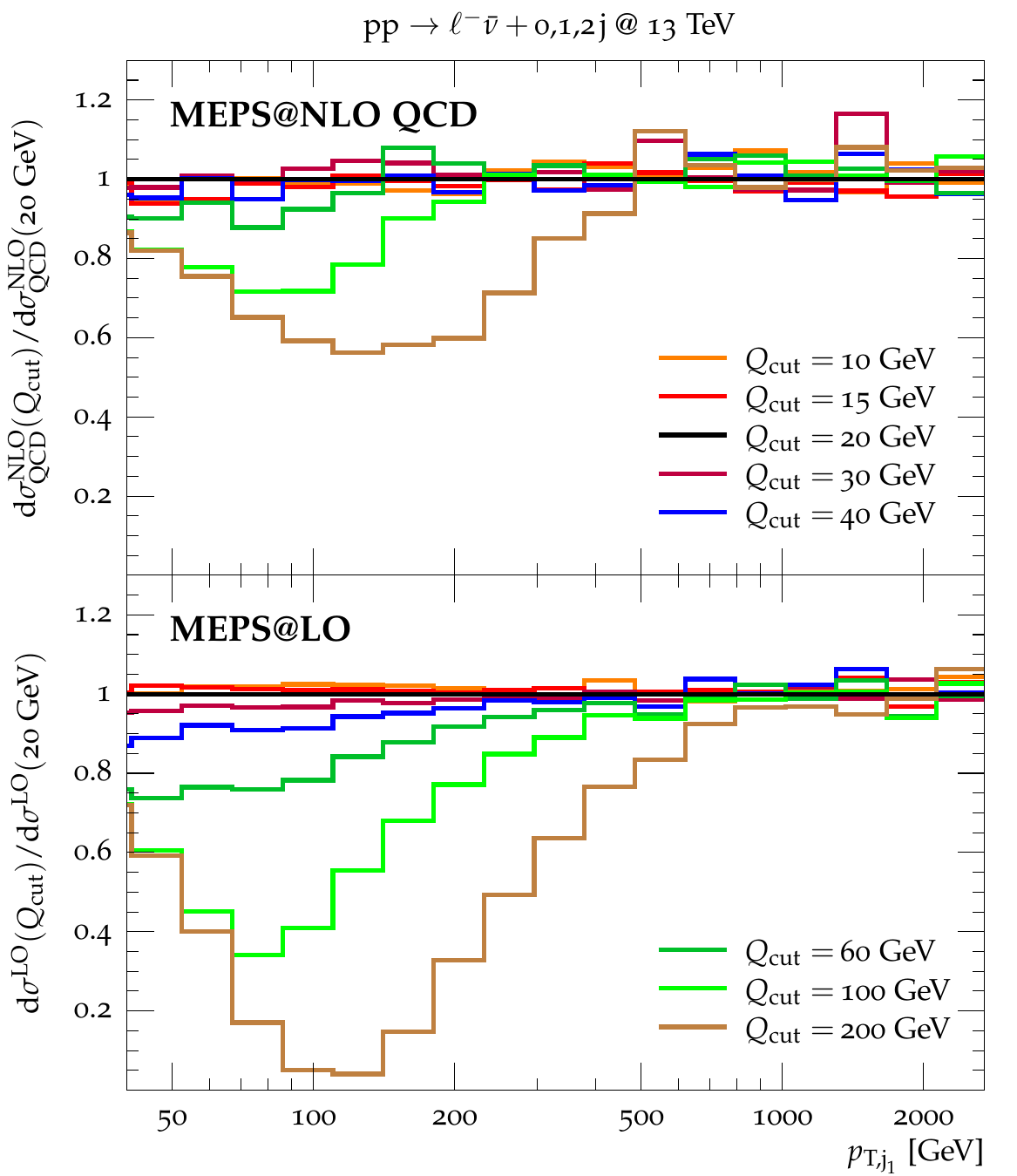}\\
   \includegraphics[width=\relplotwidth\textwidth]{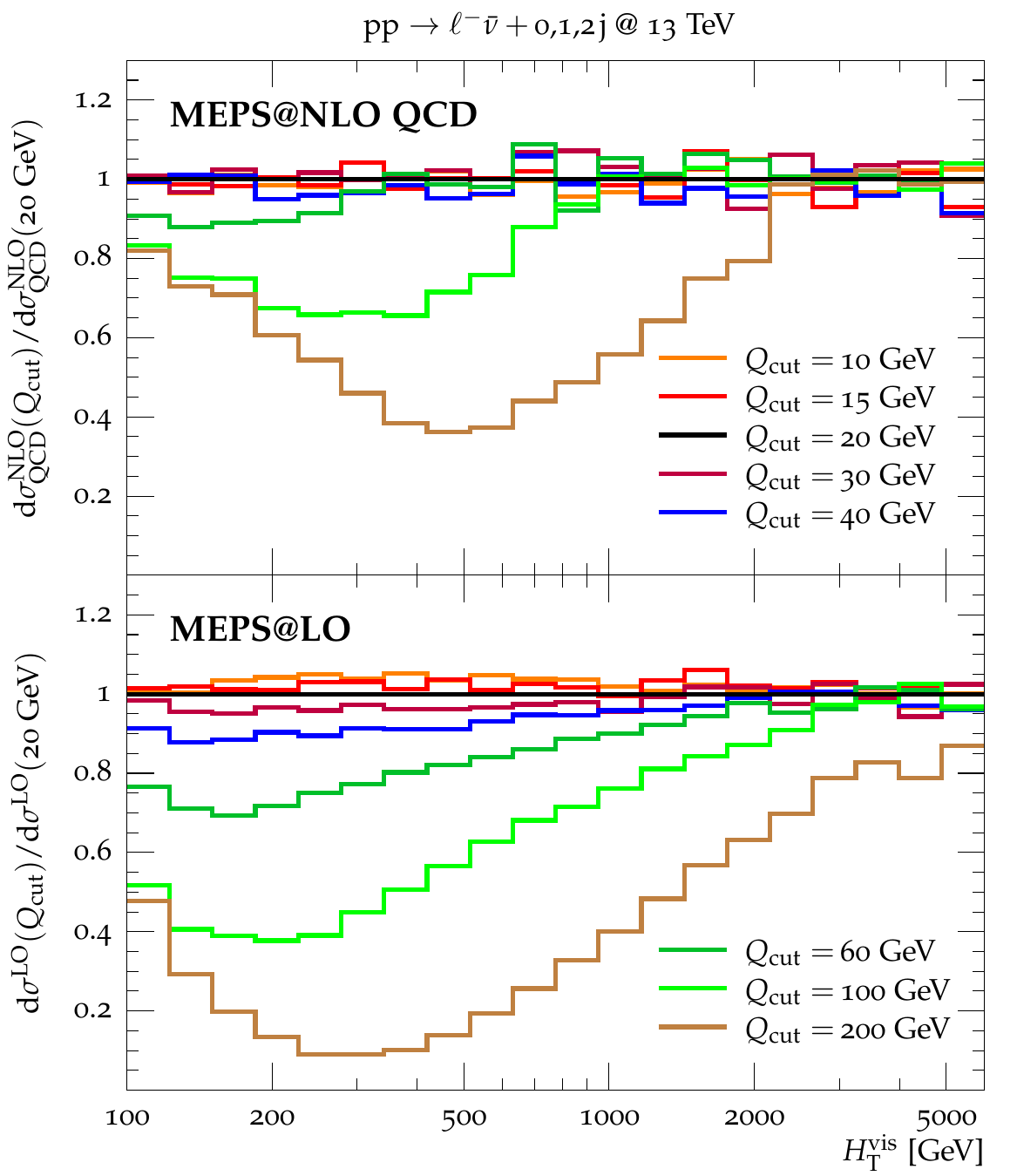}
\caption{
	  Multijet merging systematics for transverse momentum of the 
	  reconstructed vector boson(top left), the hardest jet (top right) 
	  and the total visible transverse energy $\HTvis$ (bottom) with 
	  standard cuts.
	}
\label{fig:MEPS_Wen_systs_obs}
\end{figure*}

This appendix aims at quantifying uncertainties associated with the 
merging scale dependence of the results presented in \refse{se:results_merged}.  This
dependence has never been studied in the literature for observables in the
multi-TeV range, where the combination of very high energies and a small merging
scale could result in large spurious logarithms.
Indeed, the power counting of \cite{Hoeche:2012yf, Gehrmann:2012yg} 
shows that inclusive \MEPS predictions involve uncancelled logarithms of
the generic form
\begin{equation}
\label{eq:qcutdependence}
  \begin{split}
    \text{\MEPSatLO}:\;\frac{1}{N_C}\,\alphaS\log\frac{\mu_Q}{\qcut}
    \qquad\text{and}\qquad
    \text{\MEPSatNLO}:\;\frac{1}{N_C}\,\alphaS^2\log^3\frac{\mu_Q}{\qcut}\;,
  \end{split}
\end{equation}
where $\mu_Q$ and $\qcut$ denote the resummation and the merging scale,
respectively.
They arise in $n$-jet observables as a result of the partitioning 
of the phase space of the extra emission in an unresolved region ($Q_{n+1}<\qcut$) and a
resolved region ($Q_{n+1}>\qcut$), which are described, respectively, in
terms of $n$-jet and $n+1$-jet (N)LO matrix elements combined with the
parton shower in the \MEPS approach.  When the $(n+1)$-th emission is integrated
out, the logarithms of $\qcut$ that originate from both regions cancel to a
large extent, but the limited logarithmic accuracy of the parton shower
results in left-over contributions of type \refeq{eq:qcutdependence}.  For moderate values
of $\muq/\qcut$ their impact is small.  However, when
requiring either a vector boson or a jet of 1\,TeV transverse momentum in
inclusive $V+$\,jets production, $\mu_Q$ takes values of a comparable scale,
and the numerical value of the uncancelled logarithms \refeq{eq:qcutdependence} could in principle
exceed the size of renormalisation and factorisation scale variations at
NLO, thereby spoiling the claimed accuracy.

Such a scenario is clearly excluded by the quantitative analysis
of the $\qcut$ dependence of our predictions 
presented in  
\reffis{fig:MEPS_Wen_systs_dij_inc}{fig:MEPS_Wen_systs_obs} for the case of $W+$\,jets production.
Figure~\ref{fig:MEPS_Wen_systs_dij_inc} displays \MEPSatLO and \MEPSatNLO
predictions for the differential $0\to 1$ and $1\to 2$ jet-$k_\perp$ resolution scales, 
$d_{01}$ and $d_{12}$, which represent the most sensitive observables to
merging effects.  They can be regarded as the relative transverse momenta
associated with the emissions of the hardest and second-hardest jet,
respectively. The plots show the sensitivity with respect to 
variations of the merging scale in a very wide range, from 10 to
200\,GeV. In the phase space region below the minimum $\qcut$,
we observe that all computations are in good mutual agreement.
This is due to the fact that, by construction, in this region
the $n\to n+1$ jet resolution scale always corresponds to 
a parton shower emission matched to $n$-jet (N)LO matrix elements.

In the regions where $d_{01}$ and $d_{12}$ are between the minimum and the maximun 
$\qcut$ one can see a significant sensitivity to the merging scale. 
More precisely,
predictions with $\qcut\in [10,30]$\,GeV are very stable, 
while increasing $\qcut$ beyond 30\,GeV gives rise to increasingly 
pronounced and wide dips centered at $d_{ij}\sim \qcut$.
This is due to the fact that in the region above 
30\,GeV the parton shower's soft-collinear approximation ceases to be a 
sufficiently good description.
This feature is clearly more pronounced in \MEPSatLO,
where the emission is entirely given by the parton shower, 
while the problem is significantly alleviated in 
\MEPSatNLO, where the shower emission below $\qcut$
is matched to tree-level matrix elements. 
Nevertheless, merging scales well beyond 50\,GeV start to be
problematic also for NLO merging.
In this regime, it is important to realise that 
$\qcut$ variations in \MEPSatNLO 
(\MEPSatLO) amount to a comparison of LO+PS versus
NLO+PS (PS versus LO+PS) descriptions, where the latter 
are  clearly superior. 
Thus, $\qcut$ should be chosen as small as computationally feasible. 
Note also that, in principle, the deficit of the parton shower in the intermediate 
$k_\perp$ regions could be attenuated by increasing the 
resummation scale $\mu_Q$. However this would alter the
resummation of large logarithms in the region of very small $k_\perp$.
Thus, in order to avoid large shower uncertainties,
it is preferable to keep $\qcut$ below 30--40\,GeV in \MEPSatNLO.
In this case the $d_{01}$ and $d_{12}$ distributions
turn out to be very robust with respect to $\qcut$ variations 
in the whole range from $\ord(1\,\GeV)$ to $\ord(1\,\TeV)$.

The $\qcut$ sensitivity in the TeV region is investigated in more detail in 
\reffi{fig:MEPS_Wen_systs_dij_tev}, which shows distributions in the
$1\to 2$ $k_\perp$ resolution scale in presence of a lower cut of
1\,TeV on the transverse momentum of the $W$ boson or, alternatively, of the
first jet. We find that the $d_{12}$ spectra are remarkably stable---especially 
in \MEPSatNLO but also in \MEPSatLO---with respect to $\qcut$ variation from 10 to 200\,GeV, i.e.~in
a region where $\muq$ and $\qcut$ differ by up to two orders of magnitude.

This high quality of the \MEPSatNLO merging procedure is due, among other things,
to the fact that the implementation of Sudakov effects through
the truncated vetoed parton shower guarantees exactly the same
logarithmic resummation on both sides of the merging cut.
In the domain above $\qcut$ the NLO calculation supplements 
fixed-order terms beyond the parton shower accuracy, which result, upon integration,
in potentially troubling terms of the type \refeq{eq:qcutdependence}.
The fact that these extra contributions remain 
comparably small can be understood by considering the 
de facto quality of 
logarithmic resummation in the parton shower despite its formally 
limited accuracy. For instance, although not fully under control,
subleading colour NLL contributions are 
largely captured by replacing 
$\left.C_{A/F}\right|_{N_C\to\infty}\to \left.C_{A/F}\right|_{N_C=3}$ 
in the parton shower's resummation, otherwise performed in the $N_C\to\infty$-limit. 
Further, the usage of CMW scales in the parton shower~\cite{Catani:1990rr} 
includes dominant contributions of NNLL accuracy through the running of 
$\alphaS$, giving a good numerical reproduction of higher logarithmic 
terms.

Finally, \reffi{fig:MEPS_Wen_systs_obs} examines the $\qcut$-dependence of 
the observables studied in \refse{se:results_merged}, now 
integrating over additional emissions. The uncertainties displayed here 
are dominated by statistical fluctuations 
for a reasonable variation in the range of $\qcut\in[10,40]$\,GeV. Taking 
these fluctuations into account the merging systematics are on a level of 
5\% and are thus not included in the uncertainty estimate of 
\refse{se:results_merged}. As can be seen, if only the TeV range is 
to be studied, $\qcut$ values of up to 200\,GeV can be chosen without 
introducing a large uncertainty in the results. However, only a small 
merging cut as the one used in \refse{se:results_merged} ensures
a reliable prediction for the whole energy range.

\end{appendix}

\clearpage

\bibliographystyle{JHEP}
\bibliography{vjets}

\begin{thebibliography}{100}
\bibitem{munich}
{\sc Munich} is the abbreviation of ``MUlti-chaNnel Integrator at Swiss (CH)
  precision''---an automated parton level NLO generator by S.~Kallweit. In
  preparation.
\end{thebibliography}

\end{document}